\documentclass{article}

\usepackage{arxivsimplified}

\usepackage{moreverb}
\usepackage{cite}
\usepackage{amsmath,amsfonts,amssymb}
\usepackage{subfigure}
\usepackage{float}
\usepackage{color}
\usepackage{multirow}
\usepackage{graphicx}
\graphicspath{{Figures/}}
\usepackage{mathtools}
\usepackage[ruled,vlined]{algorithm2e}
\usepackage{enumitem}

\usepackage{booktabs}

\usepackage[colorlinks,bookmarksopen,bookmarksnumbered,citecolor=red,urlcolor=red]{hyperref}

\newcommand\BibTeX{{\rmfamily B\kern-.05em \textsc{i\kern-.025em b}\kern-.08em
T\kern-.1667em\lower.7ex\hbox{E}\kern-.125emX}}

\usepackage[utf8]{inputenc} 
\usepackage[T1]{fontenc}    
\usepackage{hyperref}       
\usepackage{url}            
\usepackage{booktabs}       
\usepackage{amsfonts}       
\usepackage{nicefrac}       
\usepackage{microtype}      
\usepackage{lipsum}

\usepackage{graphicx}
\usepackage{cite}
\usepackage[dvipsnames]{xcolor}
\usepackage{amsmath}
\usepackage{blindtext}
\usepackage{amsfonts}
\usepackage{multirow}
\usepackage{array}
\usepackage{mathtools}
\usepackage{listings}
\usepackage{xcolor}
\definecolor{codegreen}{rgb}{0,0.6,0}
\definecolor{codegray}{rgb}{0.5,0.5,0.5}
\definecolor{codepurple}{rgb}{0.58,0,0.82}
\definecolor{backcolour}{rgb}{0.95,0.95,0.92}

\lstdefinestyle{mystyle}{
    backgroundcolor=\color{backcolour},   
    commentstyle=\color{codegreen},
    keywordstyle=\color{magenta},
    numberstyle=\tiny\color{codegray},
    stringstyle=\color{codepurple},
    basicstyle=\ttfamily\footnotesize,
    breakatwhitespace=false,         
    breaklines=true,                 
    captionpos=b,                    
    keepspaces=true,                 
    numbers=left,                    
    numbersep=5pt,                  
    showspaces=false,                
    showstringspaces=false,
    showtabs=false,                  
    tabsize=2
}

\usepackage{scalerel,stackengine}
\stackMath
\newcommand\reallywidehat[1]{%
\savestack{\tmpbox}{\stretchto{%
  \scaleto{%
    \scalerel*[\widthof{\ensuremath{#1}}]{\kern-.6pt\bigwedge\kern-.6pt}%
    {\rule[-\textheight/2]{1ex}{\textheight}}
  }{\textheight}%
}{0.5ex}}%
\stackon[1pt]{#1}{\tmpbox}%
}

\title{Sampling and resolution characteristics in reduced order models of shallow water equations: intrusive vs non-intrusive}

\author{
  Shady~E.~Ahmed \\
  School of Mechanical \& Aerospace Engineering,\\
  Oklahoma State University, \\
  Stillwater, Oklahoma - 74078, USA.\\
  \texttt{shady.ahmed@okstate.edu} \\
   \And
 Omer San \\
 School of Mechanical \& Aerospace Engineering,\\
  Oklahoma State University, \\
  Stillwater, Oklahoma - 74078, USA.\\
  \texttt{osan@okstate.edu} \\
  \And
  Diana~A.~Bistrian  \\
  Department of Electrical Engineering and \\Industrial Informatics,\\
  University "Politechnica" of Timisoara,\\
  331128, Hunedoara, Romania.\\
  \texttt{diana.bistrian@fih.upt.ro}\\
  \And
  Ionel~M.~Navon \\
  Department of Scientific Computing,\\
  Florida State University, Tallahassee,\\
  Florida 32306 USA.\\
  \texttt{inavon@fsu.edu}\\
}

\begin{document}
\maketitle

\begin{abstract}
We investigate the sensitivity of reduced order models (ROMs) to training data resolution as well as sampling rate. In particular, we consider proper orthogonal decomposition (POD), coupled with Galerkin projection (POD-GP), as an intrusive reduced order modeling technique. For non-intrusive ROMs, we consider two frameworks. The first is using dynamic mode decomposition (DMD), and the second is based on artificial neural networks (ANNs). For ANN, we utilized a residual deep neural network, and for DMD we have studied two versions of DMD approaches; one with hard thresholding and the other with sorted bases selection. Also, we highlight the differences between mean-subtracting the data (i.e., centering) and using the data without mean-subtraction. We tested these ROMs using a system of 2D shallow water equations for four different numerical experiments, adopting combinations of sampling rates and resolutions. For these cases, we found that the DMD basis obtained with hard threshodling is sensitive to sampling rate, performing worse at very high rates. The sorted DMD algorithm helps to mitigate this problem and yields more stabilized and converging solution. Furthermore, we demonstrate that both DMD approaches with mean subtraction provide significantly more accurate results than DMD with mean-subtracting the data. On the other hand, POD is relatively insensitive to sampling rate and yields better representation of the flow field. Meanwhile, resolution has little effect on both POD and DMD performances. Therefore, we preferred to train the neural network in POD space, since it is more stable and better represents our dataset. Numerical results reveal that an artificial neural network on POD subspace (POD-ANN) performs remarkably better than POD-GP and DMD in capturing system dynamics, even with a small number of modes.
\end{abstract}


\section{Introduction} \label{INTRO}
Despite the gigantic progress in computational power and numerical simulations accuracy, dealing with complex realistic situations is still prohibitive. This is particularly important in geophysical flows (e.g., atmosphere and oceans) characterized by a wide span of spatio-temporal scales. Resolving all of these scales on a fine grid covering the globe is just impossible, even with the largest supercomputer in the world nowadays \cite{kalnay2003atmospheric,powers2017weather}. Reduced order modeling (ROM) offers an alternative to handle these complex simulations in a more efficient way. In general, ROM presents approximate substitutes to the original system. Those substitutes should be cheaper to solve and analyze while retaining similar dynamics to the original system \cite{quarteroni2015reduced, taira2017modal}. Fortunately, this is feasible given the fact that these complex systems are often dominated by a few number of underlying manifolds controlling most of the mass, momentum, and energy transfer. The idea of ROM is simple; \textit{instead of `watching' each grid point, we may just follow a small number of features and still be able to know equivalent amount of information}.

In other words, rather than simulating every single point (which represents a degree of freedom in computations), ROM enables us to detect the major directions over which most variations occur. By superposing the contributions of a small number of these directions, an approximate `numerical twin' of the original system is obtained. The remaining directions are cut-off assuming their contributions are minimum. Dealing with that truncated numerical twin saves a large portion of computational power, time, and storage. This is especially evident when multiple forward simulations are required. Examples include data assimilation \cite{daescu2007efficiency,navon2009data,cao2007reduced,he2011use,houtekamer1998data,houtekamer2001sequential,bennett2005inverse,evensen2009data,law2012evaluating,buljak2011inverse}, optimal control \cite{ito2001reduced,mcnamara2004fluid,bergmann2008optimal,ravindran2000reduced,graham1999optimala,graham1999optimalb,brunton2015closed}, parameter identification \cite{fu2018pod,boulakia2012reduced,kramer2017feedback}, and uncertainty quantification \cite{galbally2010non,biegler2011large,mathelin2005stochastic,smith2013uncertainty,najm2009uncertainty}.

Model order reduction was introduced to fluid dynamics community many decades ago through the work of Lumley \cite{lumley1967structure} as a mathematical technique to extract coherent structures from turbulent flow fields. Different ROM techniques vary in the way they define these underlying structures, either growing/decaying with a specific rate, oscillating with a constant frequency or containing a significant amount of kinetic energy. Living in the era of big data, snapshot-based projection methods have gained the largest popularity, where the response of a system, either recorded from experiments or high-fidelity numerical simulations given a certain input, is assumed to contain the essential behavior of that system.

Proper orthogonal decomposition (POD), also known as principal component analysis \cite{pearson1901lines}, computes a set of orthonormal spatial basis vectors which describe the main directions (modes), by which the given flow field is characterized, in the $L_2$ sense \cite{berkooz1993proper}. Each of these modes contains a proportion of the system's total energy. The most energetic POD modes are selected to generate the reduced order twin. Often, POD is coupled with Galerkin projection (GP) to generate ROMs for linear and nonlinear systems (POD-GP) \cite{ito1998reduced,iollo2000stability,rowley2004model,pinnau2008model, stankiewicz2008reduced,sachs2010pod,akhtar2009stability,barone2009stable}. A recent mathematical discussion on POD-based ROMs is given in Ref.\cite{grassle2019model} for nonlinear parametric and time-dependent PDEs. In brief, GP makes use of the orthonormality of the POD modes to reduce the full order model (FOM) (i.e., governing equations) from partial differential equations (PDEs) to a simpler set of ordinary differential equations (ODEs) involving the POD temporal coefficients. These ODEs can be solved with much more simplicity than the original set of PDEs. In practice, ROM approximation can lead to speedups of several orders of magnitude as well as great reduction in memory requirements \cite{milk2016pymor,puzyrev2019pyrom} with minor impact on accuracy of numerical solution.

The spatiotemporal biorthogonal decomposition, a variant of POD, has been introduced to recover the temporal information associated with the POD modes \cite{aubry1991spatiotemporal,aubry1991hidden}. A spectral POD (SPOD) has been applied to extract low-dimensional compression of the full order model (FOM) extending POD to account for temporal dynamics in addition to energetic optimality \cite{sieber2016spectral}. Although the spatial SPOD modes are no longer orthonormal, it was shown that the norm of the spatial modes gives further insights into the dataset and reduces POD in the limiting case. The name `SPOD' was also given to the frequency-based POD where standard POD algorithm is applied to a set of realizations of the temporal Fourier transform of the flow field \cite{picard2000pressure,taira2017modal,towne2018spectral}. Loiseau et al. \cite{loiseau2018sparse} have also introduced a manifold modeling approach as a potentially viable compression model. They showed that a two-dimensional manifold is more accurate than an expansion using 50 POD modes for the transient cylinder wake. Another variation of POD in linear cases is balanced proper orthogonal decomposition, which extracts two sets of modes for specified inputs and outputs, corresponding to the controllable states, and the observable states, respectively. Balanced truncation of linear systems has been presented by combining POD and balanced realization theory \cite{willcox2002balanced}. In ROM, the aim is to retain both the most controllable modes and the most observable modes \cite{moore1981principal}. However, for some systems, states that have very small controllability might have very large observability, and vice versa. Balancing involves determining a coordinate system in which the most controllable directions in state space are also the most observable directions \cite{rowley2005model,singler2009proper,singler2012balanced}. The modes that are least controllable/observable are then truncated. 

Projection-based techniques (e.g., POD-GP) belong to a family of intrusive reduced order modeling, also called physics-informed ROM, since they require knowledge of the high-order governing equations, and projecting them onto the reduced basis to obtain a set of ordinary differential equations (ODEs). This limits their use to systems which we fully understand and can formulate in a mathematical representation \cite{lassila2014model}. Also, as the projection depends on governing equations, the source codes have to be modified for even slightly different cases, making the process inconvenient for complex nonlinear problems. Moreover, in fluid applications, POD-GP results in a dense system consisting of triadic interactions due to the quadratic nonlinearity with an order of $O(R^3)$ computational load, where $R$ refers to the retained number of modes. This poses a constraint on the number of modes to be selected. Several efforts have been devoted to minimize the number of selected modes through closure modeling to reduce the amount of information lost due to truncation \cite{Rahman2019dynamic,osth2014need,san2013proper,wang2012proper,borggaard2011artificial}. Stefanescu and Navon \cite{cstefuanescu2013pod} employed the discrete empirical interpolation method (DEIM) to reduce the computational complexity of the POD-GP reduced order model caused by its dependence on the nonlinear full dimension model. Significant CPU gain was obtained using POD/DEIM approach compared to only POD for both implicit and explicit numerical schemes. Also, Wang et al. \cite{wang20162d} studied POD/DEIM approach on the 2D Burgers equation at high Reynolds number and found a speed-up by a factor of $O(10)$ in CPU time compared to conventional POD-GP. To recover the missing information and account for the small-scale dissipation effects of the truncated POD modes, they introduced a Tikhonov regularization to build a calibrated ROM. In \cite{grassle2019model}, linearizing and projecting the nonlinearity onto the POD space was proposed, allowing for solving the resulting evolution equations explicitly without spatial interpolation.

Another class for ROM which is gaining popularity in recent years, is the purely data-driven, non-intrusive ROMs. That is a family of methods that solely access datasets to identify system's dynamics, with little-to-no knowledge of the governing equations, sometimes called physics-agnostic modeling. Dynamic mode decomposition (DMD) is one of the very popular techniques in this area, relying on Koopman theory. Mezic \cite{mezic2005spectral} was the first to apply the Koopman theory for the purposes of reduced order modelling. Later, Schmid and Sesterhenn \cite{schmid2008dynamic} introduced the DMD algorithm which is considered the numerical implementation of the Koopman modal decomposition. In DMD, time-resolved data are decomposed into modes, each having a single characteristic frequency of oscillation and growth/decay rate. There are a number of variants for DMD algorithms. The monograph written by Kutz et al. \cite{kutz2016dynamic} collects various DMD methods and their implementations. Hemati et al. \cite{hemati2014dynamic} proposed an online-DMD algorithm that is capable of handling large and streaming datasets. The algorithm reads consecutive pairs of snapshots, whenever available, and updates the DMD modes accordingly, while keeping track of historical datasets with minimal storage requirements. In order to achieve a desirable trade-off between the quality of approximation and the number of modes that are used to approximate the given fields, Jovanovi{\'c} et al. \cite{jovanovic2014sparsity} developed a sparsity-promoting DMD. In their algorithm, sparsity is induced by regularizing the least-squares deviation between the data matrix and a linear combination of DMD modes with an additional term that penalizes the $L_1$-norm of the vector of DMD amplitudes. A multiresolution DMD (mrDMD) was proposed by Kutz et al. \cite{kutz2016multiresolution} employing one-level separation recursively to separate background (low-rank) and foreground (sparse) dynamics. This technique has been recently extended to treat a broader variety of multiscale systems and more faithfully reconstruct their isolated components \cite{dylewsky2019dynamic}.

An optimized DMD method which tailors the decomposition to a desired number of modes was introduced by Chen et al. \cite{chen2012variants}. Dang et al. \cite{dang2018optimized} proposed a denoising and feature extraction algorithm for multi-component coupled noisy mechanical signals by combining multiscale permutation entropy with sparse optimization based on non-convex regularization. Bistrian and Navon \cite{bistrian2017randomized,bistrian2018efficiency} utilized an adaptive randomized DMD to obtain a reduced basis in the offline stage, and the temporal values in the online stage through an interpolation using radial basis functions. Also, they provided an improved selection method for the DMD modes and associated amplitudes and Ritz values, that retains the most influential modes and reduces the size of the model \cite{bistrian2015improved,bistrian2017method}. In their method, modes are arranged in descending order of the energy of the DMD modes weighted by the inverse of the Strouhal number. Insignificant modes are truncated based on the conservation of quadratic integral invariants by the reduced order flow. Higham et al. \cite{higham2018implications} suggested that DMD is more favorable than POD in situations where that nonlinear dynamics can arise in the transition of underlying structures to a quasi-two-dimensional behaviour. However, Taira et al. \cite{taira2017modal} referred that DMD often yields less reliable results when the system's nonlinearity is significant unless more measurements are included in the snapshots. Theoretical work on DMD has centered mainly on exploring connections with other methods, such as Koopman spectral analysis \cite{rowley2009spectral,mezic2013analysis,bagheri2013koopman}, POD \cite{schmid2010dynamic}, and Fourier analysis \cite{chen2012variants}. Lu and Tartakovsky \cite{lu2019predictive} provides a theoretical error estimator of DMD predictability for  linear and nonlinear parabolic equations.

There have been many efforts to hybridize and bridge the gap between POD (energy-based decomposition) and DMD (frequency-based decomposition). A single-frequency POD method (F-POD) has been proposed by Zhao et al. \cite{zhao2019modified} to re-decompose the POD modes from the perspective of frequency by utilizing discrete Fourier transform to enable dynamic analysis. The associated modal growth rate information is obtained through a band of adjacent frequencies centered about some isolated peaks. A combination of POD and DMD was proposed by Cammilleri et al. \cite{cammilleri2013pod} performing DMD on the temporal coefficients of the POD modes to force purely harmonic temporal modes, while keeping the main energy optimality of the POD. Noack et al. \cite{noack2016recursive} used a recursive DMD approach to extract oscillatory modes from the data sequence, a feature of DMD, while ensuring the orthogonality of the modes and minimization of the truncation error, features of POD. First, DMD is employed on the dataset, and a normalized DMD mode is chosen such that it minimizes the averaged error. The following modes are constructed recursively from the orthogonal residual of the Galerkin expansion with the preceding modes. This results in an orthonormal set of modes while retaining the monochromatic feature of DMD. Alla and Kutz \cite{alla2017nonlinear} proposed the use of DMD for the approximation of nonlinear terms in POD-based projection framework. Mendez et al. \cite{mendez2019multi} proposed a multi-scale POD (mPOD) that combines multi-resolution analysis with POD. The multi-resolution analysis splits the correlation matrix into the contributions of different scales. Then, the temporal mPOD basis functions are computed by setting frequency constraints to the classical POD. Williams et al. \cite{williams2013hybrid} proposed a hybrid PDE/ROM integrator to allow for online ROM generation based on a comparison between POD-GP and DMD results. When they agree to some limit, the forward simulation is pursued in the reduced space. Otherwise, the original PDE is solved in full space for a specific time interval to update the POD and DMD basis functions.

Recently, machine learning (ML) tools, especially artificial neural networks (ANNs) have been playing a significant role in different fields of science and engineering \cite{brunton2019data,kutz2017deep,al2003truncated}. The merits of machine learning algorithms have been demonstrated for prototypical fluid mechanics applications, including flow modeling, reconstruction, and control \cite{milano2002neural,fukami2019super,erichson2019shallow,morton2018deep,kim2019deep}. This is due to their superior capability to identify the linear and non-linear maps between input and output data as well as extracting underlying patterns \cite{bishop2006pattern,bishop1995neural,ripley1996pattern,narendra1990identification}. For model order reduction purposes, neural networks can be used as an auto-encoder \cite{baldi1989neural,erichson2019physics,otto2019linearly,lee2018model}, composed mainly of two parts; encoder and decoder. For the encoder, the input is the full order flow field, while the output is a reduced order representation of the field (i.e., using a smaller number of neurons). This represents the projection step in ROM paradigm. The decoder does the opposite job, taking the reduced representation and recovering back the flow field (i.e., reconstruction). However, it has been noted that the autoencoder space is not optimal in terms of orthonormality, order, and conserving physical quantities. Alternatively, ANNs capability of identifying maps can be hybridized with POD basis optimality (POD-ANN) \cite{narayanan1999low,san2019artificial,pawar2019deep}. San et al. \cite{san2019artificial} proposed a supervised machine learning framework for the non-intrusive model order reduction of unsteady fluid flows to provide accurate predictions of state variables while varying the control parameter values. Instead of using the flow field as an input, a pre-processing step is conducted by projecting the field onto the POD modes to obtain temporal modal coefficients, reducing memory requirements. Then, ANNs can be exploited to predict the evolution of these coefficients with time. Finally, a post-processing step is employed to reconstruct the flow field. In other words, based on a POD compression, ANN is used to bypass Galerkin projection and solving the resulting ODEs. This allows the use of POD in a non-intrusive ROM framework, eliminating the need to access the governing equations. Also, neural networks can be utilized for closure modeling and correction of ROMs \cite{san2018neural,san2018machine,rahman2018hybrid,mohebujjaman2019physically,hijazi2019data}.

Since data amounts witness tremendous expansion, it is increasingly important to assess which dataset better represents the system. This is particularly important in snapshot-based ROM since these datasets are used to distill the main dynamics and influences further use of the obtained model. Therefore, in the present paper, we investigate ROM sensitivity to the training data, in terms of resolution and sampling rate. We choose the two-dimensional (2D) shallow water equations (SWEs) as our testbed. This should provide more insights and guidelines when formulating ROMs, especially for convective, nonlinear flows. Four numerical experiments were performed using two different resolutions and sampling rates. POD and DMD basis functions are computed for these four experiments, and ROM performance is assessed accordingly. For DMD, we studied two approaches for basis computation and selection. One of them is based on hard thresholding without further sorting and selection, and the other one uses a sorting criteria to select the most influential DMD modes. For the present dataset, hard thresholding DMD was found to suffer from overfitting issues when using a relatively high sampling rate. Sorted DMD algorithm helps in reducing this problem by truncating the modes with low amplitudes and low growth/decay rates. On the other hand, POD does not suffer from this problem and gave almost matching results for different sampling rates. Resolution (i.e., full order 2D field) was found to have negligible effect as long as it carries sufficient information about the flow, although lower resolution is observed to mitigate the overfitting problems resulting from unnecessarily high sampling rate. We investigated the effect of data preparation and centering on ROM performances. We repeated the same four experiments with centered datasets (through mean-subtraction) and uncentered datasets (without mean-subtraction). Although POD-based ROMs gave similar results for both cases, DMD performed significantly better for data without mean-subtraction.

The rest of the paper is organized as follows. In Section~\ref{sec:SWE}, our testbed, the two-dimensional shallow water equations, is described. Governing equations as well as initial and boundary conditions are introduced. Also, numerical schemes used to solve these equations are presented briefly. Reduced order modeling techniques are given in Section~\ref{sec:ROM}, where POD and DMD basis computations are explained. Also, the use of POD with either Galerkin projection as an intrusive ROM or artificial neural networks as a non-intrusive ROM is discussed. Results are demonstrated in Section~\ref{sec:results} by illustrating the geopotential flow field obtained after a one-day prediction using POD-GP, POD-ANN, and DMD, compared to the true projection for different number of modes. Also, time series prediction of POD modal coefficients are presented to compare the ANNs prediction capability with its POD-GP counterpart. Finally, concluding remarks are drawn in Section~\ref{sec:conclusion}

\section{Shallow Water Equations}
\label{sec:SWE}
Most atmospheric and oceanic flows feature horizontal length scales which are much larger than the vertical length scale (i.e., layer depth) \cite{vallis2006atmospheric}. Therefore, representing those systems using shallow water equations (SWEs) has been a common practice in geophysical fluid dynamics studies. Conventionally, SWEs describe a thin layer of constant density fluid in hydrostatic balance, bounded from below by a rigid surface and from above by a free surface. Above that free surface, another fluid of negligible inertia is assumed to exist. 

Conceptually, to derive SWEs, the classic Navier-Stokes equations are integrated, and averaged across the vertical height, and pressure term is substituted by potential height, through the hydrostatic approximation. As for most geophysical flows, Coriolis term is included to account for forces due to Earth's rotation. This results in a set of simplified one-dimensional equations (in the case of 1D SWEs) or two-dimensional equations (as in 2D SWEs). In the present study, we consider the latter case in inviscid form. SWEs have been used as a testbed for many geophysical fluid dynamics studies, since they provide valuable physical and numerical insights, while avoiding the computational burden of the general primitive equations. For a single-layer inviscid shallow water equations, also called Saint-Venant equations, the momentum and mass conservations can be written as follows, \cite{bistrian2015improved}
\begin{align}
\dfrac{\mbox{D}\mathbf{u}}{\mbox{D}t} + \mathbf{f} \times \mathbf{u} &= -\nabla \eta , \\
\dfrac{\mbox{D}\eta}{\mbox{D}t} + \eta \nabla \cdot \mathbf{u} &=0 ,
\end{align}
where $\mathbf{u}$ is the horizontal velocity vector (i.e., $\mathbf{u}=u \hat{i}  + v \hat{j}$), $\eta$ is the geopotential height, $\eta=gh$, and $h$ is fluid depth. Here, $\mathbf{f}$ represents the Coriolis factor, and $g$ is the gravitational acceleration. The material derivative, $\mbox{D}/\mbox{D}t$ is defined as
\begin{equation}
\dfrac{\mbox{D}}{\mbox{D}t}=\dfrac{\partial}{\partial t} + \mathbf{u}\cdot \nabla 
\end{equation}
Considering a rectangular 2D domain, $\Omega=[0, L_x]\times[0,L_y]$, SWE can be rewritten in Cartesian form,
\begin{subequations}\label{eq:SWE}
\begin{align}
\dfrac{\partial u}{\partial t} + u\dfrac{\partial u}{\partial x} + v \dfrac{\partial u}{\partial y} + \dfrac{\partial \eta}{\partial x} - fv&=0 ,\\
\dfrac{\partial v}{\partial t} + u\dfrac{\partial v}{\partial x} + v \dfrac{\partial v}{\partial y} + \dfrac{\partial \eta}{\partial y} + fu&=0 ,\\
\dfrac{\partial \eta}{\partial t} + \dfrac{\partial(\eta u)}{\partial x} + \dfrac{\partial(\eta v)}{\partial y} &=0 .
\end{align}
\end{subequations}

\begin{figure}[ht]
	\centering
	\includegraphics[width=0.3\linewidth]{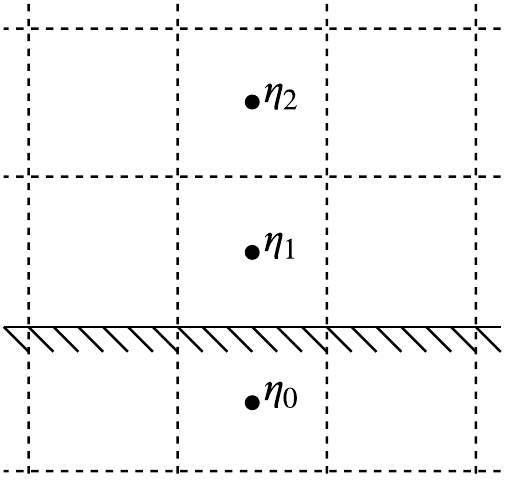}
	\caption{Boundary, inside and ghost cells at the bottom wall}
	\label{fig:ybound}
\end{figure}

Adopting a $\beta$-plane assumption, the effect of the Earth's sphericity is modeled by a linear variation in the Coriolis factor in the spanwise direction,
\begin{equation} \label{eq:corio}
f=f_0 + \dfrac{\beta}{2}(2y-L_y) ,
\end{equation}
where $f_0$ and $\beta$ are constants. The framework used here consists of the nonlinear SWEs in a channel on the rotating earth, associated with periodic boundary conditions in the $x$-direction and slip-wall boundary condition in the $y$-direction, which can be mathematically formulated as follows,

\begin{eqnarray}
u(x,y,t)=u(x+L_x,y,t) ,& v(x,y,t)=v(x+L_x,y,t), & \eta(x,y,t)=\eta(x+L_x,y,t) , \nonumber \\
\dfrac{\partial u}{\partial y}\bigg|_{(x,0,t)}= 0 ,& v(x,0,t)=0 ,& \dfrac{\partial \eta}{\partial y}\bigg|_{(x,0,t)}= constant  ,\\
\dfrac{\partial u}{\partial y}\bigg|_{(x,L_y,t)}= 0  ,& v(x,L_y,t)=0 ,& \dfrac{\partial \eta}{\partial y}\bigg|_{(x,L_y,t)}= constant . \nonumber
\end{eqnarray}
In the above expressions, it follows that $ \dfrac{\partial^2 \eta}{\partial y^2}\bigg|_{(x,0,t)}=0$, $\dfrac{\partial^2 \eta}{\partial y^2}\bigg|_{(x,L_y,t)}= 0$, which allows the following approximation (i.e., also known as linear extrapolation),
\begin{equation}
    \eta_0 = 2\eta_1 - \eta_2,
\end{equation}
where $\eta_0$, $\eta_1$, and $\eta_2$ are the geopotential heights defined discretely at cell centers shown in Figure~\ref{fig:ybound}. 

For initial state, Grammeltvedt \cite{grammeltvedt1969survey} introduced several stable initial conditions for geopotential heights, and here we adopt their initial condition I1 given as,
\begin{equation}\label{eq:init_h}
\eta_0(x,y)=g H_0 + g H_1 \tanh{\left( \dfrac{9\left(L_y/2 - y\right)}{2L_y}\right)} + g H_2 \dfrac{\sin{\left( \dfrac{2\pi x}{L_x}\right)} } { \cosh^2{\left(  \dfrac{9\left(L_y/2 - y\right)}{L_y} \right)}  }.
\end{equation}

\begin{figure}[ht]
	\centering
	\includegraphics[width=\linewidth]{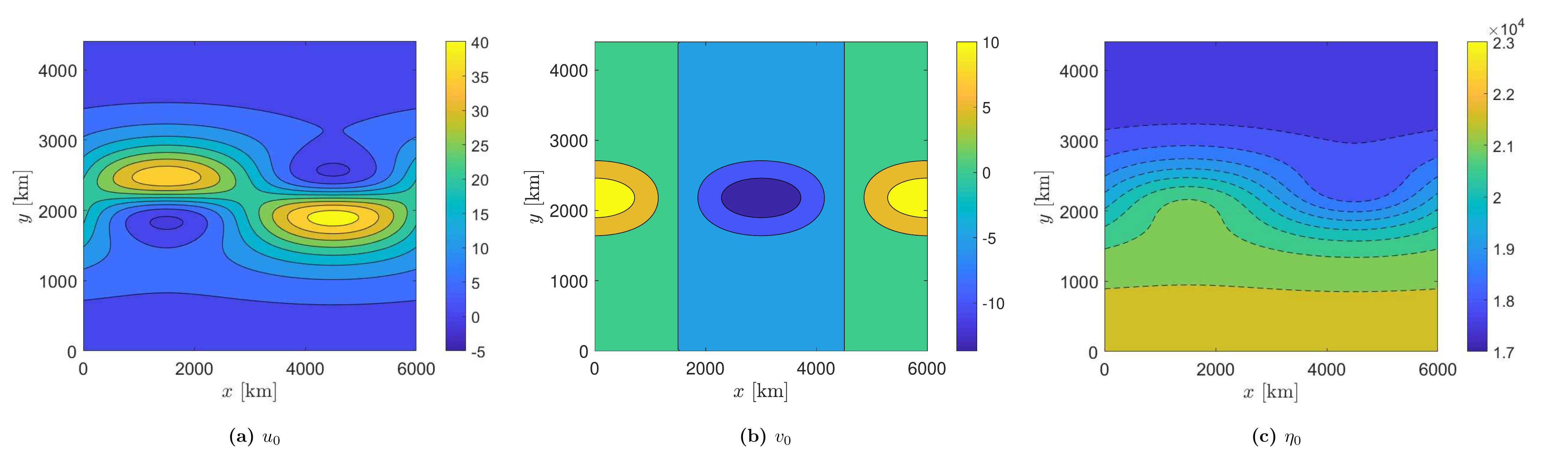}
	\caption{Initial conditions for $u$, $v$, and $\eta$}
	\label{fig:init}
\end{figure}

Using the geostrophic relation, where the Coriolis force is balanced by the pressure force, the dominant balance at the initial state becomes,
\begin{subequations}
\begin{align}
u_0(x,y)&=-\dfrac{1}{f}\dfrac{\partial \eta_0(x,y)}{\partial y}, \\
v_0(x,y)&=\dfrac{1}{f}\dfrac{\partial \eta_0(x,y)}{\partial x}.
\end{align}
\end{subequations}
Hence, the initial velocity field can be obtained by differentiating Eq.~\eqref{eq:init_h}. Initial $u$, $v$, and $\eta$ fields are shown in Figure~\ref{fig:init}.


\subsection{Numerical methods} \label{Num}
\subsubsection{Spatial discretization for full order model} \hfill 

\begin{figure}[H]
	\centering
	\includegraphics[width=.6\linewidth]{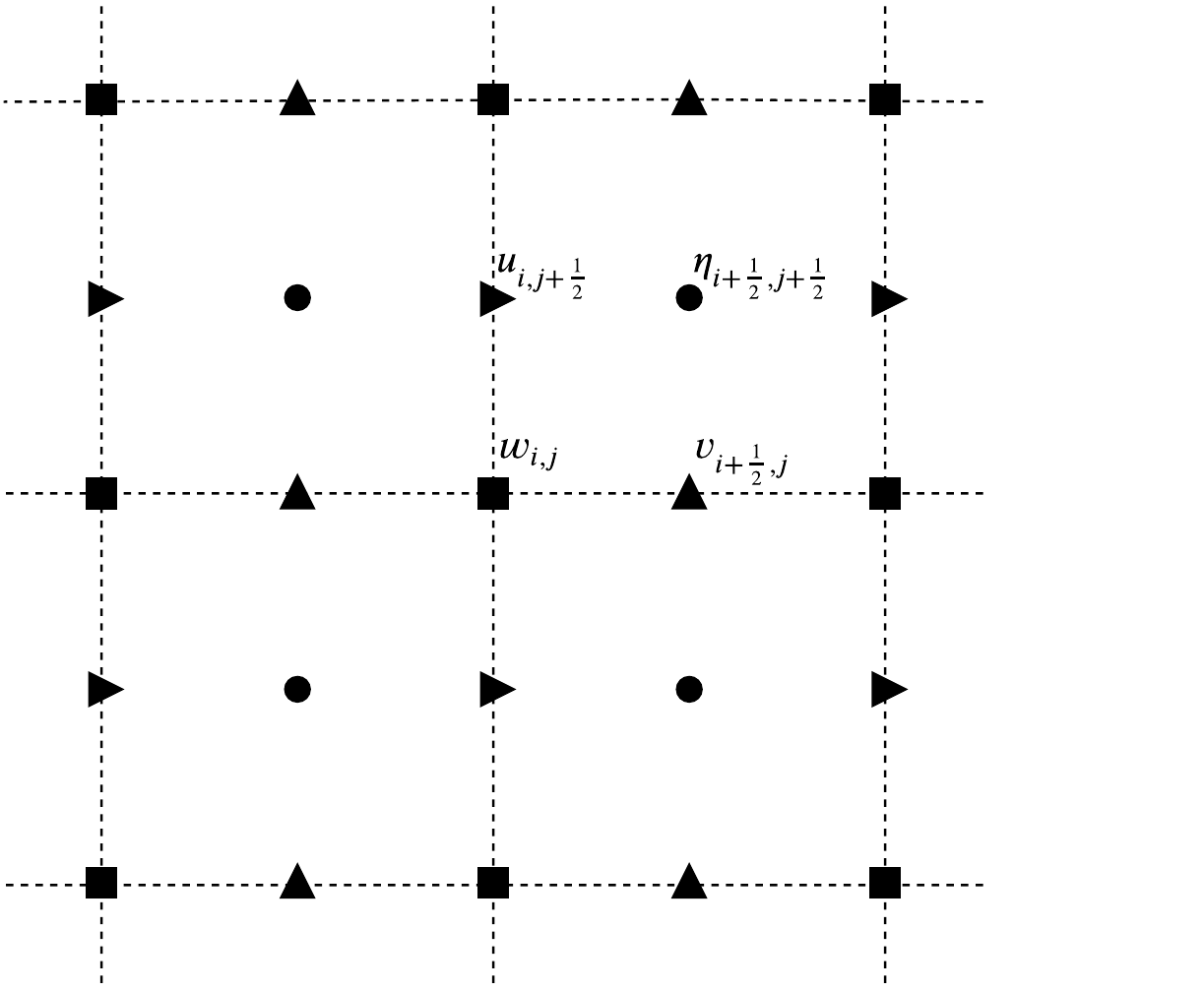}
	\caption{Staggered grid (Arakawa C-grid) arrangement for mass, energy, and enstrophy conserving discretization}
	\label{fig:stag}
\end{figure}

In order to solve the SWEs represented before, staggered rectangular grid is adopted, specifically Arakawa C-grid \cite{arakawa1966computational} as shown in Figure~\ref{fig:stag}. Staggered grid is often used in computational fluid dynamics (CFD) as a way of conserving total mass, energy and enstrophy through numerics. Also, it prevents the odd-even decoupling problem that is usually encountered when using collocated grid. Use of a staggered grid necessitates some attention, since interpolations and/or approximations are required to communicate data between different locations.
For first-order spatial derivatives, standard second-order-accurate centered-finite-difference scheme is utilized. This yields the following semi-discretized SWEs,

\newcommand{\half}{\frac{1}{2}}
\newcommand{\halff}{\frac{3}{2}}

\begin{subequations}\label{eq:SWEstag}
\begin{align}
\dfrac{\partial u}{\partial t}\bigg| _{i,j+\half} = &- u_{i,j+\half} \dfrac{ u_{i+1,j+\half}- u_{i-1,j+\half}}{2\Delta x} \nonumber \\
&- \dfrac{1}{4} \left( v_{i+\half,j} + v_{i-\half,j} + v_{i+\half,j+1} + v_{i-\half,j+1} \right) \dfrac{ u_{i,j+\halff} - u_{i,j-\half}}{2\Delta y} \nonumber\\
&- \dfrac{\eta_{i+\half,j+\half} - \eta_{i-\half,j+\half} }{\Delta x} \nonumber\\
&+ \dfrac{f_{i,j+\half}}{4}  \left( v_{i+\half,j} + v_{i-\half,j} + v_{i+\half,j+1} + v_{i-\half,j+1} \right) ,\\
\dfrac{\partial v}{\partial t}\bigg| _{i+\half,j} &= - \dfrac{1}{4}\left( u_{i,j+\half} + u_{i,j-\half} + u_{i+1,j+\half} + u_{i+1,j-\half}  \right) \dfrac{ v_{i+\halff,j} - v_{i-\half,j} }{2\Delta x} \nonumber\\
&- v_{i+\half,j} \dfrac{ v_{i+\half,j+1} - v_{i+\half,j-1} }{2\Delta y} \nonumber \\
&- \dfrac{ \eta_{i+\half,j+\half} - \eta_{i+\half,j-\half} }{\Delta y} \nonumber \\
&- \dfrac{f_{i+\half,j}}{4} \left( u_{i,j+\half} + u_{i,j-\half} + u_{i+1,j+\half} + u_{i+1,j-\half}  \right) ,\\
\dfrac{\partial \eta}{\partial t}\bigg|_{i+\half,j+\half} &= - \dfrac{  (\eta_{i+\half,j+\half} + \eta_{i+\halff,j+\half})u_{i+1,j+\half}- (\eta_{i+\half,j+\half} + \eta_{i-\half,j+\half})u_{i,j+\half} }{2\Delta x} \nonumber \\
&- \dfrac{(\eta_{i+\half,j+\half} + \eta_{i+\half,j,j+\halff} )v_{i+\half,j+1} - (\eta_{i+\half,j+\half} + \eta_{i+\half,j,j-\half})v_{i+\half,j} }{2\Delta y} .
\end{align}
\end{subequations}

\begin{figure}[H]
	\centering
	\includegraphics[width=1\linewidth]{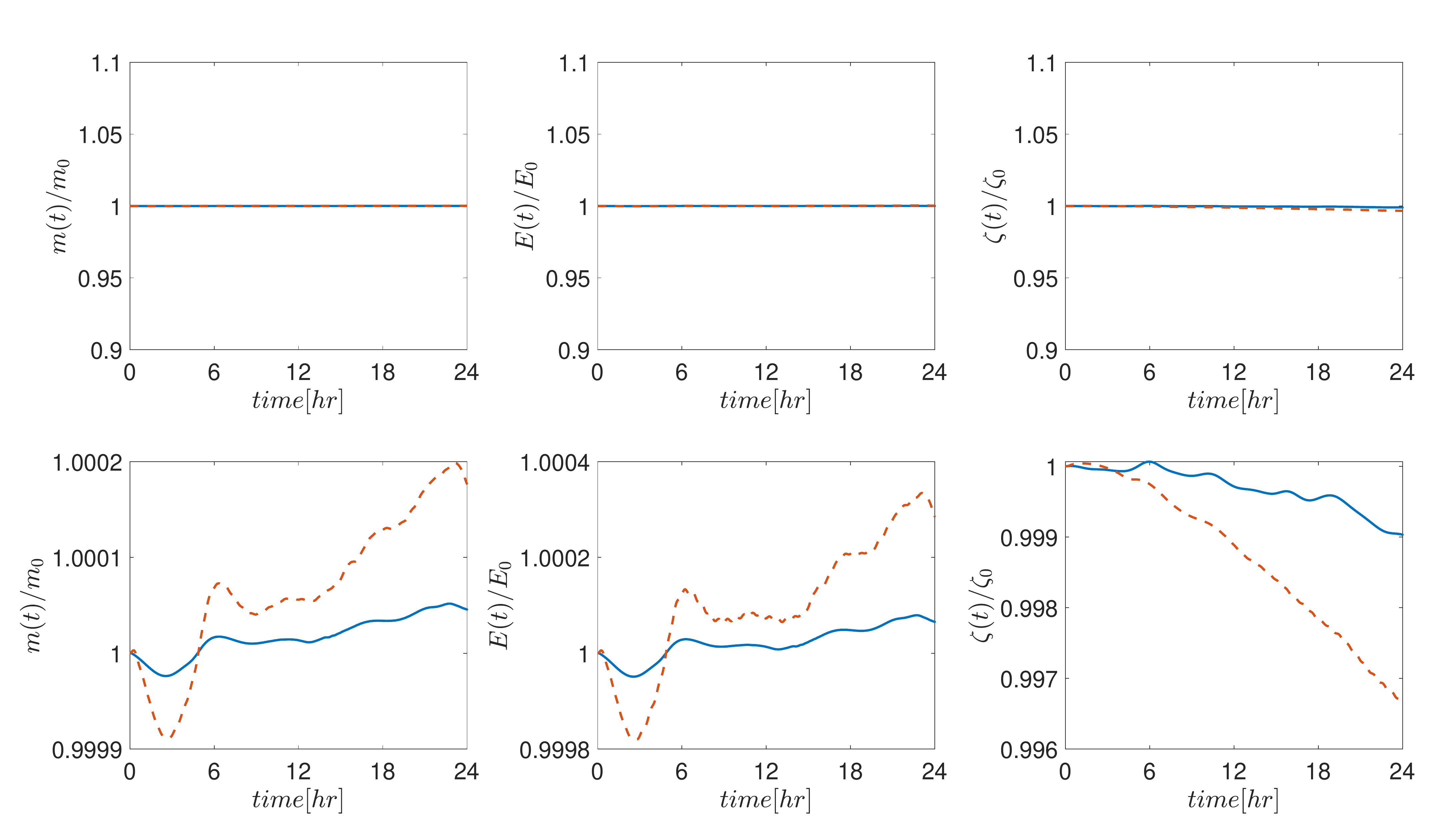}
	\caption{Time histories of SWE system's total mass, energy, and enstrophy. Above row is zoom-out scale and below is zoom-in scale. Solid line is for fine grid ($240\times176$) and dashed line is for coarse grid ($60\times44$) }
	\label{fig:invariant}
\end{figure}

The system's total mass, $m$, energy, $E$, and enstrophy, $\zeta$, can be computed at any time using the following relations,
\begin{align}
    m(t) &= \int_{\Omega}{h(x,y,t)dxdy}, \\
    E(t) &= \int_{\Omega}{\half (u^2(x,y,t) + v^2(x,y,t)) + \half g h^2(x,y,t)dxdy}, \\
    \zeta(t) &=  \int_{\Omega}{\dfrac{(\xi(x,y,t)+f(x,y))^2}{h(x,y,t)}dxdy},
\end{align}
where $\xi$ is the vorticity field defined as $\xi=\dfrac{\partial v}{\partial x} - \dfrac{\partial u}{\partial y}$. The time history of these three invariants, with respect to their initial values are given in Figure~\ref{fig:invariant} for the two resolutions studied in this paper. From this figure, we can easily verify that this staggered grid satisfies the conservation of total mass, energy and enstrophy.

\subsubsection{Temporal discretization} \hfill 

For temporal discretization, third-order total variational diminishing (TVD) Runge-Kutta (TVD-RK3) scheme \cite{gottlieb1998total} is used. We apply a conservative finite difference formulation for our numerical framework in the current study. We cast the governing equations in the following semi-discretized ordinary differential equations form,
\begin{align}\label{ODE1}
\frac{d \vec{q}}{dt} = \vec{\pounds},
\end{align}
where
\begin{align}
    \vec{q} = \begin{bmatrix} u \\ v \\ \eta    \end{bmatrix}, \qquad
    \vec{\pounds}= \begin{bmatrix}
-u\dfrac{\partial u}{\partial x} - v \dfrac{\partial u}{\partial y} - \dfrac{\partial \eta}{\partial x} + fv\\
-u\dfrac{\partial v}{\partial x} - v \dfrac{\partial v}{\partial y} - \dfrac{\partial \eta}{\partial y} - fu \\
-\dfrac{\partial(\eta u)}{\partial x} - \dfrac{\partial(\eta v)}{\partial y}
    \end{bmatrix} .
\end{align}
Then, the TVD-RK3 algorithm consists of the following three steps,
\begin{eqnarray} \label{RK3}
&\vec{q}^{(1)} = \vec{q}^{(n)} + \Delta t \vec{\pounds}^{(n)}, \nonumber \\
&\vec{q}^{(2)}  = \frac{3}{4}  \vec{q}^{(n)} + \frac{1}{4} \vec{q}^{(1)} + \frac{1}{4} \Delta t \vec{\pounds}^{(1)}, \\
&\vec{q}^{(n+1)} =  \frac{1}{3} \vec{q}^{(n)} + \frac{2}{3} \vec{q}^{(2)} + \frac{2}{3} \Delta t \vec{\pounds}^{(2)}. \nonumber 
\end{eqnarray}

\subsubsection{Data storage} \hfill 

Although a staggered grid is used within the solver and solution is calculated at shifted positions, flow fields ($u$, $v$, and $\eta$) are finally stored at the collocated grid positions, to ensure consistency in the constituted ROMs. To do so, the following simple interpolations are utilized,
\begin{align} \label{eq:interpolation}
    u_{i,j} &= \dfrac{1}{2} \left( u_{i,j-1/2} + u_{i,j+1/2} \right) ,\\
    v_{i,j} &= \dfrac{1}{2} \left( u_{i-1/2,j} + u_{i+1/2,j} \right) ,\\
    \eta_{i,j} &= \dfrac{1}{4} \left( \eta_{i-1/2,j-1/2} + \eta_{i+1/2,j-1/2}  + \eta_{i+1/2,j+1/2}  + \eta_{i-1/2,j+1/2}  \right).
\end{align}

\subsection{Numerical integration}\hfill 

Integration over the whole domain might be required for checking conservative quantities. Also, it will be used in proper orthogonal decomposition to calculate the POD modes and their coefficients as will be seen in Section~\ref{sec:POD} as well as Galerkin projection in Section~\ref{sec:GP}. For the two-dimensional numerical integration of any quantity $r(x,y)$ over the domain $\Omega$, we use the Simpson's $1/3^{rd}$ rule \cite{moin2010fundamentals,hoffman2001numerical}. This integration goes as follows
\begin{align}
\int_{\Omega}{r(x,y)dxdy} =\dfrac{\Delta y}{3}\sum_{j=1}^{N_y/2}{\left( \hat{r}_{2j-1} + 4\hat{r}_{2j} + \hat{r}_{2j+1}   \right)},
\end{align}
where
\begin{align}
\hat{r}_{j} & =  \dfrac{\Delta x}{3} \sum_{i=1}^{N_x/2}{\left( r_{2i-1,j} + 4r_{2i,j} + 4r_{2i+1,j}  \right)}, \quad \mbox{for} \quad j=1,2, ..., N_y+1.
\end{align}
Here, $N_x$ and $N_y$ are the number of grid intervals in both $x$ and $y$ directions, respectively. For valid implementation of Simpson's $1/3^{rd}$ rule, $N_x$ and $N_y$ should be even. Also, uniform interval sizes in each coordinate are assumed in the above formulae.

\section{Reduced Order Modeling} \label{sec:ROM}
\subsection{Proper orthogonal decomposition} \label{sec:POD}

Despite being introduced a few decades ago, POD-based ROM is still the state-of-the-art in model order reduction, especially when coupled with Galerkin projection. In POD, the dominant spatial subspaces are extracted from a given dataset. In other words, POD computes the dominant coherent directions in an infinite space which best represent the spatial evolution of a system. POD-ROM is closely-related to either singular value decomposition or eigenvalue decomposition of snapshot matrix. However, in most fluid flow simulations of interest, the number of degrees of freedom (number of grid points) is usually orders of magnitude larger than number of collected datasets. This results in a tall and skinny matrix, which makes the conventional decomposition inefficient, as well as time and memory consuming. Sirovich \cite{sirovich1987turbulence} proposed an alternative efficient technique for generating the POD basis, called `Method of Snapshots'. A number of snapshots, $N_s$, of the 2D flow field, denoted as $w(x,y,t_n)\in \{u(x,y,t_n), v(x,y,t_n), \eta(x,y,t_n) \}$, are stored at consecutive times $t_n$ for $n=1,2,\dots,N_s$. The time-averaged field, called `base flow', can be computed as
\begin{equation}
\bar{w}(x,y)=\dfrac{1}{N_s}\sum_{n=1}^{N_s}w(x,y,t_n).
\end{equation}
The mean-subtracted snapshots, also called anomaly or fluctuation fields, are then computed as the difference between the instantaneous field and the mean field
\begin{equation}
w'(x,y,t_n)=w(x,y,t_n)-\bar{w}(x,y).
\end{equation}
This subtraction has been common in ROM community, and it guarantees that ROM solution would satisfy the same boundary conditions as full order model \cite{chen2012variants}. This anomaly field procedure can be also interpreted as a mapping of snapshot data to its origin.  

Then, an $N_s\times N_s$ snapshot data matrix $\mathbf{A}=[a_{ij}]$  is computed from the inner product of mean-subtracted snapshots
\begin{equation}
a_{ij}=\langle w'(x,y,t_i); w'(x,y,t_j)\rangle,
\end{equation}
where the angle-parenthesis denotes the inner product defined as 
\begin{equation}
    \langle q_1(x,y); q_2(x,y)\rangle = \int_{\Omega}{q_1(x,y)q_2(x,y) dxdy}.
\end{equation}
Then, an eigen-decomposition of $\mathbf{A}$ is performed,
\begin{equation}
\mathbf{A} \mathbf{V} = \mathbf{V} \mathbf{\Lambda},
\end{equation} 
where $\mathbf{\Lambda}$ is a diagonal matrix whose entries are the eigenvalues $\lambda_k$ of $\mathbf{A}$, and $\mathbf{V}$ is a matrix whose columns $\mathbf{v}_k$ are the corresponding eigenvectors. It should be noted that these eigenvalues need to be arranged in a descending order (i.e., $\lambda_1\ge\lambda_2\ge\dots\ge\lambda_{N_s}$), for proper selection of the POD modes. In general, the eigenvalues, $\lambda_k$, represent the respective POD mode contribution to the total variance. In case of velocity time series, it represents the turbulent kinetic energy. The POD modes $\phi_{k}$ are then computed as
\begin{equation}
\phi_{k}(x,y)=\dfrac{1}{\sqrt{\lambda_k}}\sum_{n=1}^{N_s} v^{n}_{k} w'(x,y,t_n)
\end{equation}
where $v^{n}_{k}$ is the $n$-th component of the eigenvector $\mathbf{v}_k$. The scaling factor, $1/\sqrt{\lambda_k}$, is to guarantee the orthonormality of POD modes, i.e., $\langle \phi_i; \phi_j\rangle = \delta_{ij}$, where $\delta_{ij}$ is the Kronecker delta. The algorithm used to calculate the POD modes can be summarized in Algorithm~\ref{alg:POD}.

\begin{algorithm}[H]
	\SetAlgoLined
	\KwResult{POD modes, $\phi_k$, and eigenvalues, $\lambda_k$ , for $k=1,2,\dots,R$} 
	Read snapshots: $w(x,y,t_n)\in \{u(x,y,t_n), v(x,y,t_n), \eta(x,y,t_n) \}$ \\
	(1) $\bar{w}(x,y) = \dfrac{1}{N_s}\sum_{n=1}^{N_s}w(x,y,t_n)$;  ! calculate mean field\\
	(2) $w'(x,y,t_n)=w(x,y,t_n)-\bar{w}(x,y)$; ! calculate mean-subtracted field\\
	(3) $a_{ij} = \langle w'(x,y,t_i); w'(x,y,t_j)\rangle$;  ! calculate correlation matrix $A^{N_s\times N_s}$ \\
	(4) $\mathbf{A} \mathbf{v}_k = \lambda_k \mathbf{v}_k $; ! eigenvalue decomposition \\
	(5) $\phi_{k}(x,y)=\dfrac{1}{\sqrt{\lambda_k}}\sum_{n=1}^{N_s} v^{n}_{k} w'(x,y,t_n)$; ! calculate POD modes for largest $R$ eigenvalues
	\caption{POD algorithm with mean-subtraction}
	\label{alg:POD}
\end{algorithm}

\subsection{Galerkin projection} \label{sec:GP}

As stated in Section~\ref{INTRO}, POD is usually coupled with Galerkin projection to generate intrusive ROMs. This requires the projection of main governing equations on the truncated POD space. To do so, we first have to rewrite the SWEs Eq.~\eqref{eq:SWE} in the following way,
\begin{subequations}\label{eq:SWE2}
\begin{align}
\dfrac{\partial u}{\partial t} &= -u\dfrac{\partial u}{\partial x} - v \dfrac{\partial u}{\partial y} - \dfrac{\partial \eta}{\partial x} + fv ,\\
\dfrac{\partial v}{\partial t} &= -u\dfrac{\partial v}{\partial x} - v \dfrac{\partial v}{\partial y} - \dfrac{\partial \eta}{\partial y} - fu ,\\
\dfrac{\partial \eta}{\partial t} &= -\dfrac{\partial(\eta u)}{\partial x} - \dfrac{\partial(\eta v)}{\partial y} .
\end{align}
\end{subequations}
The flow variables, $u$, $v$, and $\eta$, are approximated using the most energetic $R_u$, $R_v$, and $R_\eta$ POD modes, respectively as,
\begin{subequations}\label{eq:PODuvh}
\begin{align}
u(x,y,t)&=\bar{u}(x,y) + \sum_{k=1}^{R_u}a_k(t)\phi_k^u(x,y) ,\\
v(x,y,t)&=\bar{v}(x,y) + \sum_{k=1}^{R_v}b_k(t)\phi_k^v(x,y) ,\\
\eta(x,y,t)&=\bar{\eta}(x,y) + \sum_{k=1}^{R_\eta}c_k(t)\phi_k^\eta(x,y) ,
\end{align}	
\end{subequations}
where $\bar{u}(x,y)$, $\bar{v}(x,y)$, and $\bar{\eta}(x,y)$ represent the time-averaged flow field variables across the training snapshots, $\phi_{k}^{u}(x,y)$, $\phi_{k}^{v}(x,y)$, $\phi_{k}^{\eta}(x,y)$ are the $k$-th POD modes for the fluctuating part of the flow field variables, and $a_k(t)$, $b_k(t)$, and $c_k(t)$ are the corresponding $k$-th time dependent coefficients. Although the formulae in this section are generally applicable for any number of modes $R_u$, $R_v$, and $R_{\eta}$, we use equal number of modes in the present study (i.e., $R=R_u=R_v=R_{\eta}$).
Substituting Eq.~\eqref{eq:PODuvh} into Eq.~\eqref{eq:SWE2}, an orthonormal Galerkin projection can be subsequently performed by multiplying Eq.~\eqref{eq:SWE2} by the spatial POD modes $\phi_{k}^{u}(x,y)$,  $\phi_{k}^{v}(x,y)$, $\phi_{k}^{\eta}(x,y)$, then integrating over the entire domain $\Omega$, and making use of the orthonormality of POD modes. The resulting dense dynamical system for $a_k$, $b_k$, and $c_k$ can be written as
\begin{subequations}\label{eq:SWE_GP}
\begin{align}
\dfrac{d a_k}{dt}&= \mathfrak{B}_{k}^{(u)} + \sum_{i=1}^{R_u} {\mathfrak{L}_{ik}^{(u,u)} a_i} + \sum_{i=1}^{R_v} {\mathfrak{L}_{ik}^{(u,v)}b_i} +
\sum_{i=1}^{R_\eta} {\mathfrak{L}_{ik}^{(u,\eta)}c_i} + \sum_{i=1}^{R_u} \sum_{j=1}^{R_u} {\mathfrak{N}_{ijk}^{(u,u,u)}a_ia_j}  \nonumber \\ 
&+ \sum_{i=1}^{R_v}  \sum_{j=1}^{R_u}{\mathfrak{N}_{ijk}^{(u,v,u)}b_ia_j} , \qquad k=1,2,\dots,R_u, \\
\dfrac{d b_k}{dt}&= \mathfrak{B}_{k}^{(v)} + \sum_{i=1}^{R_u} {\mathfrak{L}_{ik}^{(v,u)} a_i} + \sum_{i=1}^{R_v} {\mathfrak{L}_{ik}^{(v,v)}b_i} +
\sum_{i=1}^{R_\eta} {\mathfrak{L}_{ik}^{(v,\eta)}c_i} + \sum_{i=1}^{R_u} \sum_{j=1}^{R_v} {\mathfrak{N}_{ijk}^{(v,u,v)}a_ib_j} \nonumber \\ 
&+ \sum_{i=1}^{R_v} \sum_{j=1}^{R_v}{\mathfrak{N}_{ijk}^{(v,v,v)}b_ib_j} , \qquad k=1,2,\dots,R_v, \\
\dfrac{d c_k}{dt}&= \mathfrak{B}_{k}^{(\eta)} + \sum_{i=1}^{R_u} {\mathfrak{L}_{ik}^{(\eta,u)} a_i} + \sum_{i=1}^{R_v} {\mathfrak{L}_{ik}^{(\eta,v)}b_i} +
\sum_{i=1}^{R_\eta} {\mathfrak{L}_{ik}^{(\eta,\eta)}c_i} + \sum_{i=1}^{R_u} \sum_{j=1}^{R_{\eta}} {\mathfrak{N}_{ijk}^{(\eta,u,\eta)}a_ic_j} \nonumber \\
& + \sum_{i=1}^{R_v} \sum_{j=1}^{R_{\eta}}{\mathfrak{N}_{ijk}^{(\eta,v,\eta)}b_ic_j} , \qquad k=1,2,\dots,R_{\eta},
\end{align}
\end{subequations}
where the predetermined model coefficients can be computed by the following numerical integration during an offline stage (i.e., they need to be computed only once)
\begin{subequations}
	\begin{align}
	\mathfrak{B}_k^{(u)}&= \big\langle -\bar{u}\dfrac{\partial \bar{u}}{\partial x}  -\bar{v}\dfrac{\partial \bar{u}}{\partial y} -\dfrac{\partial \bar{\eta}}{\partial x} + f\bar{v}; \phi_k^{u} \big\rangle ,\\
	\mathfrak{B}_k^{(v)}&= \big\langle -\bar{u}\dfrac{\partial \bar{v}}{\partial x}  -\bar{v}\dfrac{\partial \bar{v}}{\partial y} -\dfrac{\partial \bar{\eta}}{\partial y} - f\bar{u}; \phi_k^{v} \big\rangle ,\\
	\mathfrak{B}_k^{(\eta)}&= \big\langle -\dfrac{\partial \bar{u}\bar{\eta}}{\partial x}  -\dfrac{\partial \bar{v}\bar{\eta}}{\partial y}; \phi_k^{\eta} \big\rangle  ,
	\end{align}
\end{subequations}
while the coefficients of linear terms are computed as
\begin{subequations}
	\begin{align}
	\mathfrak{L}_{ik}^{(u,u)}&= \big\langle -\bar{u}\dfrac{\partial \phi_i^u}{\partial x}  -\phi_i^u\dfrac{\partial \bar{u}}{\partial x}
	-\bar{v}\dfrac{\partial \phi_i^u}{\partial y}; \phi_k^{u} \big\rangle ,\\
	\mathfrak{L}_{ik}^{(u,v)}&= \big\langle  -\phi_i^v\dfrac{\partial \bar{u}}{\partial x}
	+ f \phi_i^v; \phi_k^{u} \big\rangle ,\\
	\mathfrak{L}_{ik}^{(u,\eta)}&= \big\langle 
	-\dfrac{\partial \phi_i^{\eta}}{\partial x}; \phi_k^{u} \big\rangle ,\\ 
	\mathfrak{L}_{ik}^{(v,u)}&= \big\langle  -\phi_i^u\dfrac{\partial \bar{v}}{\partial x} - f \phi_i^u; \phi_k^{v} \big\rangle ,\\
	\mathfrak{L}_{ik}^{(v,v)}&= \big\langle
	-\bar{u}\dfrac{\partial \phi_i^v}{\partial x} -\bar{v}\dfrac{\partial \phi_i^v}{\partial y} -\phi_i^v\dfrac{\partial \bar{v}}{\partial y}; \phi_k^{v} \big\rangle ,\\
	\mathfrak{L}_{ik}^{(v,\eta)}&= \big\langle 
	-\dfrac{\partial \phi_i^{\eta}}{\partial y}; \phi_k^{v} \big\rangle ,\\ 
	\mathfrak{L}_{ik}^{(\eta,u)}&= \big\langle 
	-\dfrac{\partial \phi_i^{u}\bar{\eta}}{\partial x}; \phi_k^{\eta} \big\rangle ,\\ 
	\mathfrak{L}_{ik}^{(\eta,v)}&= \big\langle 
	-\dfrac{\partial \phi_i^{v}\bar{\eta}}{\partial y}; \phi_k^{\eta} \big\rangle ,\\ 
	\mathfrak{L}_{ik}^{(\eta,\eta)}&= \big\langle 
	-\dfrac{\partial \bar{u}\phi_i^{\eta}}{\partial x} -\dfrac{\partial \bar{v}\phi_i^{\eta}}{\partial y}; \phi_k^{\eta} \big\rangle .
	\end{align}
\end{subequations}
Finally, the coefficients of nonlinear terms are computed as
\begin{subequations}
	\begin{align}
	\mathfrak{N}_{ijk}^{(u,u,u)}&= \big\langle   -\phi_i^u\dfrac{\partial \phi_j^u}{\partial x}; \phi_k^{u} \big\rangle ,\\
	\mathfrak{N}_{ijk}^{(u,v,u)}&= \big\langle   -\phi_i^v\dfrac{\partial \phi_j^u}{\partial y}; \phi_k^{u} \big\rangle ,\\
	\mathfrak{N}_{ijk}^{(v,u,v)}&= \big\langle   -\phi_i^u\dfrac{\partial \phi_j^v}{\partial x}; \phi_k^{v} \big\rangle ,\\ 
	\mathfrak{N}_{ijk}^{(v,v,v)}&= \big\langle   -\phi_i^v\dfrac{\partial \phi_j^v}{\partial x}; \phi_k^{v} \big\rangle ,\\
	\mathfrak{N}_{ijk}^{(\eta,u,\eta)}&= \big\langle   -\dfrac{\partial \phi_i^u \phi_j^{\eta}}{\partial x}; \phi_k^{\eta} \big\rangle ,\\
	\mathfrak{N}_{ijk}^{(\eta,v,\eta)}&= \big\langle   -\dfrac{\partial \phi_i^v \phi_j^{\eta}}{\partial y}; \phi_k^{\eta} \big\rangle .
	\end{align}
\end{subequations}
To solve the dynamical system given by Eq.~\eqref{eq:SWE_GP} using any ODE solver, the initial conditions for $a_k(t)$, $b_k(t)$, and $c_k(t)$ have to be determined which can be obtained by the following projection of the initial snapshot on the corresponding POD modes
\begin{subequations}
	\begin{align}
	a_k(t_0) &= \big \langle u(x,y,t_0)-\bar{u}(x,y); \phi_k^{u}  \big \rangle ,\\
	b_k(t_0) &= \big \langle v(x,y,t_0)-\bar{v}(x,y); \phi_k^{v}  \big \rangle ,\\
	c_k(t_0) &= \big \langle \eta(x,y,t_0)-\bar{\eta}(x,y); \phi_k^{\eta}  \big \rangle .	\label{eq:projectc}
	\end{align}
	\label{eq:project}
\end{subequations}

\subsection{Artificial neural network} \label{sec:ANN}
A neural network is simply a mapping from some input space to an output space. The two common applications of neural networks are regression and classification. Neural networks are powerful in terms of their capability to identify the underling patterns in data. First, they were inspired by the Nobel prize winning work of Hubel and Wiesel \cite{hubel1962receptive} using neuronal networks organized in hierarchical layers of cells for processing visual stimulus. A typical feed forward network consists of a sequence of layers, each having a number of nodes, or neurons. Information is carried from one layer to the next layers through connections between their respective neurons with some weighting and biasing parameters. In mathematical representation, the input $x_i$ to a neuron in the $l$-th layer can be written as
\begin{equation}
x^l_i = \sum_{k=1}^{n}w^{l-1}_{ki}y^{l-1}_{k} + b^{l}_{i},
\end{equation}
where $w^{l-1}_{ki}$ is the weight parameter from the $k$-th neuron from the $(l-1)$-th to the $i$-th neuron in the $l$-th layer, $y_k^{l-1}$ is the output of the $k$-th neuron in the $(l-1)$-th layer, and $n$ is the number of neurons at this layer. $b^{l}_i$ represent the bias that is added as an input to the $i$-th neuron in the $l$-th layer. Finally, each node, or neuron, acts as a processing unit on its input by applying some activation, or transfer, function $G$ as follows,
\begin{equation}
y^l_i = G_l(x^l_i).
\end{equation}
In general, any differentiable function may qualify as a transfer function, however a few number of standard activation functions are used in practice (e.g., Tan-Sigmoid, Log-Sigmoid, Linear, Radial-Basis). In theory, a neural network can represent any mapping function between inputs and targets, given sufficient number of hidden layers and neurons, i.e. deep neural network (DNN) \cite{hornik1989multilayer}. This is done by tuning the weights and biases through an optimization algorithm using some \emph{training} data. In brief, starting from arbitrarily random weights and biases, and given input data, the neural network (mapping) output is calculated, and compared with the true target (true labels) to calculate the error at the output layer. This error is then back-propagated through the preceding layers, and the weights and biases are corrected accordingly. This process is repeated until convergence happens. So, in concept, training a network is simply solving a minimization problem in which the objective function represents the root mean square error between the network output and the true target.

Exploiting the mapping and predictability strengths of neural networks, they can offer an effective alternative to predict the temporal coefficients of POD modes, substituting for the Galerkin projection step. More importantly, this eliminates the need to access the governing equations and makes the process purely data-driven. This is crucial when the system in-hand is still not well-understood, or its mathematical representation is prohibitive. 

In the present study, we limit our study for $4$, $8$, and $16$ modes. A feed-forward neural network consisting of 4 hidden layers/60 neurons each, is used for number of modes $R= 4$ or $8$, and 5 layers/100 neurons each for the $16$ modes case. The `ReLu' activation function is used for all hidden layers and all tests, defined as
\begin{equation}
G_l(x_i^l) = \max(0,x_i^l) .
\end{equation}

This ANN is trained in POD space, where the training data are obtained from projecting the full-order snapshots on the POD modes, to obtain the POD modal coefficients. Instead of mapping the modal coefficients at time $n$ to the coefficients at next time, $n+1$, the target is set as the difference between them. In other words, the network takes as input a set of $a_k^{(n)}$ and gives as output a set of $a_k^{(n+1)}-a_k^{(n)}$, referring to it as residual network, or ResNet. The residual between two time steps is then applied to update the current state during prediction. The learning of the residual information was observed to result in a more stabilized performance \cite{haber2017stable,san2019artificial} and also improves the accuracy of neural network prediction \cite{lu2018beyond}. Also, it is typical in machine learning community to normalize the input and output data. This is particularly effective when dealing with data of different order of magnitudes. In our case, we have chosen to use normalization from $-1$ to $+1$ using the following formula,
\begin{equation}
\tilde{z} = \dfrac{2z - \left(z_{max}+z_{min}\right)}{z_{max}-z_{min}} .
\end{equation}

\subsection{Dynamic mode decomposition} \label{DMD}
While modes in POD are computed based on energy cascade \cite{holmes2012turbulence} (i.e., the modes that contains the largest amount of energy), their interpretation and usage for feature detection might become unclear in some situations which can occur at largely distinct frequencies, with very similar energy content. On the other hand, frequency-based methods characterize each mode with a specific oscillating frequency and growth/decay rate. They are considered as data-driven adaptations of the discrete Fourier transform (DFT), with the advantage of allowing each mode to grow or decay exponentially. The most popular frequency-based decomposition method is the dynamic mode decomposition (DMD), approximating the modes of the so-called Koopman operator \cite{koopman1931hamiltonian}. The Koopman operator is a linear operator that represents nonlinear dynamics without linearization \cite{rowley2009spectral,mezic2013analysis,alekseev2016linear} taking the form $\mathbf{X}_{n+1} = \mathbf{A} \mathbf{X}_{n}$, where $\mathbf{A}$ is a time-independent linear operator that maps the data from time $t_n$ to $t_{n+1}$. The terms `DMD mode' and `Koopman mode' are often used interchangeably in the fluids literature.

The method can be viewed as computing the eigenvalues and eigenvectors (low dimensional modes) of a linear model that approximates the underlying dynamics, even if the dynamics is nonlinear \cite{tu2013dynamic}. The DMD gives the growth rates and frequencies associated with each mode. Theorems regarding the existence and uniqueness of DMD modes and eigenvalues can be found in \cite{chen2012variants}. Assuming the mean-subtracted snapshot matrix to be $\mathbf{X}= \{\mathbf{w'}_1,\mathbf{w'}_2,\mathbf{w'}_3, \dots, \mathbf{w'}_{N_s-2},\mathbf{w'}_{N_s-1},\mathbf{w'}_{N_s}\}$, we construct two matrices, $\mathbf{X}_1= \{\mathbf{w'}_1,\mathbf{w'}_2,\mathbf{w'}_3, \dots, \mathbf{w'}_{N_s-2},\mathbf{w'}_{N_s-1}\}$ and $\mathbf{X}_2= \{\mathbf{w'}_2,\mathbf{w'}_3, \dots, \mathbf{w'}_{N_s-2},\mathbf{w'}_{N_s-1},\mathbf{w'}_{N_s}\}$. Here, $\{\mathbf{w'}_n $ refers to any field data (i.e., $u$, $v$, or $\eta$) at time $t_n$ that is written as a solution on our 2D grid domain mapped into column vectors. Therefore, using the Koopman operator $\mathbf{A}$, the following can be written,
\begin{align}
\mathbf{X}_2&=\mathbf{A}\mathbf{X}_1 .
\label{eq:DMDmap}
\end{align}
So, $\mathbf{A}$ is simply $\mathbf{X}_2 \mathbf{X}_1^{+}$, where $\mathbf{X}_1^{+}$ is the Moore–Penrose pseudo-inverse of $\mathbf{X}_1$. It should be noted that the size of $\mathbf{A}$ is $N_p\times N_p$, where $N_p$ is the total number of grid points, which makes this computation practically unfeasible in fluid flows problems where $N_p \gg N_s$. Instead, we implement DMD following the method of snapshots by substituting for the SVD of $\mathbf{X}_1$ as $\mathbf{X}_1 = \mathbf{U}\mathbf{\Sigma}\mathbf{ V}^*$, where the asterisk superscript denotes the conjugate transpose,
\begin{align}
\mathbf{X}_2&=\mathbf{A}\mathbf{U}\mathbf{\Sigma}\mathbf{ V}^* .
\end{align}
Defining $ \tilde{\mathbf{A}}=\mathbf{U}^* \mathbf{A} \mathbf{U}$,
\begin{align}
\tilde{\mathbf{A}}&=\mathbf{U}^* \mathbf{X}_2 \mathbf{V} \mathbf{\Sigma}^{-1} .
\end{align}
Now, computing the eigenvalues and eigenvectors of $\tilde{\mathbf{A}}$
\begin{align}
\tilde{\mathbf{A}} \mathbf{v}_j&= \lambda_j \mathbf{v}_j .
\label{eq:Atildeig}
\end{align}
The eigenvalues $\lambda_j$ of $\tilde{\mathbf{A}}$ are the same as the eigenvalues of ${\mathbf{A}}$. The DMD modes (eigenvectors of ${\mathbf{A}}$) can be computed as follows,
\begin{align}
\boldsymbol \phi_j&= \mathbf{U} \mathbf{v}_j .
\label{eq:DMDmode}
\end{align}
In literature, there are slight variations from Eq.~\ref{eq:DMDmode} for computing the DMD modes \cite{tu2013dynamic,kutz2016dynamic}. Now, defining $\alpha_j=\dfrac{\ln{(\lambda_j)}}{\Delta t}$, the reconstructed DMD flow field can be rewritten as,
\begin{align}
\mathbf{w}^{DMD}&=\bar{\mathbf{w}} + \sum_{j=1}^{R}b_j \boldsymbol \phi_j e^{(\alpha_j t)} \label{eq:DMDconst1}\\
&=\bar{\mathbf{w}} +\boldsymbol \Phi \ \mathrm{diag}(e^{\boldsymbol \alpha t}) \ \mathbf{b} ,\label{eq:DMDconst2}
\end{align}

where $\mathbf{b}$ is a vector of the coefficients $b_j$ representing the initial amplitudes of the DMD modes. $\boldsymbol \Phi$ is a matrix whose columns $\boldsymbol \phi_j$ are the DMD modes, and $\mathrm{diag}(e^{\alpha t})$ is a diagonal matrix whose entries are $\left(e^{\alpha t}\right)$.
The initial amplitudes of DMD modes can be computed from projecting the first snapshot (i.e., at $t=0$) onto the DMD modes as,
\begin{align}
\mathbf{b}=\Phi^{+} {\mathbf{w'}}_1 ,
\label{eq:DMDamp}
\end{align}
where $\boldsymbol \Phi^{+}$ is the pseudo-inverse of $\boldsymbol \Phi$, and ${\mathbf{w'}}_1$ is the first snapshot (mean-subtracted). It should be noted here that the amplitudes vector, $\mathbf{b}$, is computed here from only the first snapshot. This is expected to be less representative of the true modal amplitudes, especially for integration over a long time interval. Alternatively, these amplitudes can be calculated from any arbitrary snapshot using Eq.~\ref{eq:DMDconst2}, or it can be calculated using an optimization algorithm that minimizes the average error between true data and DMD predictions. To summarize, the DMD is a non-intrusive ROM technique which approximates the modes of a linear operator which transforms the system state into its future state. 

\subsubsection{DMD mode selection} \hfill \\
Selection of DMD modes used in the reduced order flow model constitutes the source of many discussions. Here, we will describe two methods for computing the DMD modes. The first method is based on what is called `hard thresholding', computing a reduced rank matrix, $\tilde{\mathbf{A}}_R$. That is instead of retaining all the vectors in $\mathbf{U}$ and $\mathbf{V}$, we use only the first $R$ of them, $\mathbf{U}_R$ and $\mathbf{V}_R$, respectively. Also, only the first largest $R$ singular values in $\mathbf{\Sigma}$ are kept and the remaining are truncated. This reduces the dimension of $\tilde{\mathbf{A}}$ from an $(N_s-1)\times(N_s-1)$ to $R\times R$, giving exactly $R$ modes without any further modal selection. This algorithm, which will be denoted as DMD-HT from here onward, can be summarized in Algorithm~\ref{alg:DMDHT}.

\begin{algorithm}[H] 
	\SetAlgoLined
	\KwResult{DMD modes, $\boldsymbol \phi_k$, and eigenvalues, $\lambda_k$ for $k=1,2, ..., R$} 
	Read snapshots: $w(x,y,t_n)\in \{u(x,y,t_n), v(x,y,t_n), \eta(x,y,t_n) \}$ \\
	(1) $w(x,y,t_n) = \dfrac{1}{N_s}\sum_{n=1}^{N_s}w(x,y,t_n)$  ! calculate mean field\\
	(2) $w'(x,y,t_n)=w(x,y,t_n)-\bar{w}(x,y)$ ! calculate anomaly field snapshots\\
	(3) $\mathbf{w'}_{n} \Leftarrow w'(x,y,t_n)$, and $\bar{\mathbf{w}} \Leftarrow \bar{w}(x,y)$  ! stack fields into vector form \\ 
	(4) $\mathbf{X}_1= \{\mathbf{w'}_1,\mathbf{w'}_2,\mathbf{w'}_3, \dots,\mathbf{w'}_{N_s-2},\mathbf{w'}_{N_s-1}\}$, \\ and  
	  $ \mathbf{X}_2=  \{\mathbf{w'}_2,\mathbf{w'}_3, \dots, \mathbf{w'}_{N_s-2},\mathbf{w'}_{N_s-1},\mathbf{w'}_{N_s}\}$ ! obtain initial, and future snapshot matrices \\
	(5) $\mathbf{X}_1$ = $\mathbf{U}\mathbf{\Sigma}\mathbf{ V}^*$ ! singular value decomposition, where $U \in \mathbb{R}^{N_p\times N_s -1}$, $V \in \mathbb{R}^{N_s-1\times N_s -1}$, and $\Sigma \in \mathbb{R}^{N_s-1\times N_s -1}$ \\
	(6) $\mathbf{U}_R, \mathbf{\Sigma}_R , \mathbf{V}_R \Leftarrow \mathbf{U}, \mathbf{\Sigma} , \mathbf{ V}$ ! hard thresholding, where $U_R \in \mathbb{R}^{N_p\times R}$, $V_R \in \mathbb{R}^{N_s-1\times R}$, and $\Sigma_R \in \mathbb{R}^{R\times R}$ \\
	(7)	$\tilde{\mathbf{A}}_R=\mathbf{U}_R^*\mathbf{X}_2 \mathbf{V}_R \mathbf{\Sigma}_R^{-1}$ ! compute the lower-rank dynamics matrix\\
	(8) $\tilde{\mathbf{A}}_R \mathbf{v}_k = \lambda_k \mathbf{v}_k $ ! eigenvalue decomposition \\
	(9) $\boldsymbol \phi_k = \mathbf{U}_R \mathbf{v}_k $  ! calculate DMD modes \\
	(10) $\phi_k (x,y) \Leftarrow \boldsymbol \phi_k$ ! unstack the vector data into a field form 
	\caption{DMD [Hard Threshold] (DMD-HT) algorithm with mean-subtraction}
	\label{alg:DMDHT}
\end{algorithm} 
\medskip
The second approach introduces some sorting criteria to select the most effective modes. Without any hard thresholding, the eigendecomposition of $\tilde{\mathbf{A}}$ is computed, giving $(N_s-1)$ eigenvalues and DMD modes through Eq.~\ref{eq:Atildeig}-\ref{eq:DMDmode}, with their initial amplitudes using Eq.~\ref{eq:DMDmap}. Then, we define the contribution or the influence of any DMD mode, $\phi_j$, to ROM accuracy as
\begin{equation}
    I_j = |b_j| \left(e^{\sigma_j} + e^{-\sigma_j} \right),
    \label{eq:DMDsort}
\end{equation}
where $\sigma_j$ is the growth rate defined as $\sigma_j=\text{real}(\alpha_j)$. The reason we choose this criterion and definition is that we can have quickly decaying modes with high amplitude, or rapidly growing modes with lower amplitudes. In both cases these modes can be influential, having a significant contribution to ROM accuracy. Also, it was previously demonstrated that a selection based solely on either growth/decay rate  or the amplitudes is insufficient criterion \cite{noack2011reduced}. 

Based on this definition of modal contribution, the DMD modes are subsequently arranged in a descending order based on their respective $I$ values, and the first $R$ of them are retained in the ROM (i.e., $I_1\ge I_2 \ge I_3 \ge \dots \ge I_R$). This algorithm can be summarized in Algorithm~\ref{alg:DMDsort}, denoted as DMD-S. It should be noted that a soft threshodling (or regularization) can be also implemented before computing $\tilde{\mathbf{A}}$, truncating the vectors in $\mathbf{U}$ and $\mathbf{V}$ corresponding to singular values smaller than a predetermined threshold.

\begin{algorithm}[H] 
	\SetAlgoLined
	\KwResult{DMD modes, $\boldsymbol \phi_k$, and eigenvalues, $\lambda_k$, for $k=1,2, ..., R$} 
	Read snapshots: $w(x,y,t_n)\in \{u(x,y,t_n), v(x,y,t_n), \eta(x,y,t_n) \}$ \\
	(1) $w(x,y,t_n) = \dfrac{1}{N_s}\sum_{n=1}^{N_s}w(x,y,t_n)$  ! calculate mean field\\
	(2) $w'(x,y,t_n)=w(x,y,t_n)-\bar{w}(x,y)$ ! calculate anomaly field snapshots\\
	(3) $\mathbf{w'}_{n} \Leftarrow w'(x,y,t_n)$, and $\bar{\mathbf{w}} \Leftarrow \bar{w}(x,y)$ ! stack the field data into a vector form \\
	(4) $\mathbf{X}_1= \{\mathbf{w'}_1,\mathbf{w'}_2,\mathbf{w'}_3, \dots,\mathbf{w'}_{N_s-2},\mathbf{w'}_{N_s-1}\}$, \\ and  
	  $ \mathbf{X}_2=  \{\mathbf{w'}_2,\mathbf{w'}_3, \dots, \mathbf{w'}_{N_s-2},\mathbf{w'}_{N_s-1},\mathbf{w'}_{N_s}\}$ ! obtain initial, and future snapshot matrices \\
	(5) $\mathbf{X}_1$ = $\mathbf{U}\mathbf{\Sigma}\mathbf{ V}^*$ ! singular value decomposition \\
	(6)	$\tilde{\mathbf{A}}=\mathbf{U}^*\mathbf{X}_2 \mathbf{V} \mathbf{\Sigma}^{-1}$ ! compute the dynamics matrix\\
	(7) $\tilde{\mathbf{A}} \mathbf{v}_k = \lambda_k \mathbf{v}_k $ ! eigenvalue decomposition \\
	(8) $\boldsymbol \phi_k = \mathbf{U} \mathbf{v}_k $ ! calculate DMD modes (unsorted) \\
	(9) $\Phi = \{\boldsymbol \phi_1, \boldsymbol \phi_2, \dots, \boldsymbol \phi_{N_s-1}\} $ ! generate DMD basis matrix\\
	(10) $\mathbf{b}=\Phi^{+} {\mathbf{w'}}_1$ ! compute modal amplitudes\\
	(11) $ I_k = |b_k| \left(e^{\sigma_j} + e^{-\sigma_j} \right) $ ! compute modal contributions using the growth rate  \\
    (12) $I_1\ge I_2 \ge I_3 \ge \dots \ge I_R$ ! sorting and selecting the most influential modes\\
    (13) $\phi_k (x,y) \Leftarrow \boldsymbol \phi_k$ ! unstack the vector data into a field form (after sorting)\\
	\caption{DMD [Sorted] (DMD-S) algorithm with mean-subtraction}
	\label{alg:DMDsort}
\end{algorithm}

\section{Results} \label{sec:results}
\subsection{Numerical experiments}
\begin{figure}[ht]
	\centering
	\includegraphics[width=0.5\linewidth]{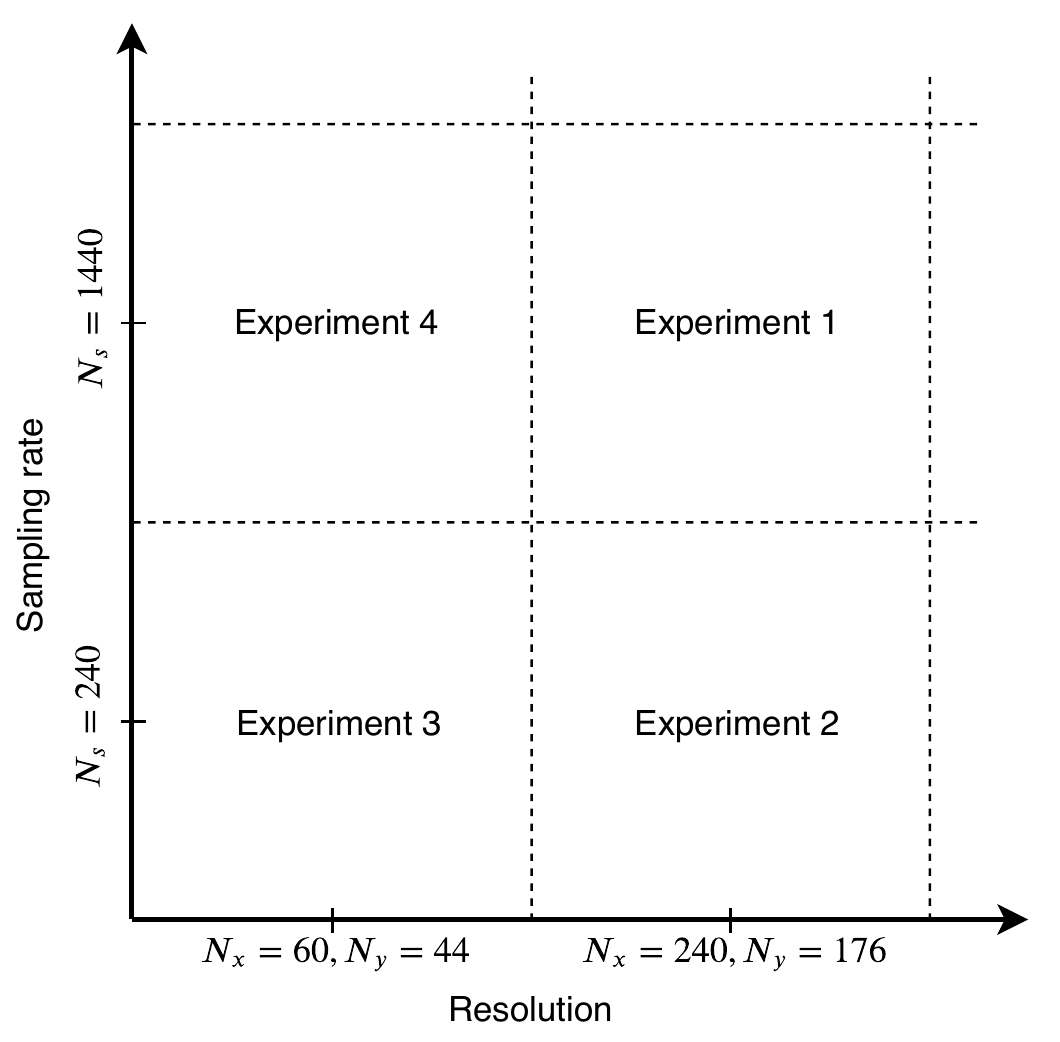}
	\caption{Four numerical experiments in present study: Experiment 1: High resolution, high sampling rate; Experiment 2: High resolution, low sampling rate; 
	Experiment 3: Low resolution, low sampling rate;
	Experiment 4: Low resolution, high sampling rate}
	\label{fig:experiments}
\end{figure}

For data collection,  we consider SWE within a rectangular 2D domain of $L_x=6000 ~\text{km}$, and $L_y=4400 ~\text{km}$. To define the initial field as in Eq.~\ref{eq:init_h}, values of $g=9.81\ \text{m/s}^2$, $H_0 = 2000$ km, $H_1 = 220$ km, and $H_2 = 133$ km are adopted. Also, Coriolis constants of $f_0 = 10^{-4}$ s$^{-1}$ and $\beta = 1.5\times 10^{-11}$ m$^{-1}$s$^{-1}$ are used in Eq.~\ref{eq:corio}. Since the primary objective of the current study is to assess and compare the performance of different ROM and field decomposition techniques and their sensitivities to data quality, we study two different resolutions, $\left\{ N_x=240, N_y=176\right\}$ and $\left\{ N_x=60, N_y=44\right\}$, and two sampling rates, $\left\{ N_s=1440\right\}$ and $\left\{ N_s=240 \right\}$, where $N_x$ and $N_y$ are the number of grid spacings in the $x$ and $y$ directions, respectively and $N_s$ is the number of snapshots collected from solving the SWEs full order model (FOM) for 24 hours. This gives us four numerical experiments as shown in Figure~\ref{fig:experiments}.

For all experiments, a time-step of $10 \text{ s}$ is used to guarantee numerical stability through the Courant-Friedrichs-Lewy (CFL) number as well as convergence. We have chosen to focus on geopotential height, $\eta$, predictions being a conservative quantity. Therefore, for DNN and DMD, only $\eta$ fields, their bases, $\phi^{\eta}$, and corresponding coefficients, $c(t)$, are used for training and prediction. However, for Galerkin projection the $u$ and $v$ fields are also considered since they are coupled with $\eta$ in the ROM ODE system of equations given in Eq.~\ref{eq:SWE_GP}.

\subsection{Eigenvalue analysis} \label{sec:eigresults}
\begin{figure}[ht]
	\centering
	\includegraphics[width=1\linewidth]{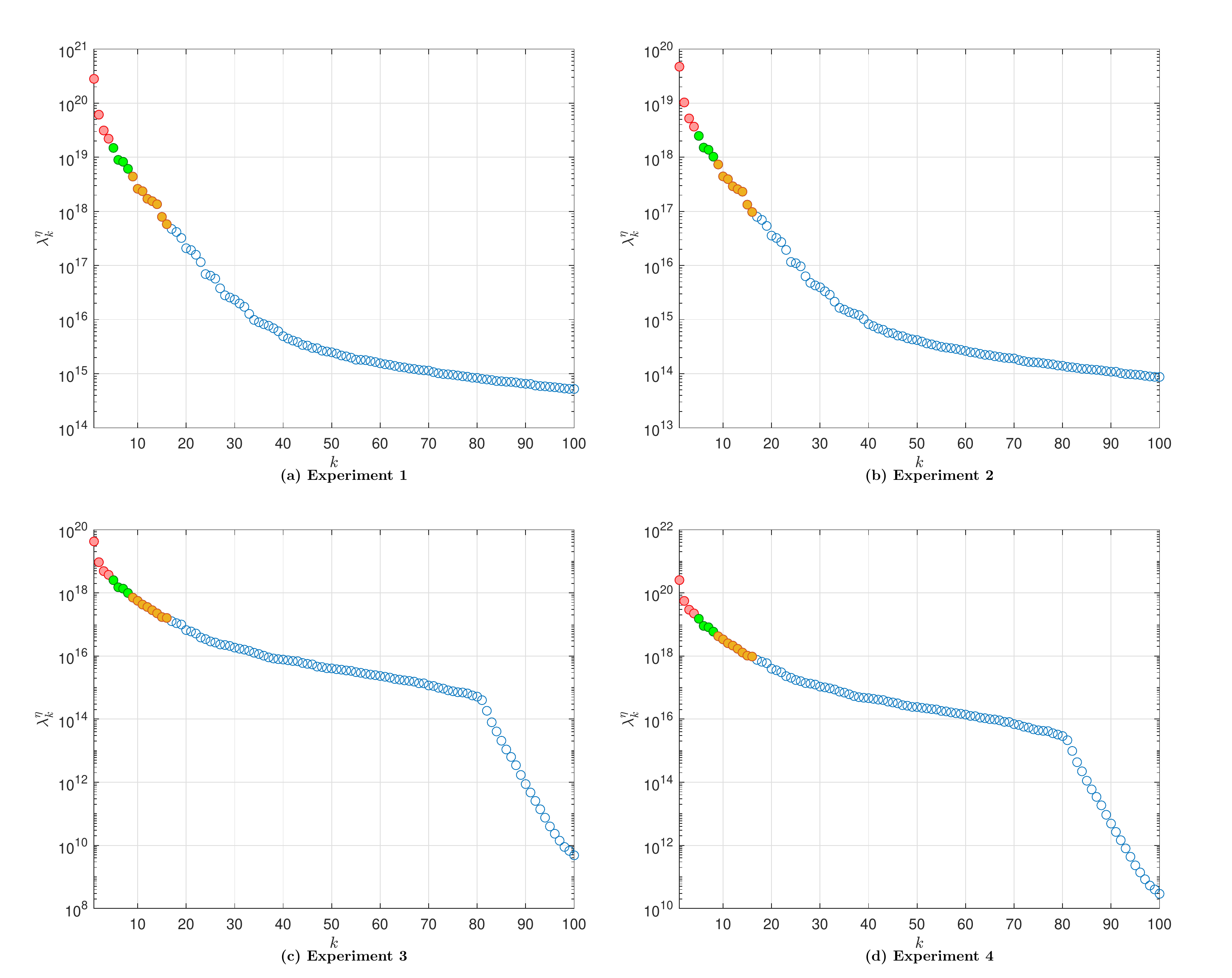}
	\caption{POD eigenvalues for geopotential fields}
	\label{fig:PODeig}
\end{figure}

\begin{figure}[ht]
	\centering
	\includegraphics[width=1\linewidth]{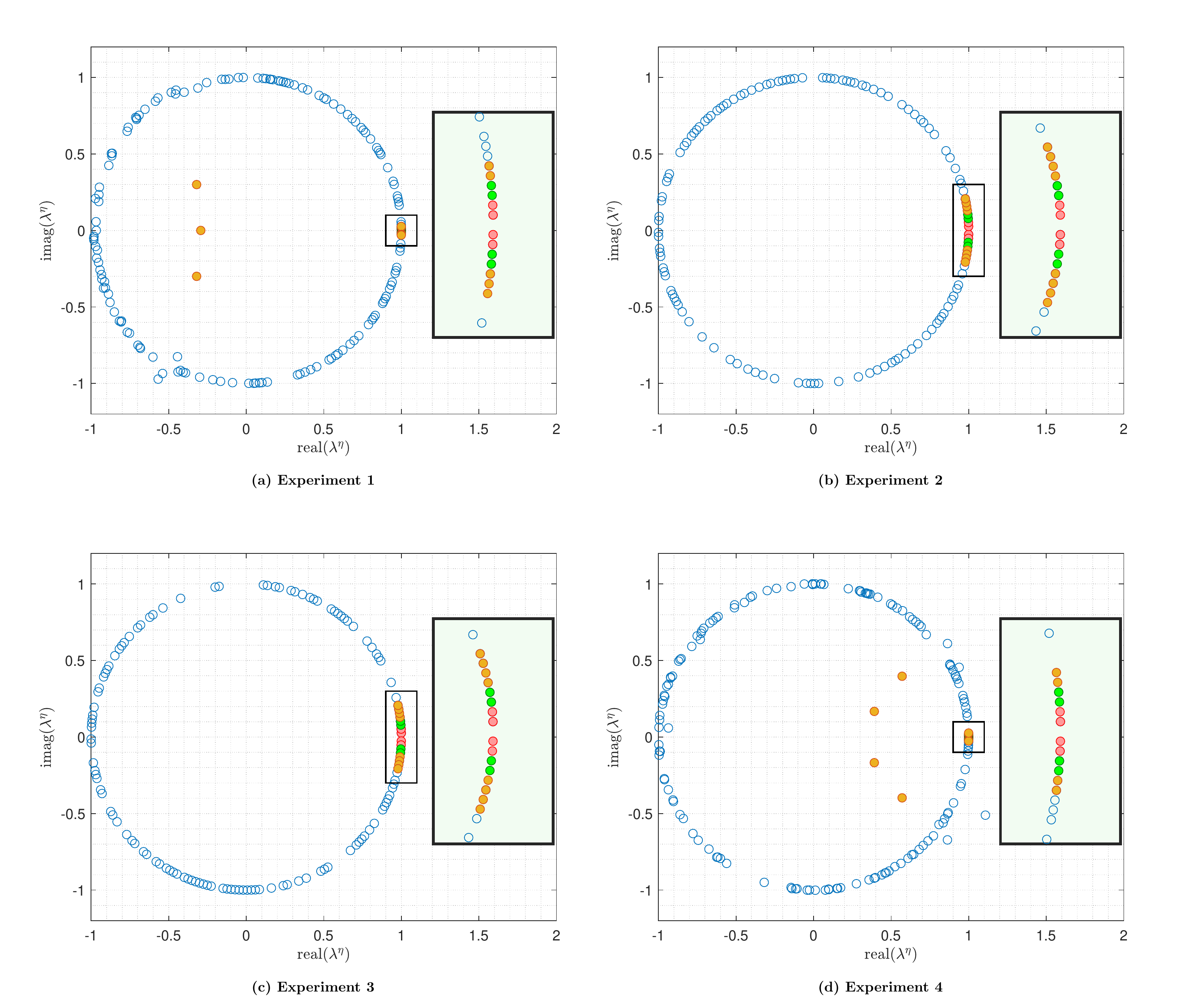}
	\caption{DMD eigenvalues for geopotential fields}
	\label{fig:DMDeig}
\end{figure}

Eigenvalues decomposition is a principal step in both POD and DMD algorithms. In POD, the modal eigenvalues gives an indication of the amount of information (or variance) contained in this respective mode, which is the main criteria in POD modal selection. In DMD, the eigenvalues are, in general, complex numbers, providing information about the temporal growth rate and frequency of DMD modes. In sorted DMD (Algorithm \ref{alg:DMDsort}), the analysis of DMD eigenvalues is important and gives a deeper insight onto the influence of DMD modes on ROM dynamical evolution. The POD eigenvalues for Experiments 1-4 are shown in Figure~\ref{fig:PODeig}, where the largest four eigenvalues (corresponding to the most four energetic POD modes) are denoted with red color. The next largest four eigenvalues are given green color, and the next eight ones (i.e., from $\lambda^{\eta}_9$ to $\lambda^{\eta}_{16}$) are given orange color. The same color code is used for DMD eigenvalues in Figure~\ref{fig:DMDeig} to denote the selected modes based on the criteria given in Eq.~\ref{eq:DMDsort}.

In POD, the percentage modal energy is computed using the following relative information content (RIC) formula \cite{gunzburger2012flow},
\begin{align}
	P(k) = \left(\frac{\sum_{j=1}^{k}\lambda_j} {\sum_{j=1}^{N_s}\lambda_j}\right) \times 100.
\end{align}

The Kolmogorov n-width \cite{kolmogoroff1936uber,pinkus2012n} provides a mathematical guideline to quantify the optimal n-dimensional linear subspace and the associated error (i.e., a measure of system's reducibility). More specifically, it quantifies how a given n-dimensional subspace compares with all other possible n-dimensional subspaces. In POD, it can be considered as a measure of how well a linear superposition of POD modes might represent the underlying dynamics. If the decay of Kolmogorov width is slow, there exists no feasible linear subspace that fits well the data. Therefore, in our context, RIC can be used as a simple quantitative metric to understand the Kolmogorov width of the systems. Table~\ref{table:RIC} presents the RIC of the flow field variables, $u$, $v$, and $\eta$.

\begin{table}[ht!]
	\caption{Relative information content using different numbers of POD modes}
	\centering
	\begin{tabular}[t]{l  c  c  c } 
		\hline
		$R$ \qquad \qquad &  \quad$P(k^u)$\quad & \quad$P(k^v)$\quad &  \quad$P(k^{\eta})$\quad \\ [0.8ex] 
		\hline
		\multicolumn{4}{l}{\emph{Experiment 1}}\\
		 4  & 98.09 & 97.70 & 87.62   \\ 	
		 8  & 99.63 & 99.35 & 96.07   \\ 
		 16 & 99.99 & 99.90 & 99.47   \medskip \\ 
		
		\multicolumn{4}{l}{\emph{Experiment 2}} \\
		4  &  98.09 & 97.70 & 87.62     \\ 	
		8  &  99.62 & 99.35 & 96.06     \\ 
		16 &  99.99 & 99.90 & 99.47     \medskip \\ 
		\multicolumn{4}{l}{\emph{Experiment 3}}\\
		4  & 97.91 & 97.40 & 85.56     \\ 	
		8  & 99.46 & 99.18 & 94.57     \\ 
		16 & 99.88 & 99.84 & 98.65     \medskip \\ 
		\multicolumn{4}{l}{\emph{Experiment 4}}\\
		4  & 97.90 & 97.40 & 85.54      \\ 	
		8  & 99.46 & 99.18 & 94.58      \\ 
		16 & 99.88 & 99.84 & 98.66      \\ 
		\hline
	\end{tabular}
	
	\label{table:RIC}
\end{table}

\subsection{Geopotential field} \label{sec:hresults}
\begin{figure}[ht]
	\centering
	\includegraphics[width=0.5\linewidth]{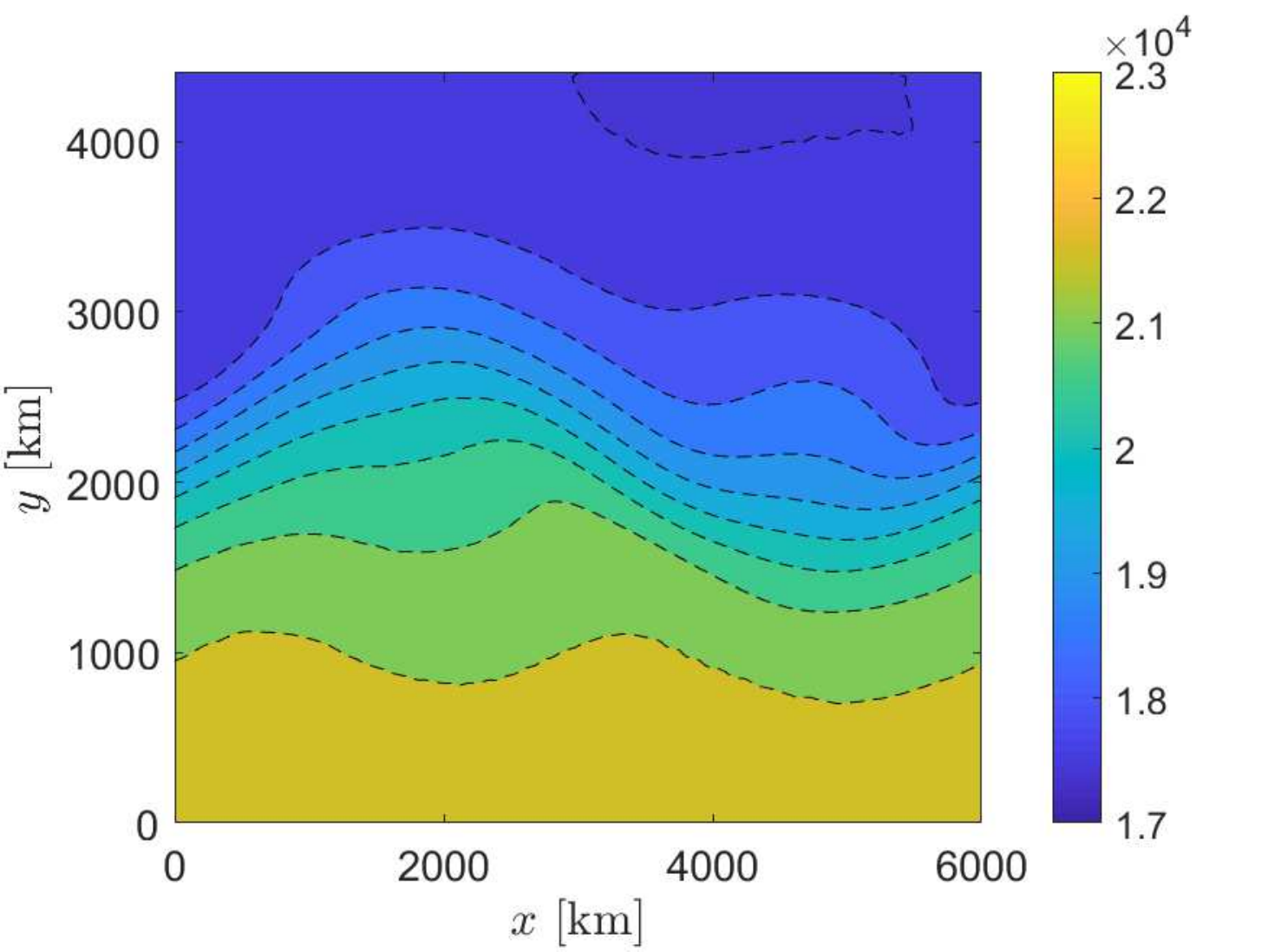}
	\caption{True geopotential field at the end of 24 hours}
	\label{fig:h24}
\end{figure}

Now, we would like to compare reconstructed full-order fields obtained from POD-GP, POD-ANN, and DMD. Here, DMD serves as a comparison with a different example of non-intrusive ROMs. The true (full-order) final field after 24 hours, using the higher resolution is given in Figure~\ref{fig:h24} for reference. Compared to the initial field from Figure~\ref{fig:init}, we can easily visualize the convective nature of the flow field. This promotes the 2D SWEs as a challenging and more realistic testing framework than many other simpler prototypical cases characterizing periodic behaviors. In order to make fair judgments of ROMs predictability, fields obtained from true field projection on POD modes are also shown. These are the fields reconstructed from only the first $R$ modes and their corresponding true coefficients, and they represent the maximum attainable accuracy from POD-based ROMs.

\begin{figure}[ht]
	\centering
    \includegraphics[width=\linewidth]{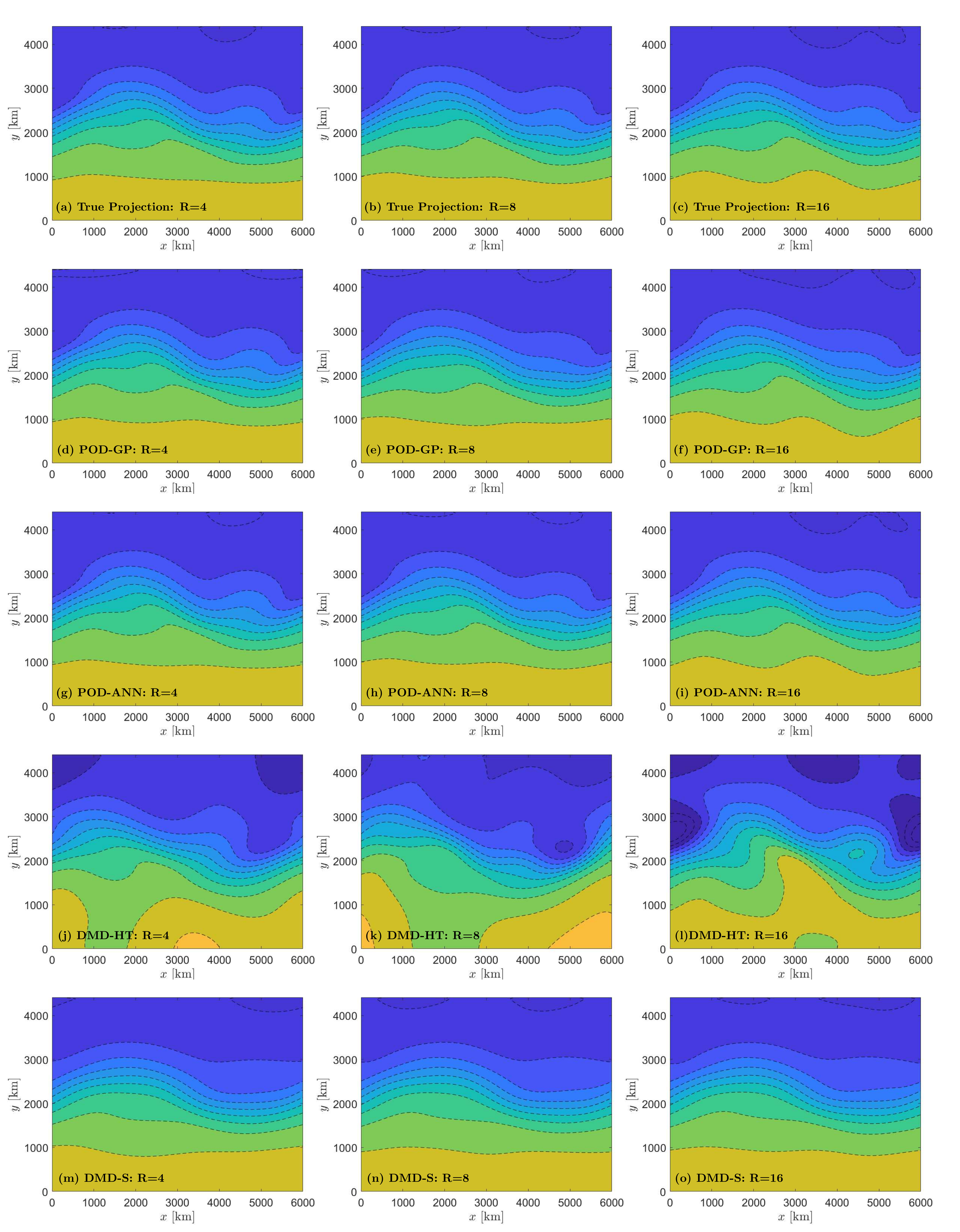}
	\caption{Geopotential field at 24 hours from Experiment 1}
	\label{fig:h_exp1}
\end{figure}

\begin{figure}[ht]
	\centering
    \includegraphics[width=\linewidth]{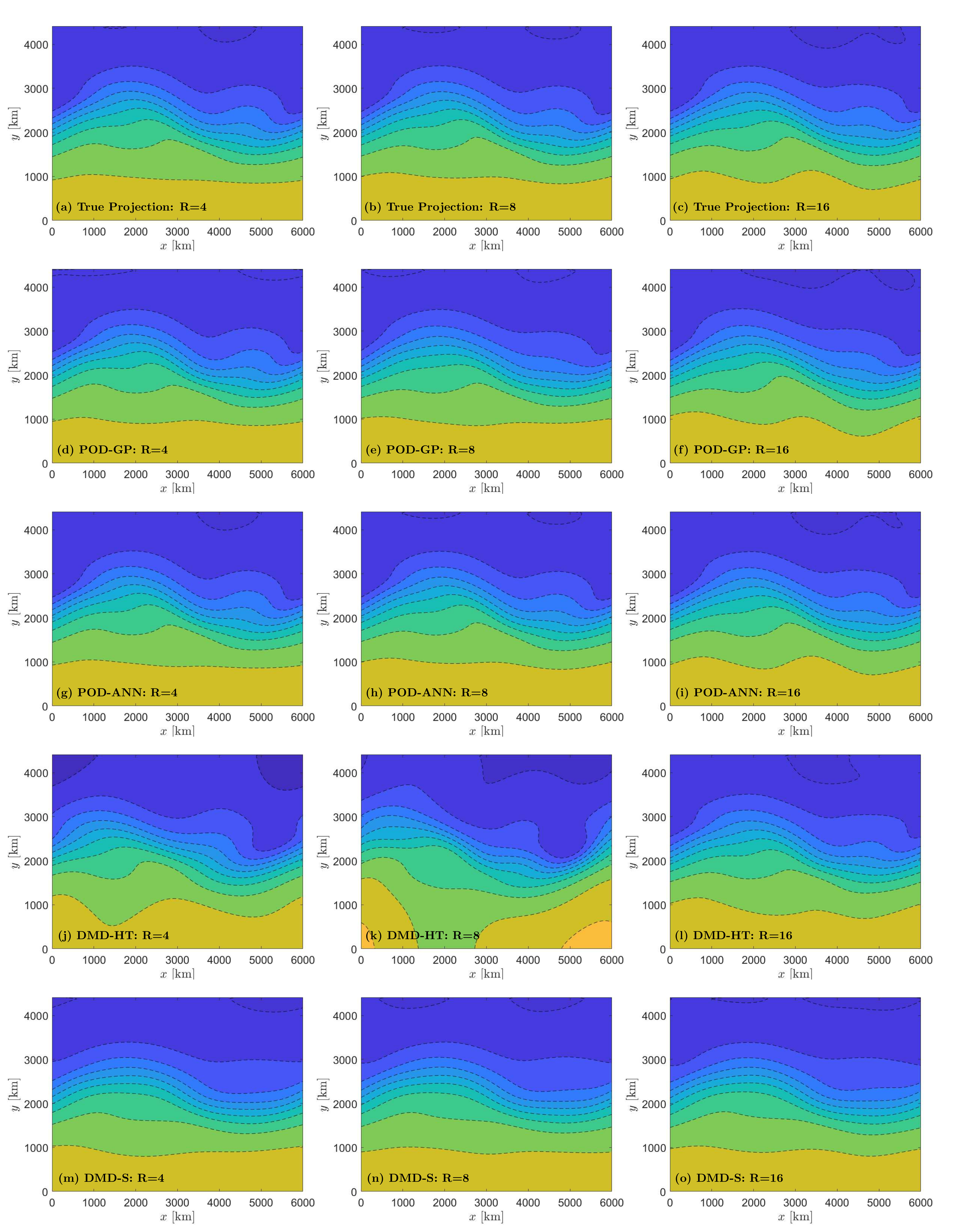}
	\caption{Geopotential field at 24 hours from Experiment 2}
	\label{fig:h_exp2}
\end{figure}

\begin{figure}[ht]
	\centering
    \includegraphics[width=\linewidth]{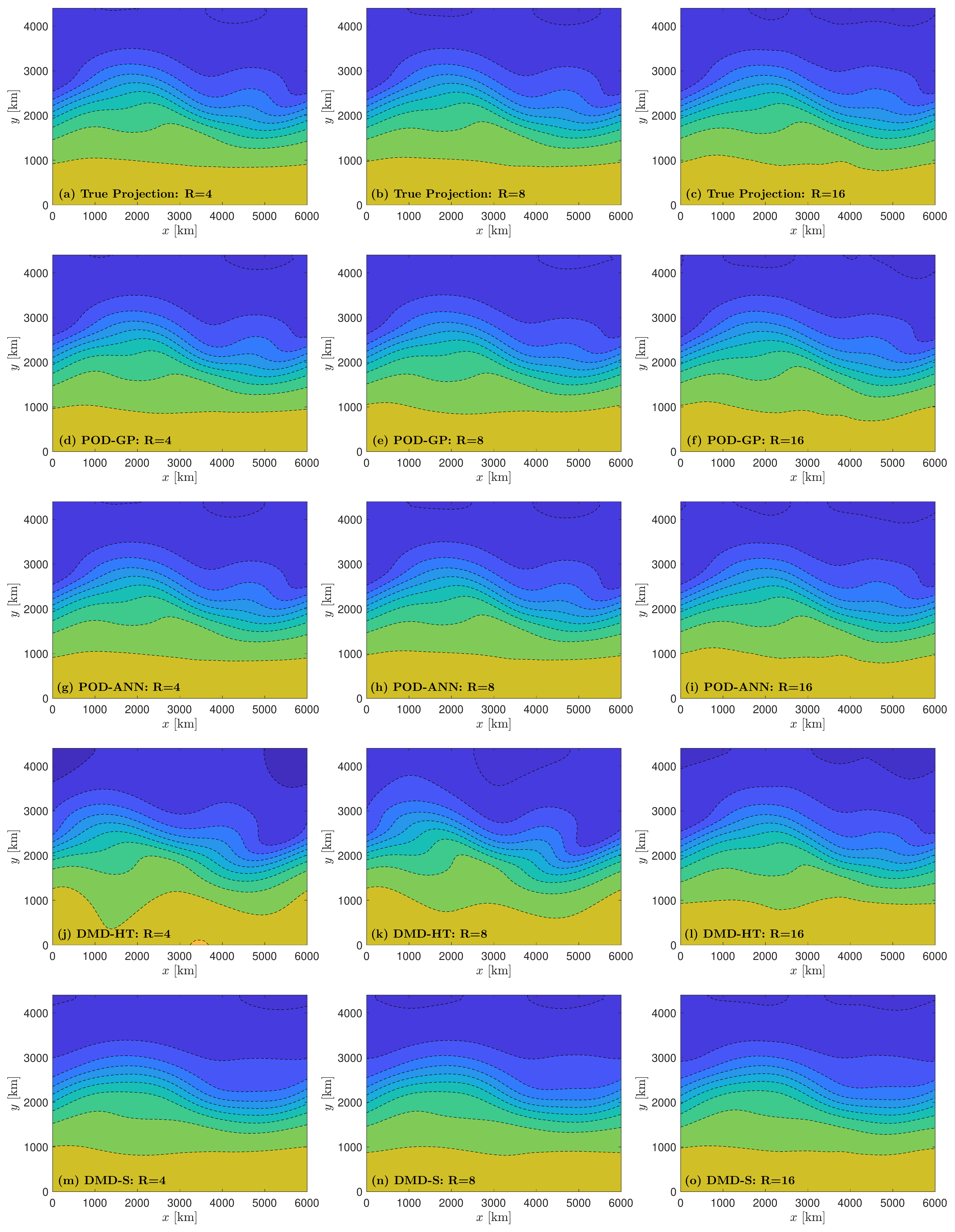}
	\caption{Geopotential field at 24 hours from Experiment 3}
	\label{fig:h_exp3}
\end{figure}

\begin{figure}[ht]
	\centering
    \includegraphics[width=\linewidth]{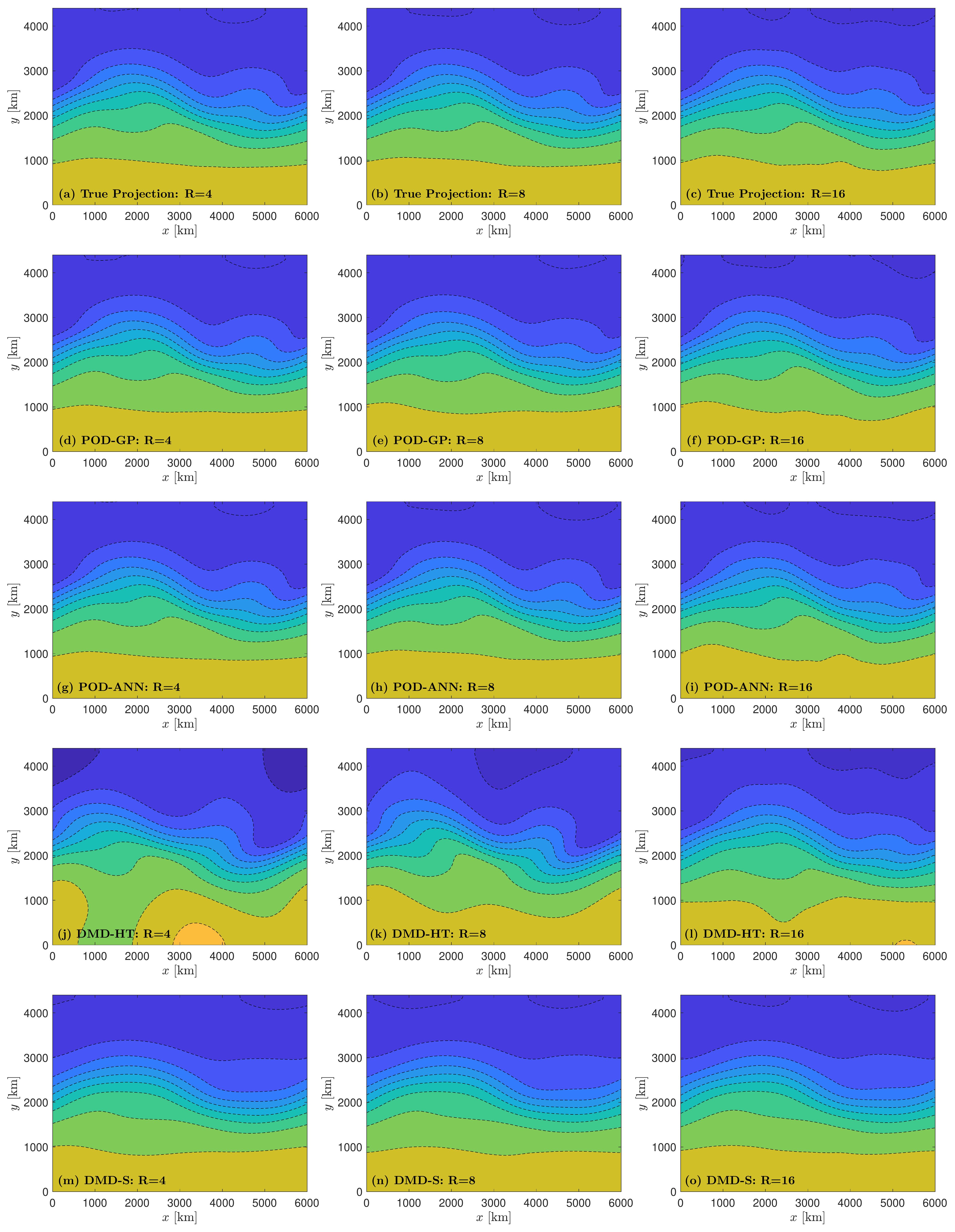}
	\caption{Geopotential field at 24 hours from Experiment 4}
	\label{fig:h_exp4}
\end{figure}

Figures~\ref{fig:h_exp1}-\ref{fig:h_exp4} show geopotential fields from the fours experiments using 4, 8, and 16 modes. It is easily observed that, for the current datasets, both POD-GP and POD-ANN fields converge to the true prediction with increasing the number of modes, irrespective of resolutions or sampling rates. On the other hand, DMD shows unsimilar behavior, especially the one with hard thresholding (DMD-HT). For Experiment 1 (with both high sampling rate and resolution), increasing the number of modes significantly deteriorates the prediction accuracy. However in Experiment 2 and Experiment 3 (with low sampling rates), using more DMD modes helps the predicted field to converge to the true one. For Experiment 4, with high sampling rates and lower resolution, the predicted flow field is slightly worse than that in Experiment 3 which uses the same resolution, but lower sampling rate. This implies that, for the present study, sampling rate plays an important role in DMD performance. 

This is, in particular, true for DMD-HT algorithm. That is unnecessarily increasing the number of snapshots, while adding no significant information about the system's underling dynamical evolution, results in a very big data matrix making the raw DMD basis computation and selection, Algorithm~\ref{alg:DMDHT}, ill-conditioned. Each couple of consecutive snapshots, in some sense, can be viewed as a constraint in the process of approximating the linear operator $\mathbf{A}$ in Eq.~\ref{eq:DMDmap}. Using too many snapshots (constraints) that do not contain additional information can lead to overfitting problems. This is significantly evident in Experiment 1 with both high resolutions and sampling rates. Also, the frequencies resolved by the DMD are still subjected to the Nyquist sampling theorem \cite{nyquist1928certain}, implying that the frequencies which are essential to the flow physics should be known in advance to adjust the sampling rate. 

Compared with DMD-HT results, the sorted algorithm, DMD-S, is found to give more stable results and be less prone to overfitting due to sampling rate issues. However, for the range of number of modes tested here (4, 8, and 16), this algorithm is found to be less capable of capturing the main convective dynamics. Regarding the effect of resolution, Experiments 2 and 3 infer that data resolution has less influence on DMD predictions as long as the number of snapshots is reasonable. Meanwhile, results from Experiment 4 suggest that using lower resolution can act as a filter to mitigate the overfitting issues, compared to Experiment 1.

\subsection{Time series predictions}
\begin{figure}[ht]
	\centering
	\includegraphics[width=\linewidth]{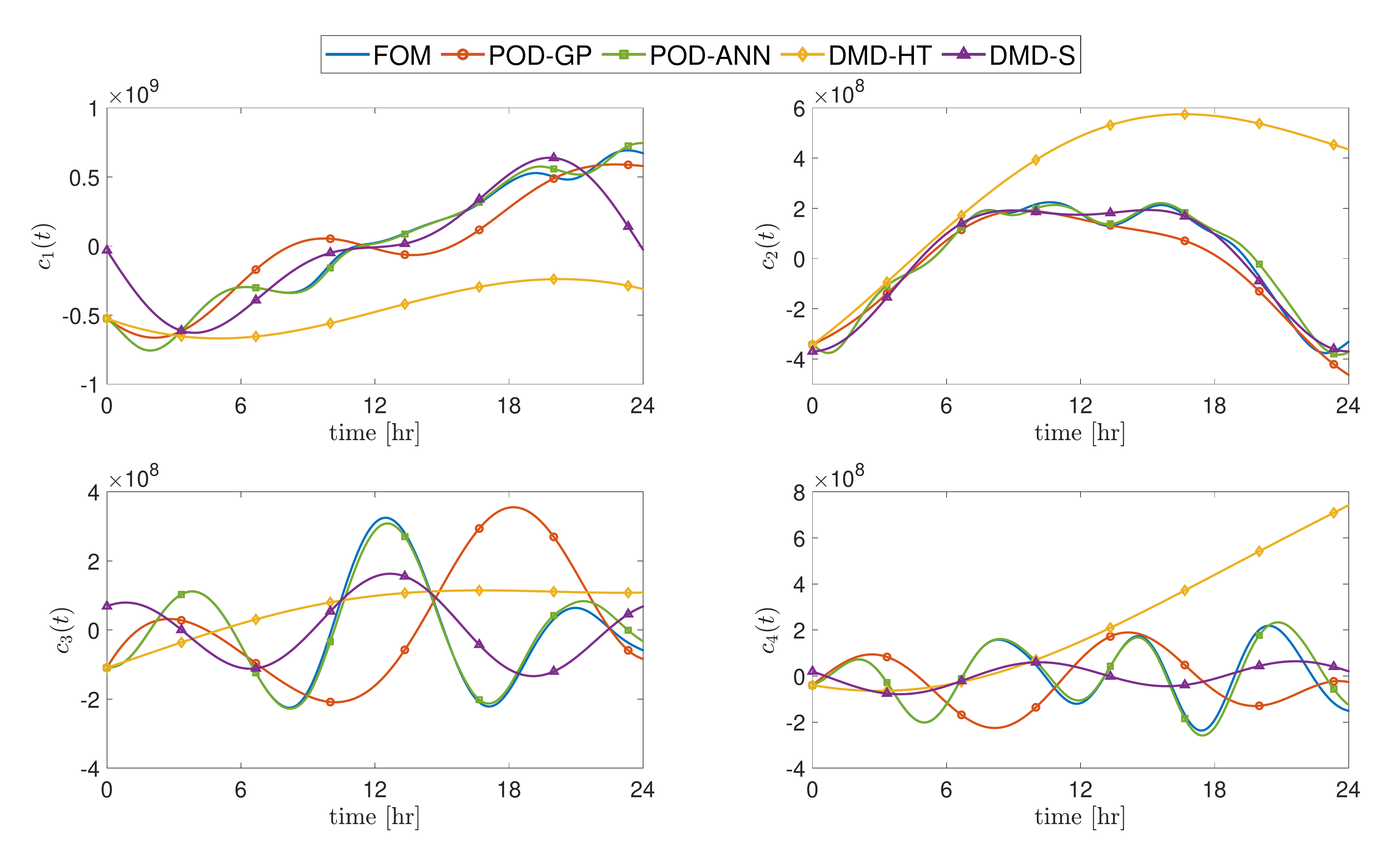}
	\caption{Time series prediction using the first 4 modes from Experiment 1}
	\label{fig:c4_exp1}
\end{figure}

\begin{figure}[ht]
	\centering
	\includegraphics[width=\linewidth]{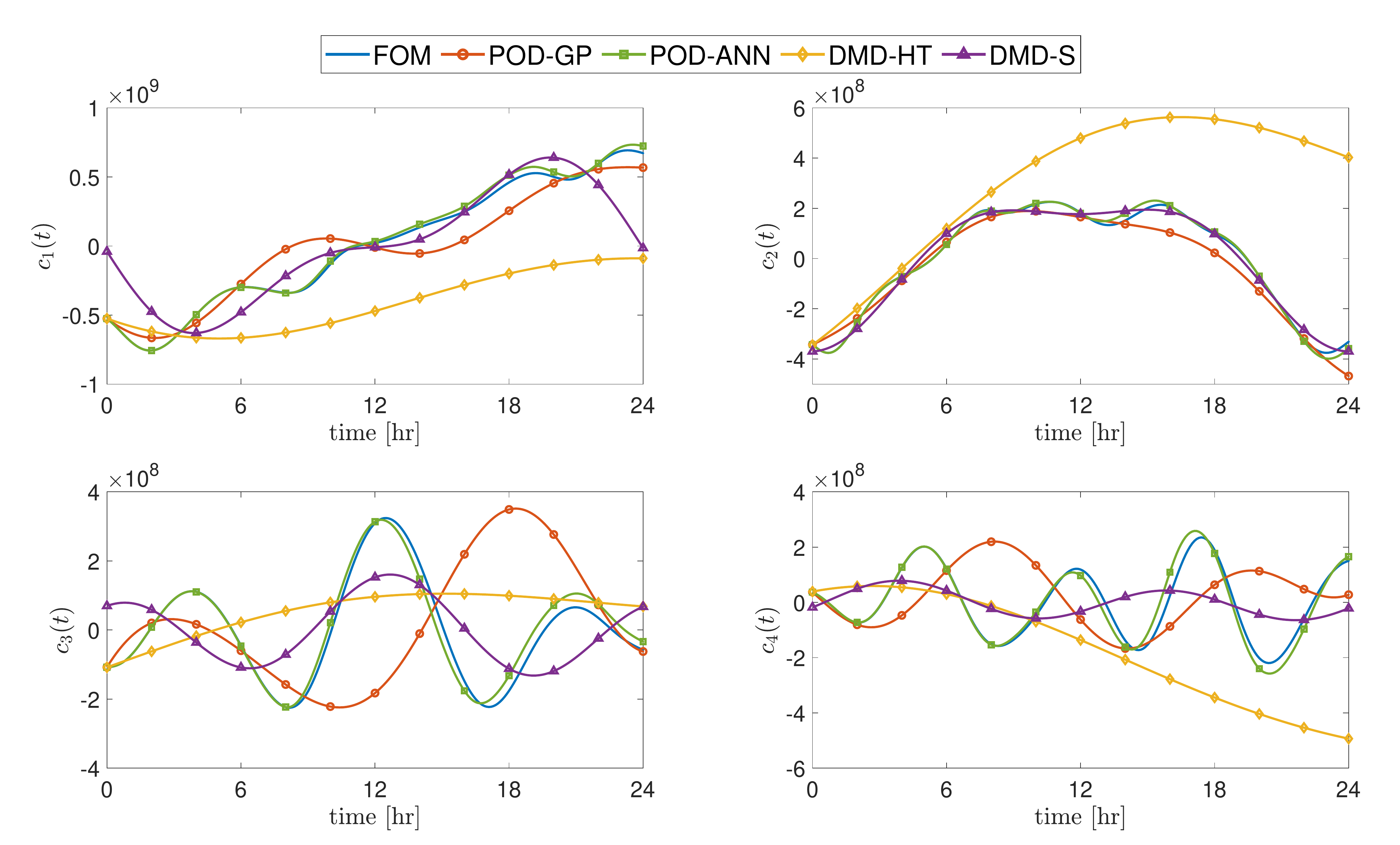}
	\caption{Time series prediction using the first 4 modes from Experiment 2}
	\label{fig:c4_exp2}
\end{figure}

\begin{figure}[ht]
	\centering
	\includegraphics[width=\linewidth]{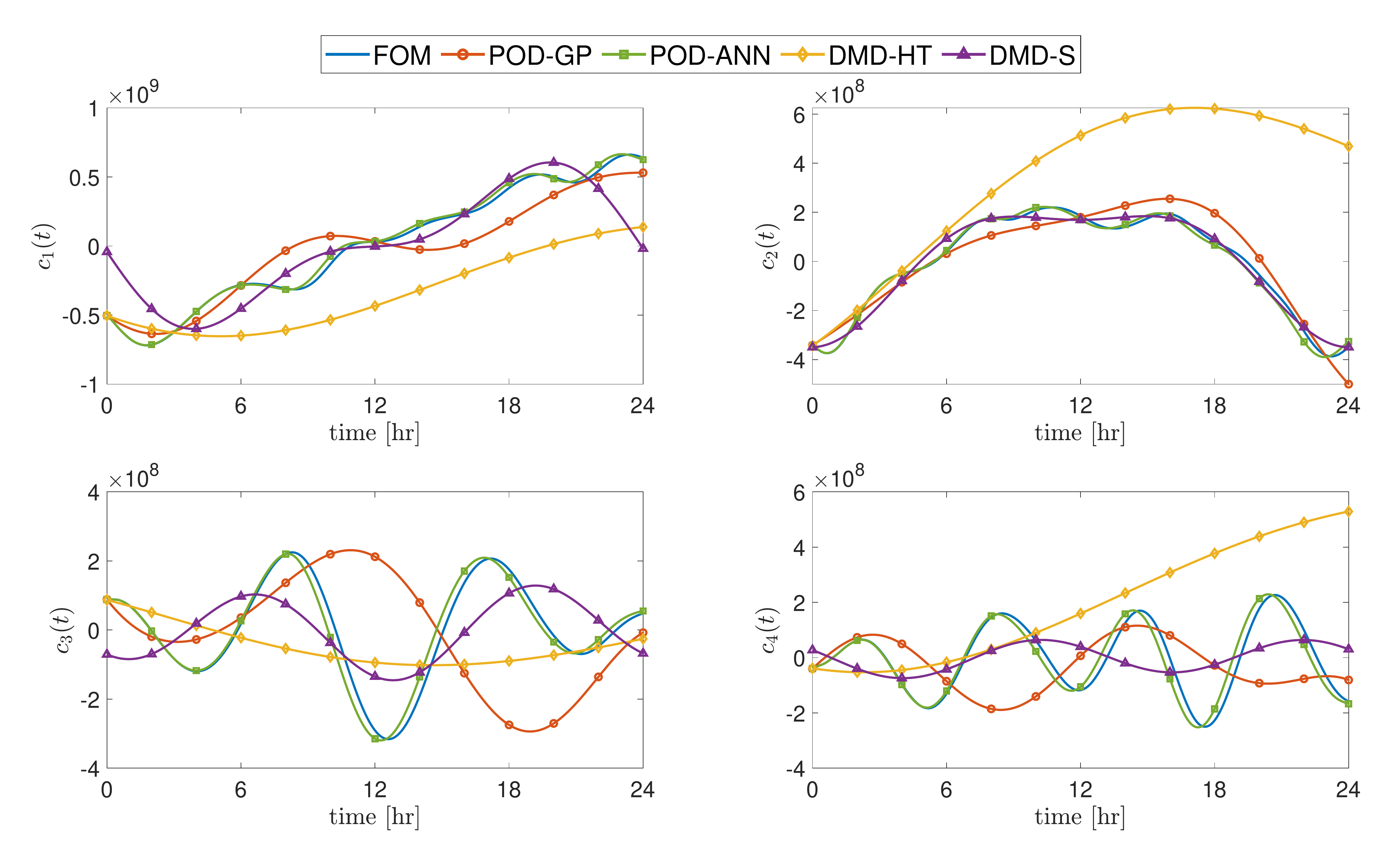}
	\caption{Time series prediction using the first 4 modes from Experiment 3}
	\label{fig:c4_exp3}
\end{figure}

\begin{figure}[ht]
	\centering
	\includegraphics[width=\linewidth]{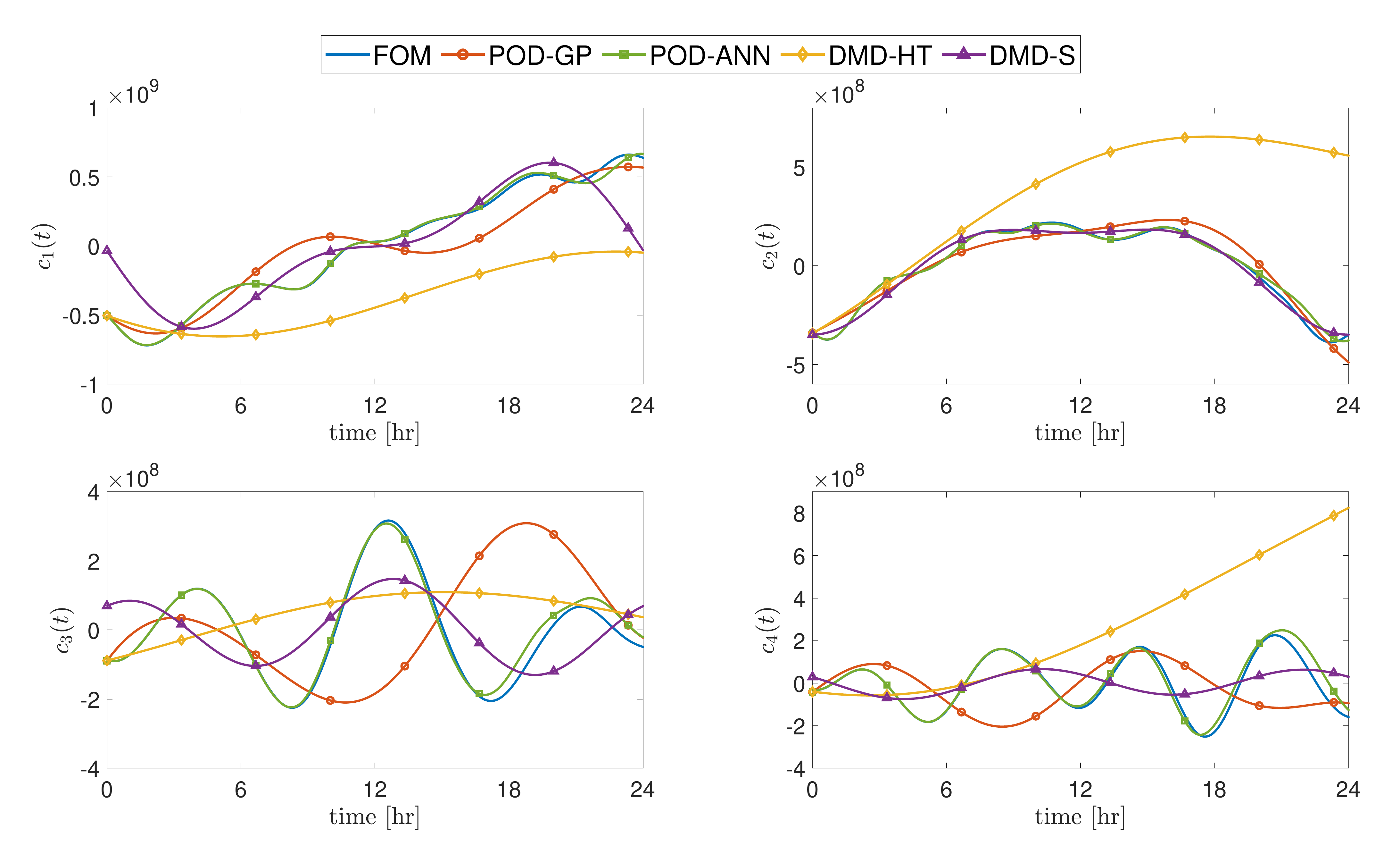}
	\caption{Time series prediction using the first 4 modes from Experiment 4}
	\label{fig:c4_exp4}
\end{figure}

Since the ANN is trained in the POD space, it is reasonable to assess its performance in that space compared to Galerkin projection. Figures~\ref{fig:c4_exp1}-\ref{fig:c4_exp4} show the performance of POD-GP, DMD and POD-ANN in different experiments using the first four modes along with the full order model (FOM) resulting field projected on these modes using Eq.~\ref{eq:projectc}. Although there is no similar time series coefficients associated with DMD algorithm, we artificially generated the DMD time series. First the flow fields are reconstructed by DMD using Eq.~\ref{eq:DMDconst1} at different times, then they are projected onto the POD space to get their corresponding modal coefficients, $c(t)$.  For the first $24$ hours, while POD-GP is incapable of capturing the main dynamics with only four modes, the ANN significantly outperforms GP predictions, being capable to better approximate the dynamical POD modal evolution. POD-ANN results are almost coincident with the true values for most of the time. As indicated in Section~\ref{sec:hresults}, resolutions and sampling rates have negligible effects on POD-based predictions, although higher sampling rates offer larger amounts of training datasets, which in turn results in a better ANN model.

\begin{figure}[ht]
	\centering
	\includegraphics[width=\linewidth]{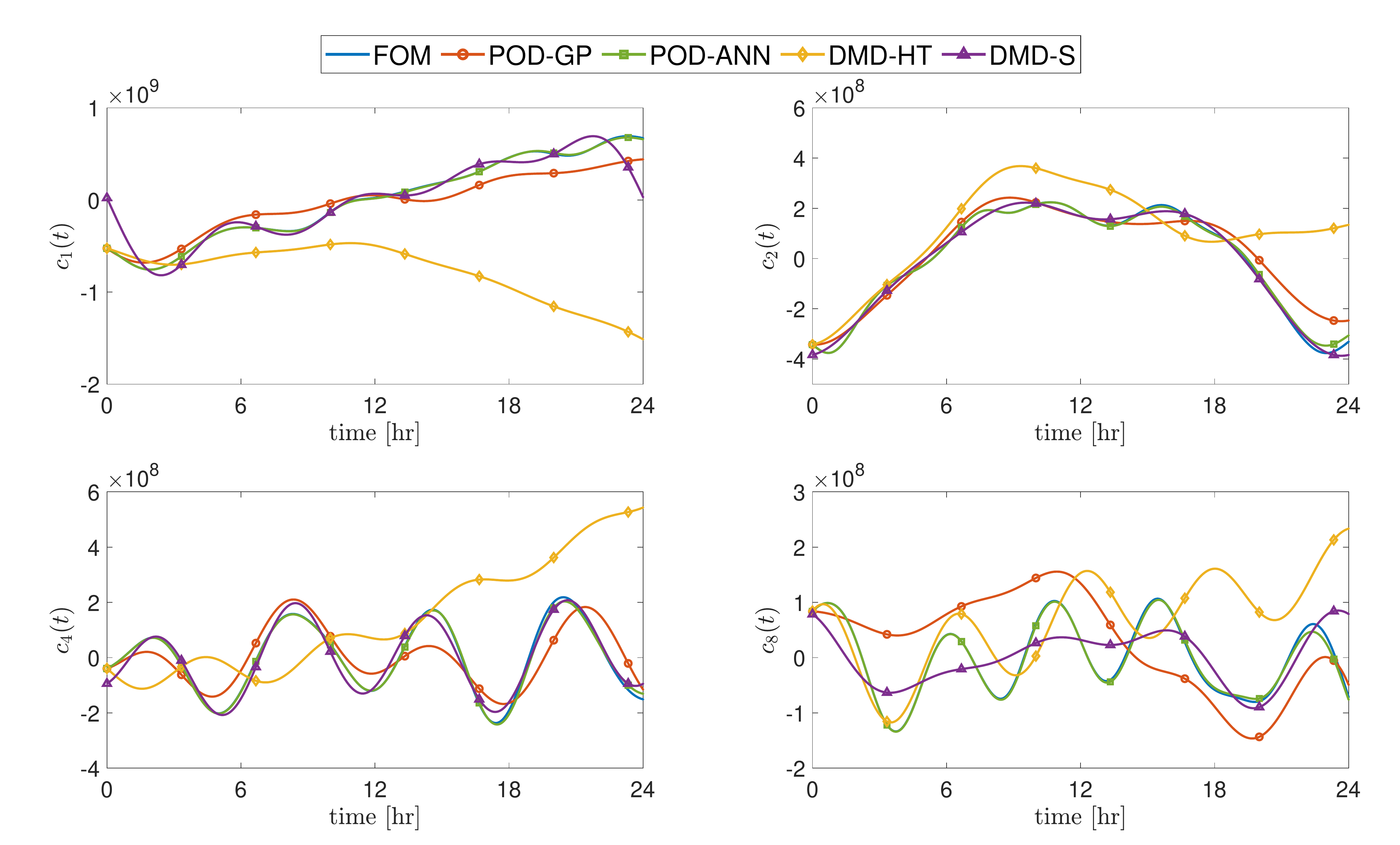}
	\caption{Time series prediction using the first 8 modes from Experiment 1}
	\label{fig:c8_exp1}
\end{figure}

\begin{figure}[ht]
	\centering
	\includegraphics[width=\linewidth]{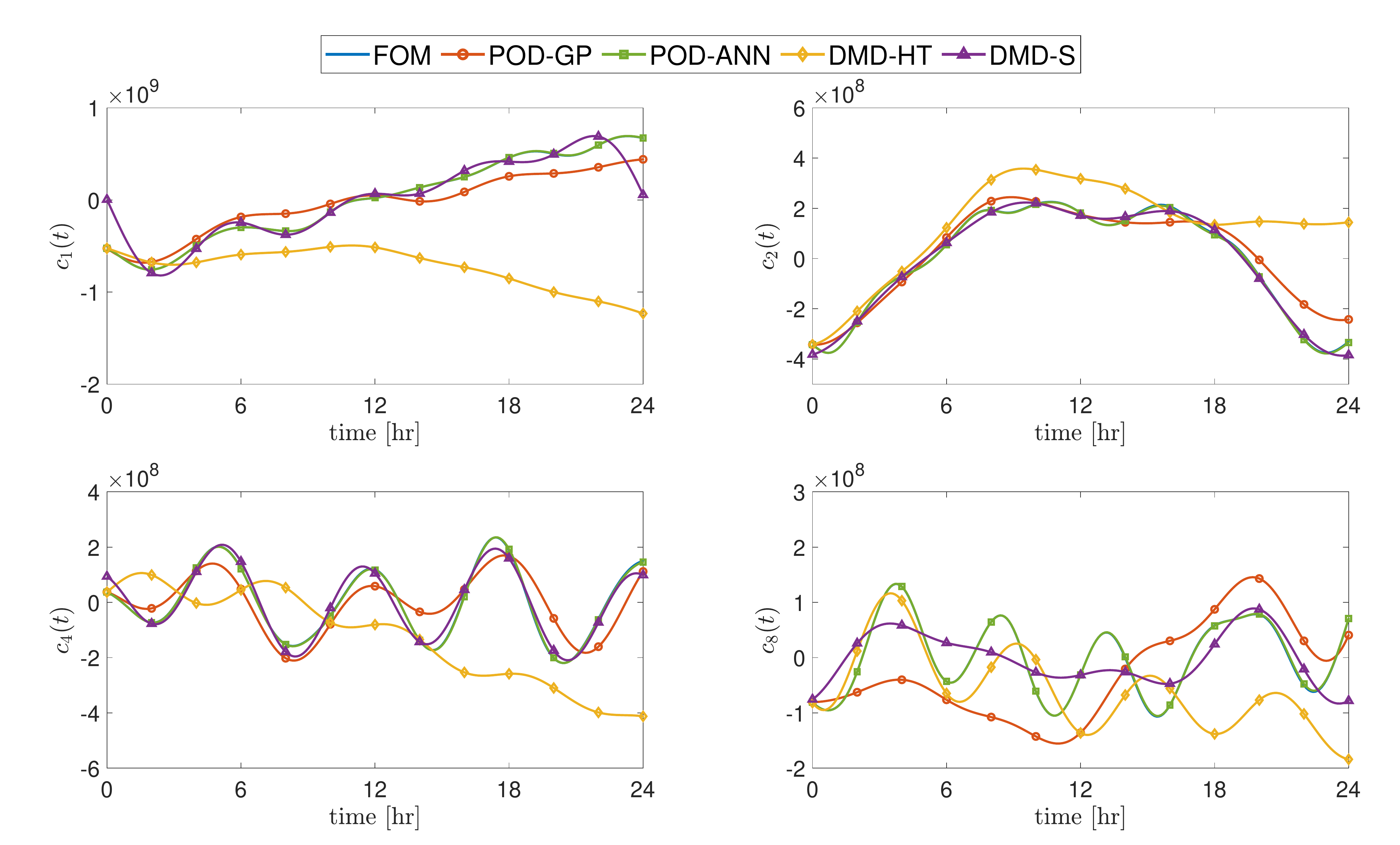}
	\caption{Time series prediction using the first 8 modes from Experiment 2}
	\label{fig:c8_exp2}
\end{figure}

\begin{figure}[ht]
	\centering
	\includegraphics[width=\linewidth]{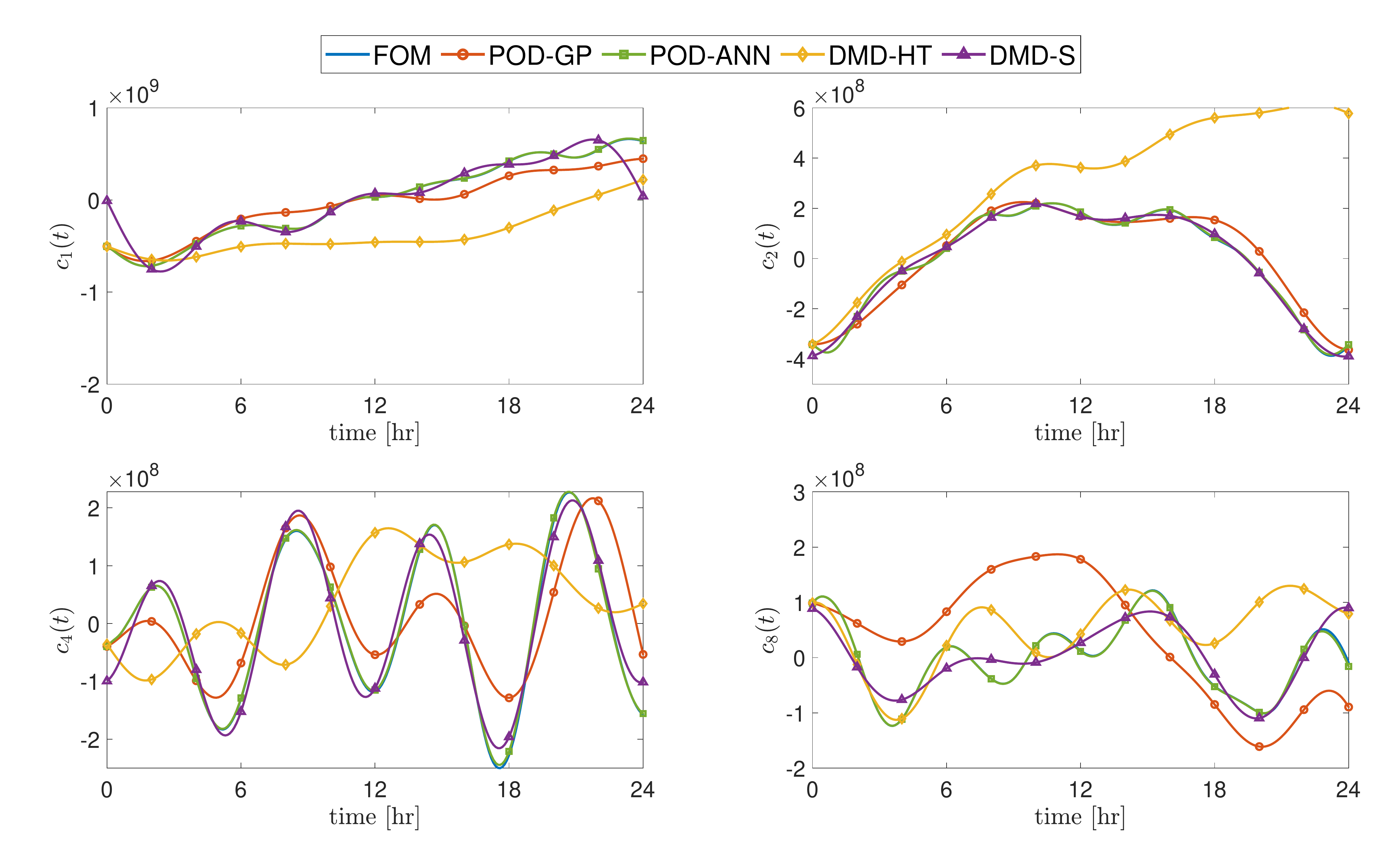}
	\caption{Time series prediction using the first 8 modes from Experiment 3}
	\label{fig:c8_exp3}
\end{figure}

\begin{figure}[ht]
	\centering
	\includegraphics[width=\linewidth]{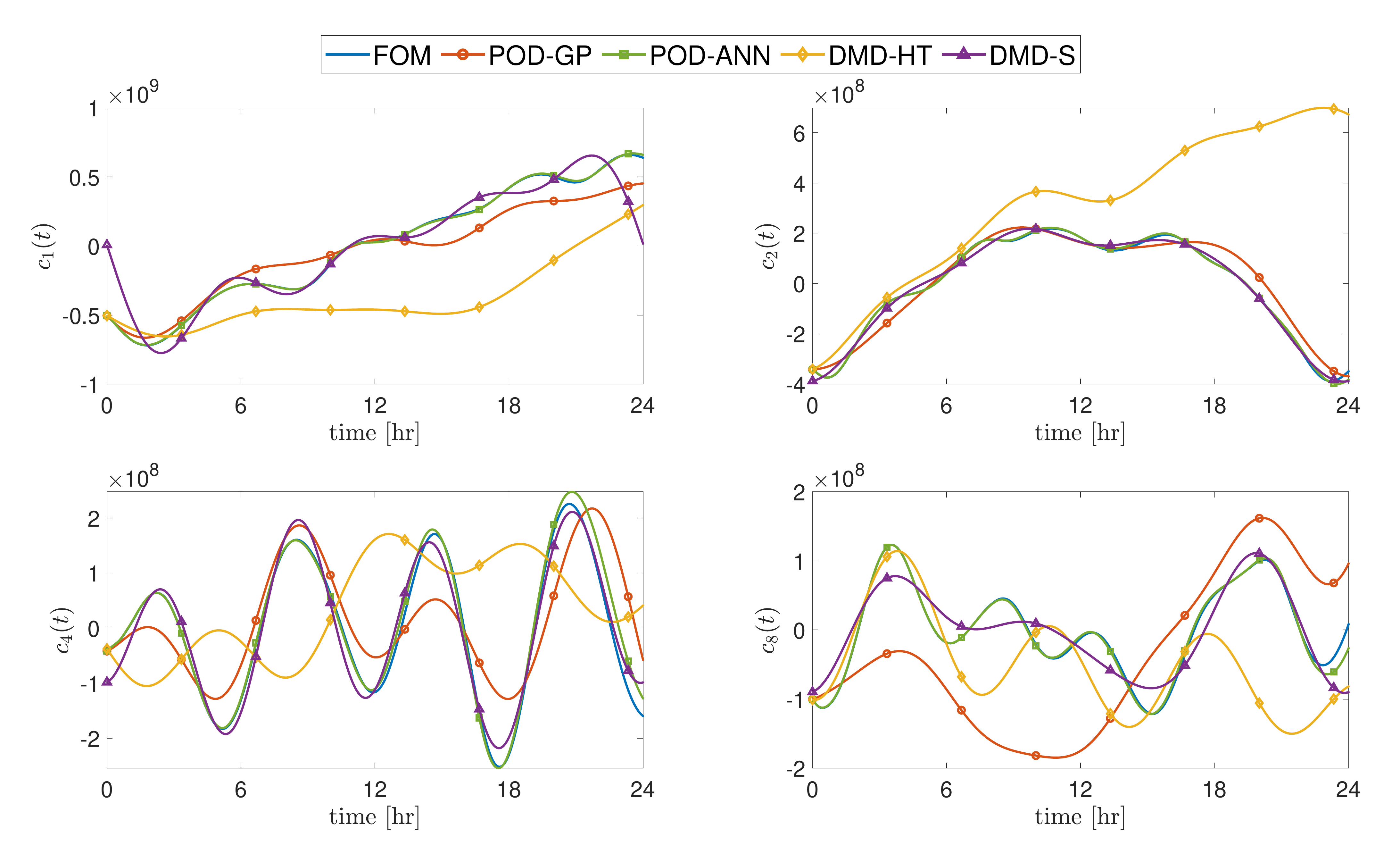}
	\caption{Time series prediction using the first 8 modes from Experiment 4}
	\label{fig:c8_exp4}
\end{figure}

\begin{figure}[ht]
	\centering
	\includegraphics[width=\linewidth]{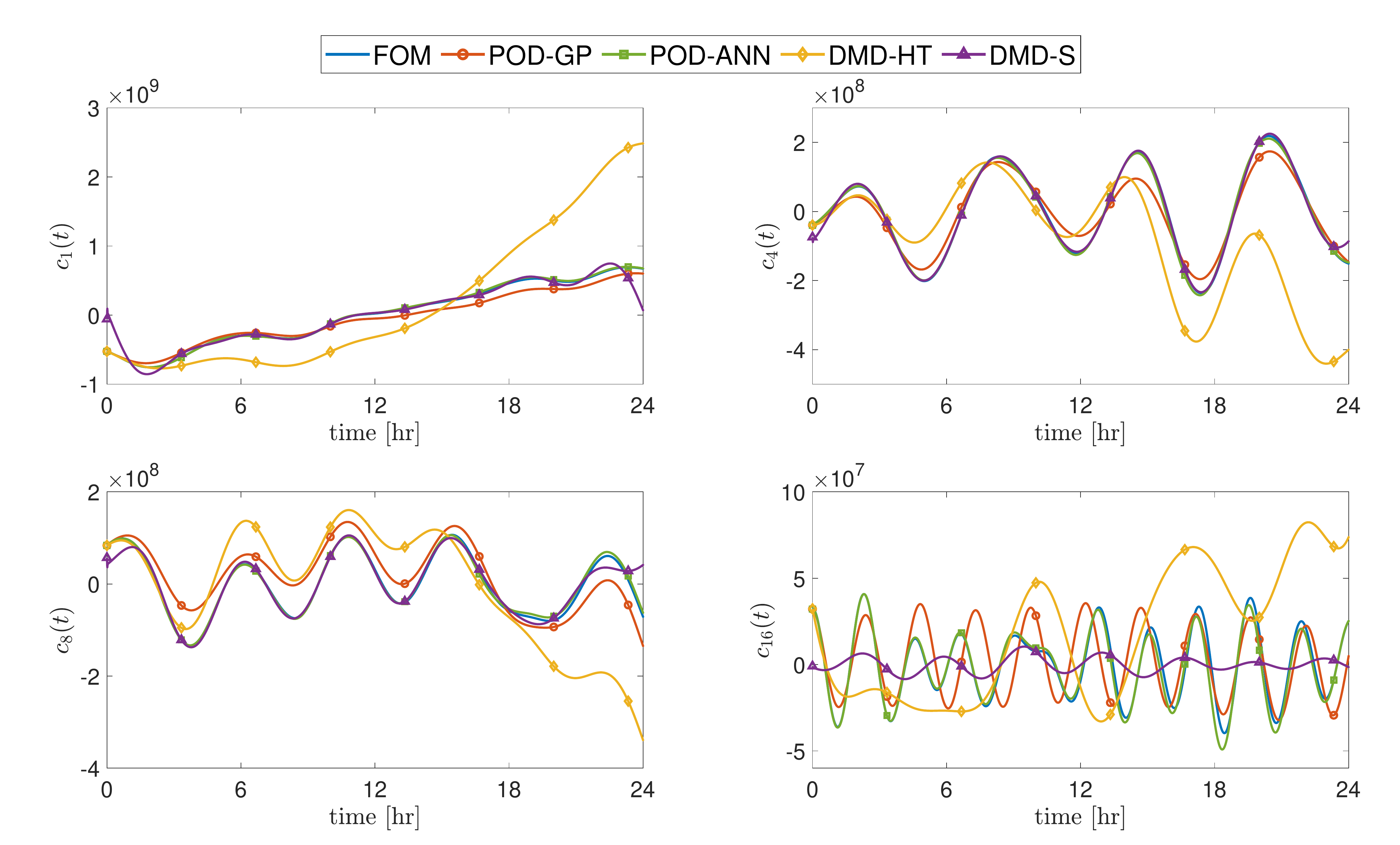}
	\caption{Time series prediction using the first 16 modes from Experiment 1}
	\label{fig:c16_exp1}
\end{figure}

\begin{figure}[ht]
	\centering
	\includegraphics[width=\linewidth]{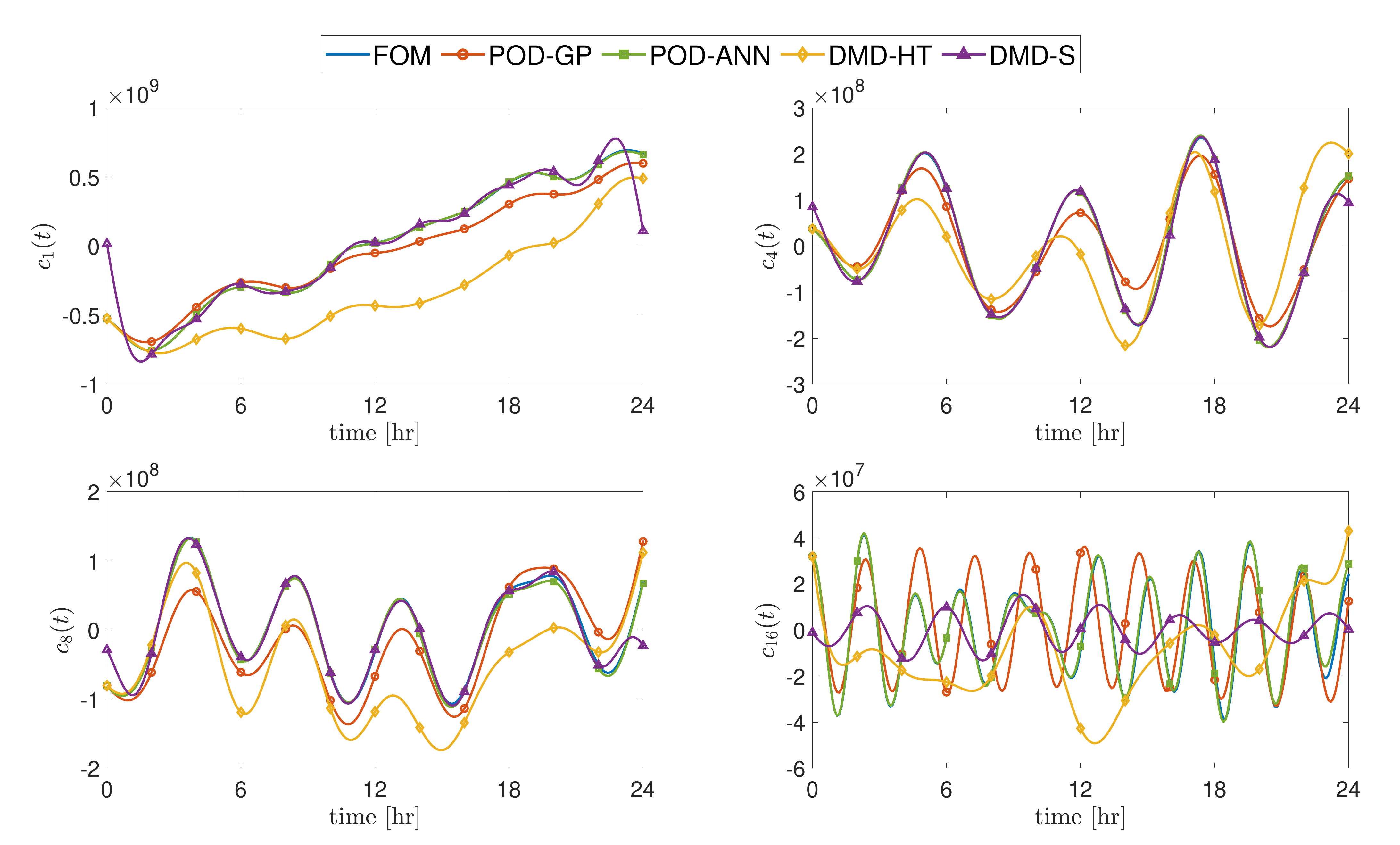}
	\caption{Time series prediction using the first 16 modes from Experiment 2}
	\label{fig:c16_exp2}
\end{figure}

\begin{figure}[ht]
	\centering
	\includegraphics[width=\linewidth]{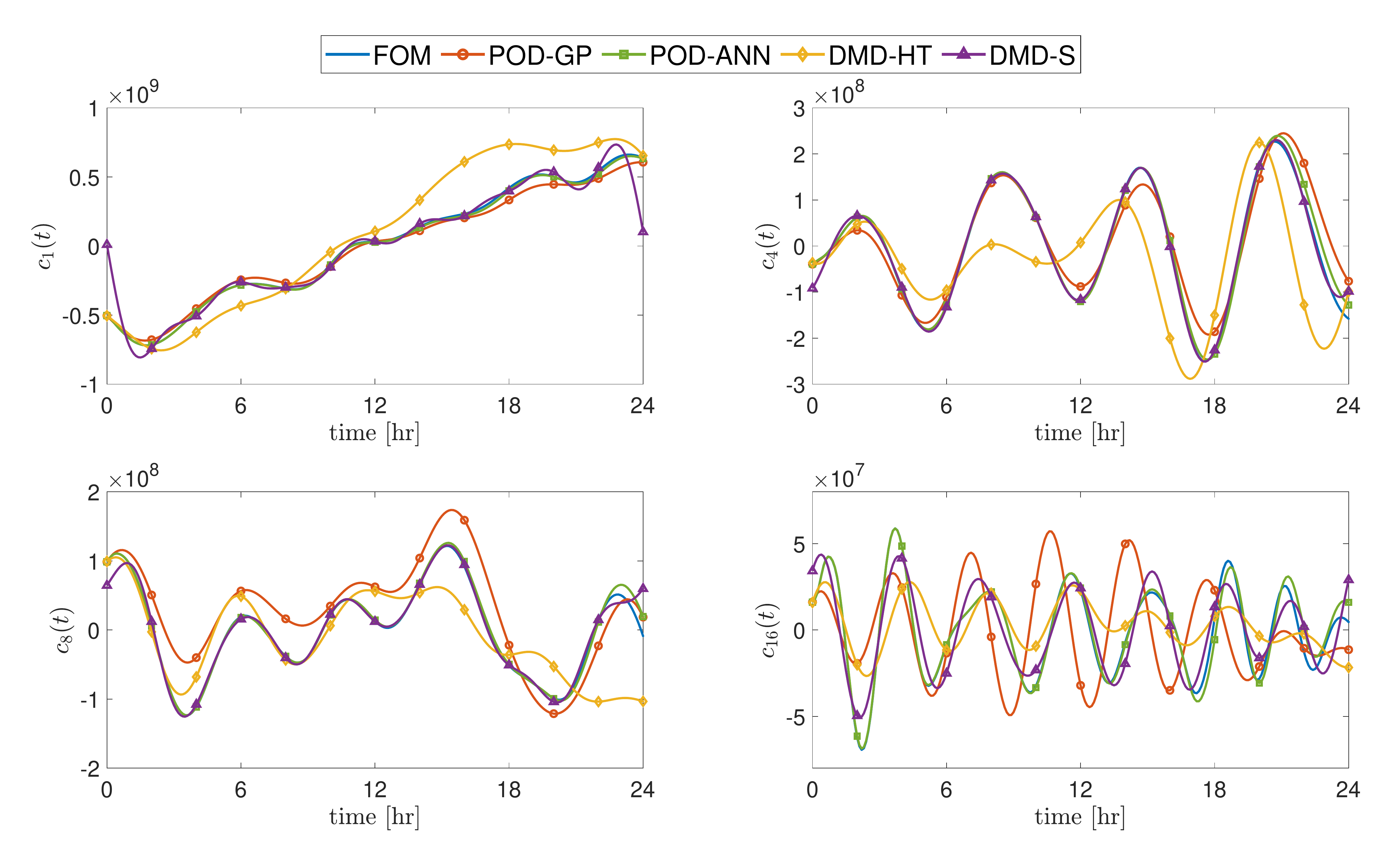}
	\caption{Time series prediction using the first 16 modes from Experiment 3}
	\label{fig:c16_exp3}
\end{figure}

\begin{figure}[ht]
	\centering
	\includegraphics[width=\linewidth]{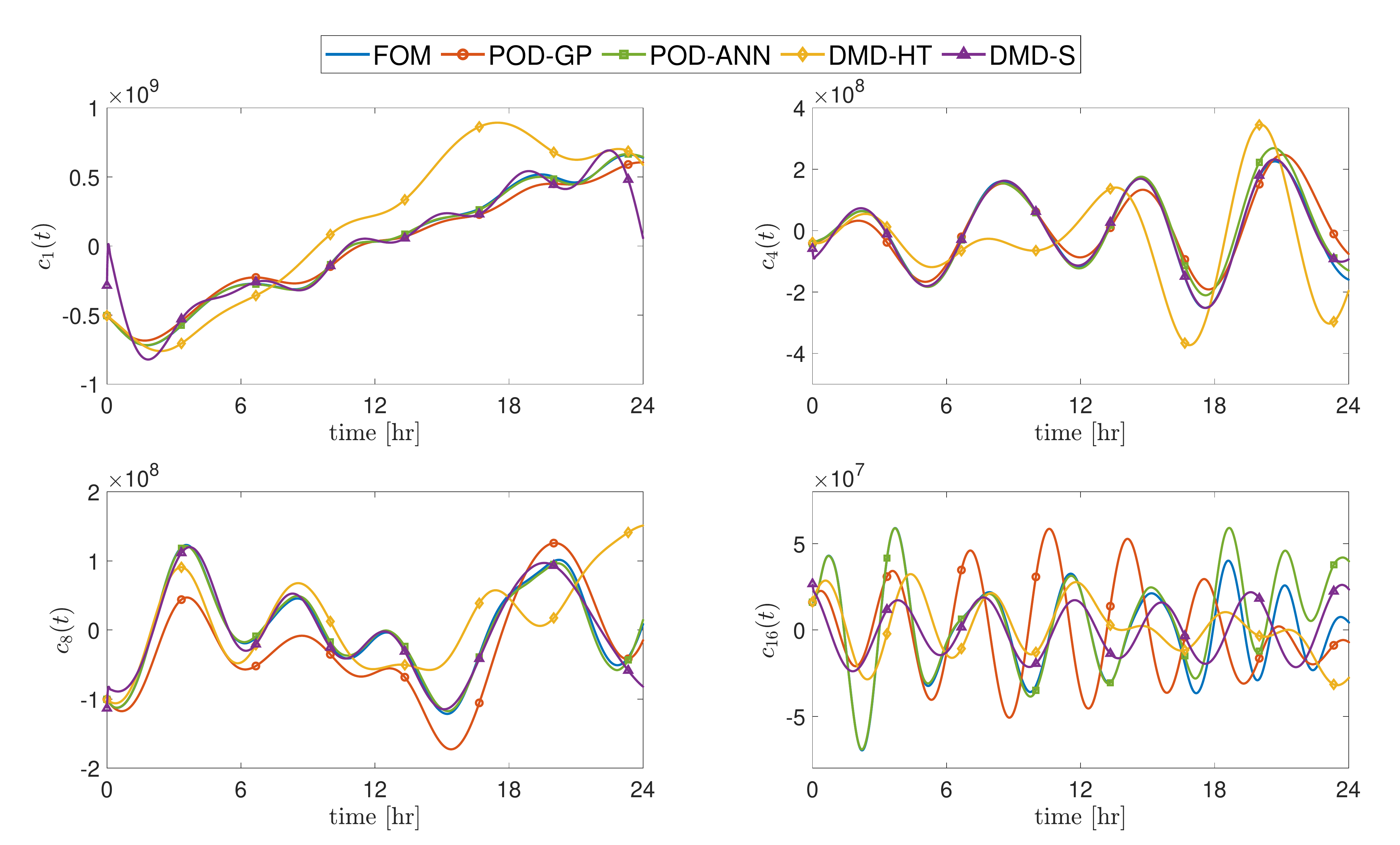}
	\caption{Time series prediction using the first 16 modes from Experiment 4}
	\label{fig:c16_exp4}
\end{figure}

Using the first 8 and 16 modes, POD-GP starts to capture the system's dynamics and follows the true trajectories, as shown in Figures~\ref{fig:c8_exp1}-\ref{fig:c16_exp4}. Meanwhile, POD-ANN performs perfectly well within the 24-hour time frame for 8 and 16 modes with cheaper cost than POD-GP, leveraging its potential as a non-intrusive ROM technique to bypass both the Galerkin projection and solving the resulting ODEs steps. It should be noted here that many other advanced and evolving machine learning-based time series prediction tools can be exploited in a similar way to the present POD-ANN framework.

\subsection{Mean subtraction}
\begin{figure}[ht]
	\centering
    \includegraphics[width=\linewidth]{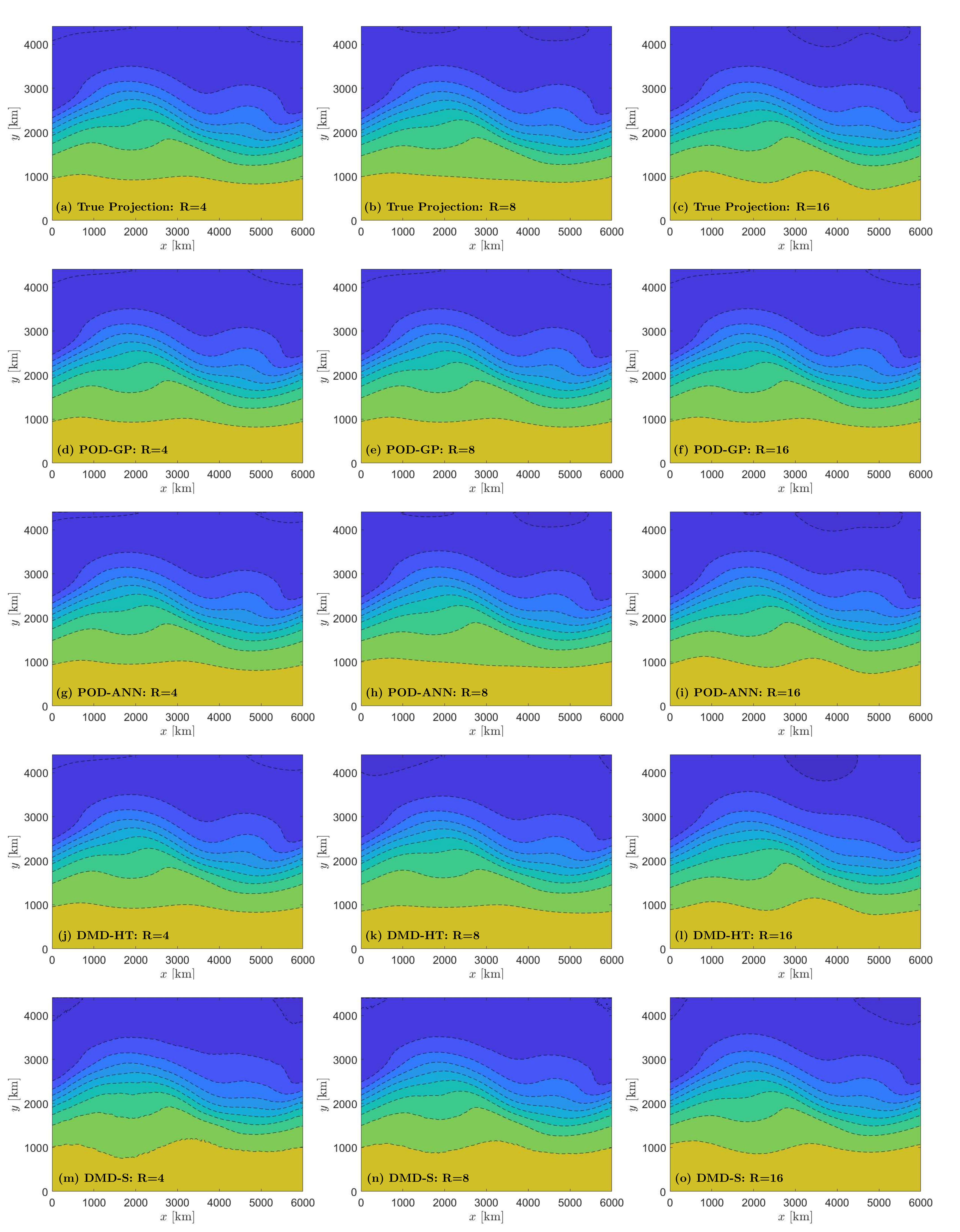}
	\caption{Geopotential field at 24 hours from Experiment 1 - obtained without mean subtraction}
	\label{fig:h_exp1no}
\end{figure}

\begin{figure}[ht]
	\centering
    \includegraphics[width=\linewidth]{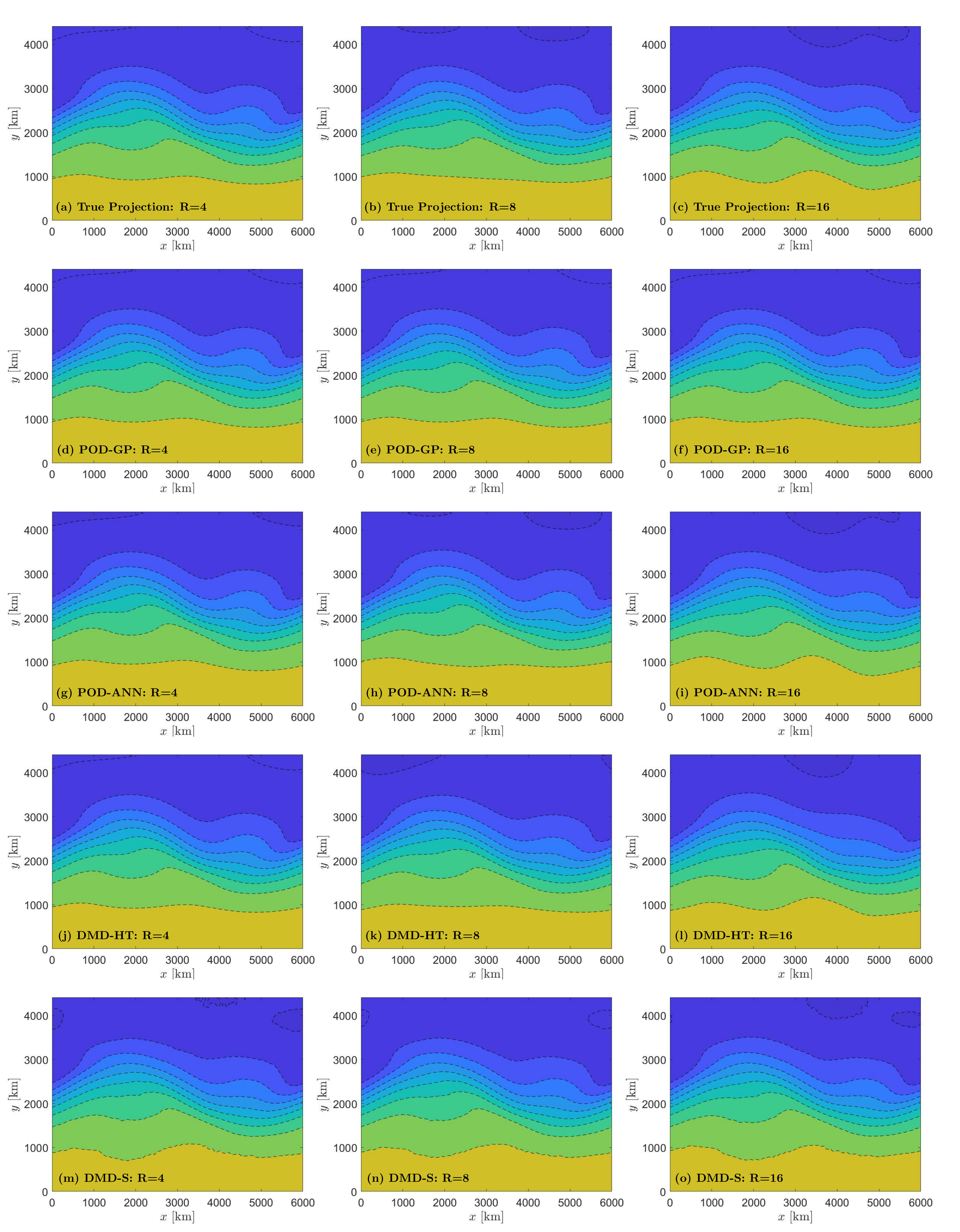}
	\caption{Geopotential field at 24 hours from Experiment 2 - obtained without mean subtraction}
	\label{fig:h_exp2no}
\end{figure}

\begin{figure}[ht]
	\centering
    \includegraphics[width=\linewidth]{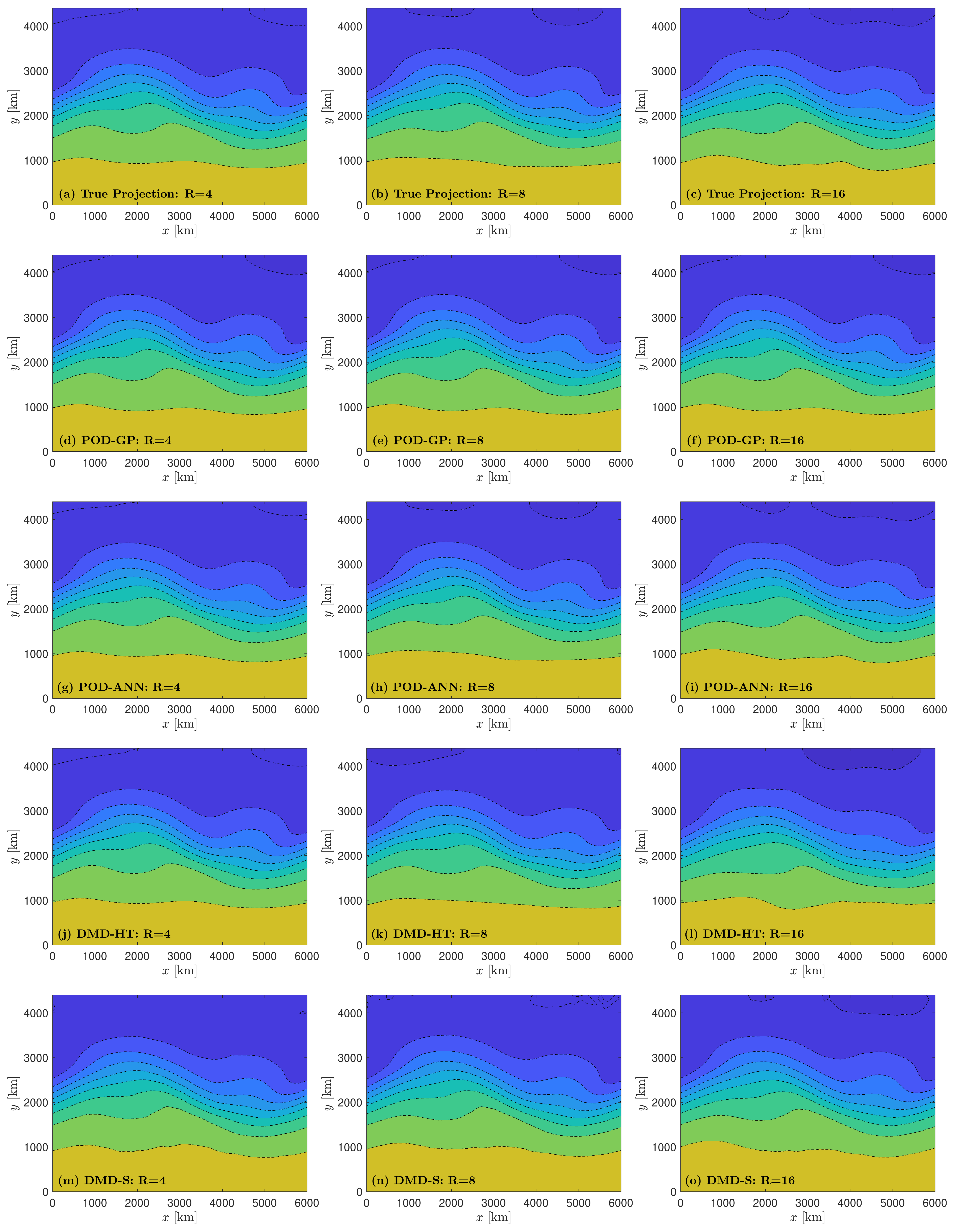}
	\caption{Geopotential field at 24 hours from Experiment 3 - obtained without mean subtraction}
	\label{fig:h_exp3no}
\end{figure}

\begin{figure}[ht]
	\centering
    \includegraphics[width=\linewidth]{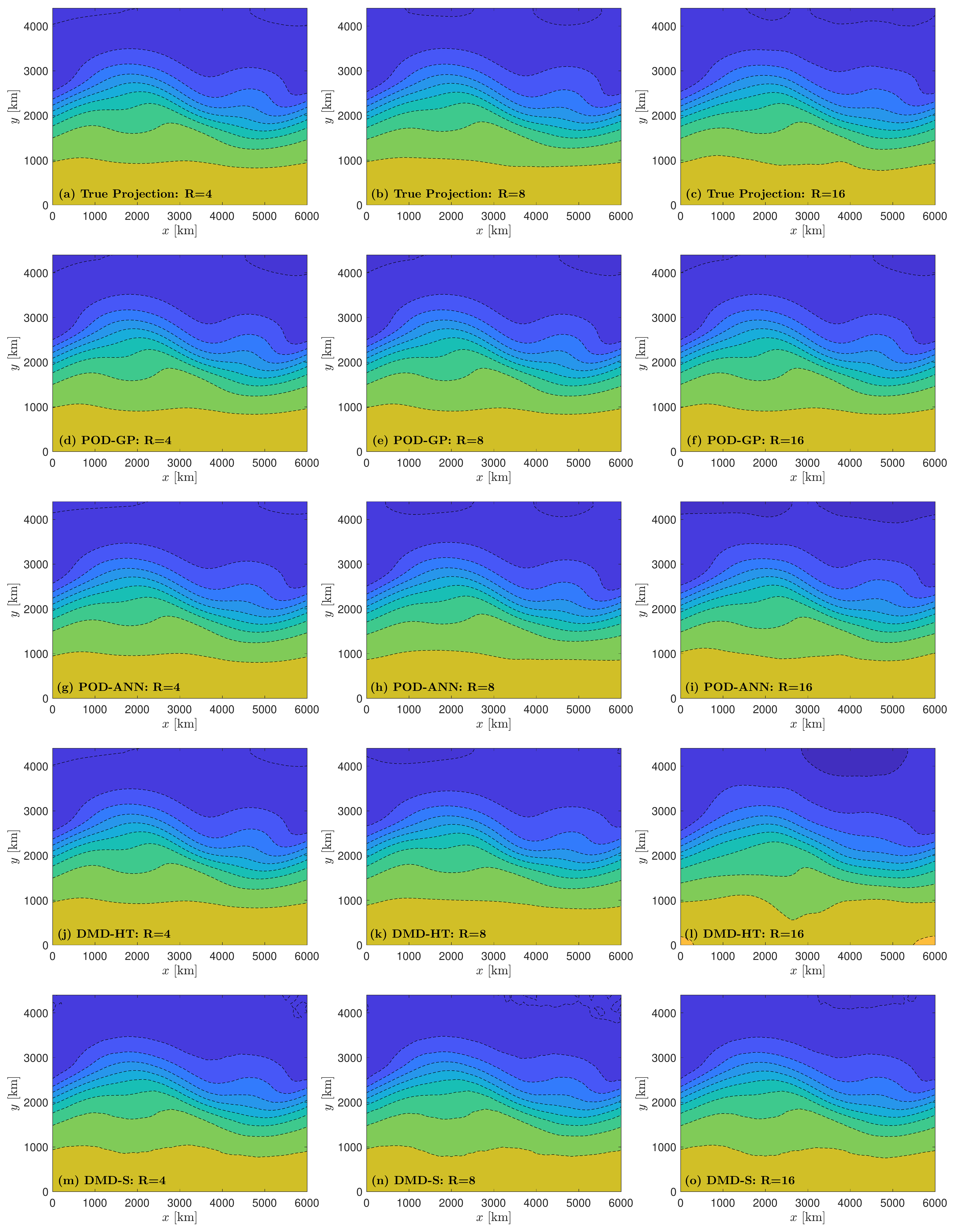}
	\caption{Geopotential field at 24 hours from Experiment 4 - obtained without mean subtraction}
	\label{fig:h_exp4no}
\end{figure}

As mentioned in Section~\ref{sec:POD}, it has been customary in ROM community to use mean-subtracted (or anomaly) fields, $w'(x_i,y_i)$, to construct ROMs. This is the case in almost every POD-based work. However, in DMD, this is not generally accurate. Chen et al. \cite{chen2012variants} showed that subtracting the mean field strips away important dynamical information, yielding a periodic trajectory, and makes the DMD algorithm incapable of determining modal growth rates. Conversely, in a very recent study, Hirsh et al. \cite{hirsh2019centering} demonstrated that centering the data through mean-subtraction is an effective pre-processing step and gives more accurate results for exact DMD spectra which might not be extracted from data without centering, especially when the dynamics matrix is full rank. Therefore, we replicated our four experiments without subtracting the mean field (i.e., using $w$ instead of $w'$) to study the effect of centering data in convective non-periodic flow dynamics. For POD, we found that final results almost matching and no major difference can be noticed, Figures~\ref{fig:h_exp1no}-\ref{fig:h_exp4no}.

To quantify the results in a more quantitative way, we use the root mean square error ($RMSE$) as an error measure, defined as
\begin{equation}
    RMSE = \sqrt{\dfrac{1}{N_xN_y} \sum_{i=1}^{N_x} \sum_{j=1}^{N_y} \left(w(x_i,y_i)-w^{FOM}(x_i,y_j) \right)^2     }.
\end{equation}

Interestingly, the $RMSE$ resulting from the projection of FOM onto the POD space is slightly smaller when using anomaly fields for decomposition, Table~\ref{table:RMSEpod}. This implies that POD modes resulting from mean-subtracted fields captures more information than those resulting from the flow field without subtraction. On the other hand, the Galerkin projection results are better in case of no mean-subtraction. This might be a result of amplification of noise retained in the mean-subtracted fields, being mapped to their origin.

\begin{table}[ht!]
	\caption{RMSE for POD at $t=24$ hours}
	\centering
	\begin{tabular}[t]{l  c  c  c  c} 
		\hline
		\multirow{2}{*}{$R$} & \multicolumn{2}{c}{\underline{\quad \quad \quad \quad \quad True Projection \quad \quad \quad \quad \quad }}  & \multicolumn{2}{c}{\underline{\quad \quad \quad \quad \quad Galerkin Projection \quad \quad \quad \quad \quad}}  \\ [0.8ex] 
		
							   & With Subtraction & Without Subtraction & With Subtraction & Without Subtraction \\  [0.8ex] 
		\hline
		\multicolumn{5}{l}{\emph{Experiment 1}}\\
		 4  & 53.99 & 61.53 & 67.20 &  62.53 \\ 	
		 8  & 43.50 & 45.71 & 71.81 & 66.37  \\ 
		 16 & 13.01 & 13.89 & 58.08 & 19.55  \medskip \\ 
		
		\multicolumn{5}{l}{\emph{Experiment 2}} \\
		4  &  53.31 & 60.86  & 67.29  &  61.91   \\ 	
		8  &  42.43 & 44.62  & 71.47  &  66.23   \\ 
		16 &  11.84 & 12.71  & 57.35  & 18.74   \medskip \\ 
		\multicolumn{5}{l}{\emph{Experiment 3}}\\
		4  &  47.21 & 56.58  &  62.15 &  59.96  \\ 	
		8  &  41.57 & 41.61  &  68.84 & 65.86   \\ 
		16 &  18.04 & 18.06  &  45.01 & 28.72  \medskip \\ 
		\multicolumn{5}{l}{\emph{Experiment 4}}\\
		4  & 47.87  &  57.31  &  58.39  & 61.09   \\ 	
		8  & 42.42  &  42.46  &  68.65  & 65.43   \\ 
		16 & 19.85  &  19.87  &  46.06  & 29.72   \\ 
		\hline
	\end{tabular}
	
	\label{table:RMSEpod}
\end{table}

\begin{table}[ht!]
	\caption{RMSE for POD-ANN at $t=24$ hours}
	\centering
	\begin{tabular}[t]{l  c  c } 
		\hline
		\multirow{2}{*}{$R$} & \multicolumn{2}{c}{\underline{\quad \quad \quad \quad \quad POD-ANN \quad \quad \quad \quad \quad }}    \\ [0.8ex] 
		
							   & With Subtraction & Without Subtraction  \\  [0.8ex] 
		\hline
		\multicolumn{3}{l}{\emph{Experiment 1}}\\
		 4  & 56.90 &  63.82 \\ 	
		 8  & 44.46 &  47.30  \\ 
		 16 & 14.33 &  18.50  \medskip \\ 
		
		\multicolumn{3}{l}{\emph{Experiment 2}} \\
		4  & 54.77  & 62.04    \\ 	
		8  & 42.46 & 53.77     \\ 
		16 & 12.41  & 14.24     \medskip \\ 
		\multicolumn{3}{l}{\emph{Experiment 3}}\\
		4  & 47.55  &  60.39   \\ 	
		8  & 41.64  &  43.24    \\ 
		16 & 22.94  &  20.48   \medskip \\ 
		\multicolumn{3}{l}{\emph{Experiment 4}}\\
		4  & 49.23  &  63.03     \\ 	
		8  & 44.48  &  50.39     \\ 
		16 & 31.59  &  35.46     \\ 
		\hline
	\end{tabular}
	
	\label{table:RMSEpodANN}
\end{table}

\begin{figure}[ht]
	\centering
	\includegraphics[width=\linewidth]{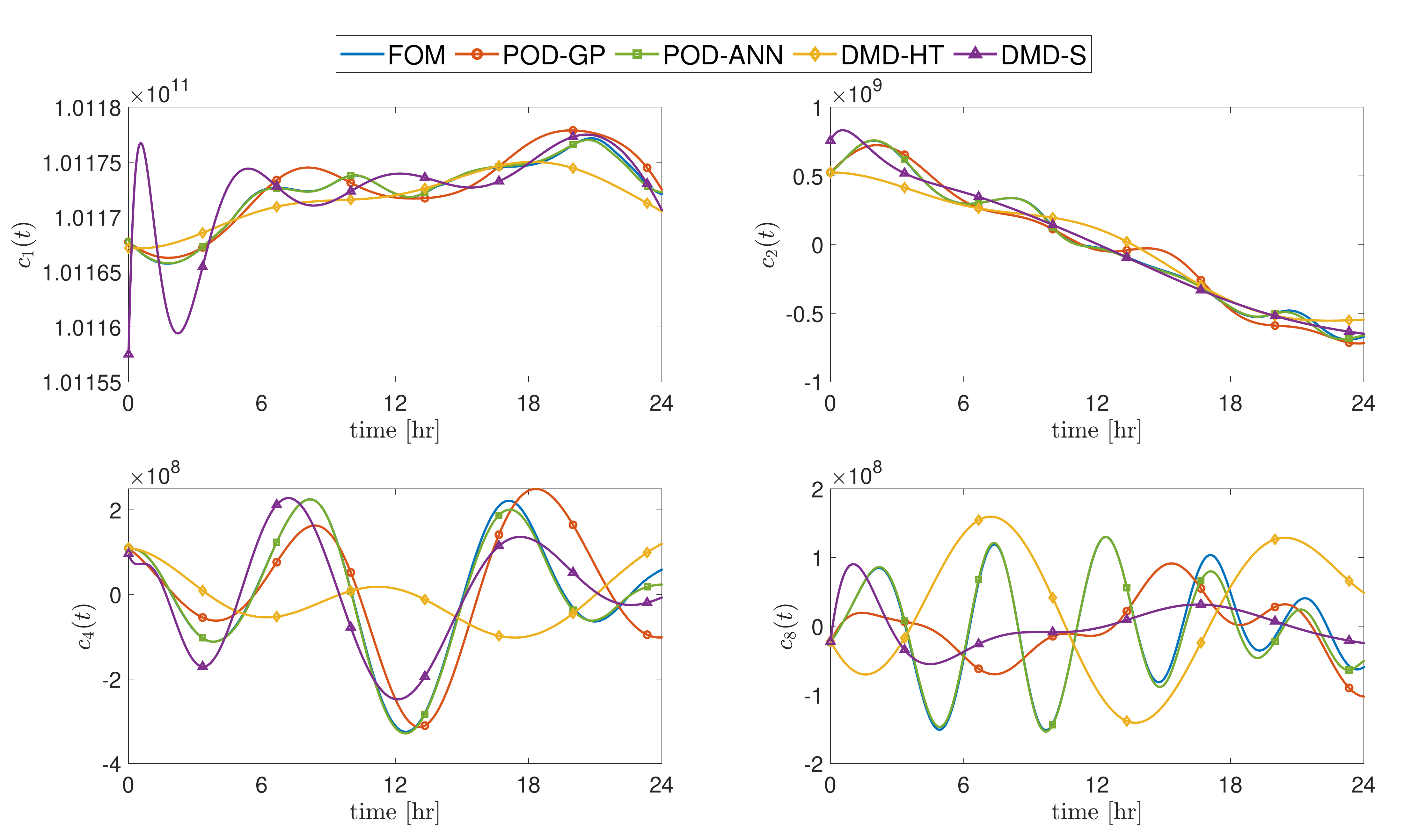}
	\caption{Time series prediction using the first 8 modes from Experiment 1 - obtained without mean subtraction}
	\label{fig:c8_exp1NO}
\end{figure}

\begin{figure}[ht]
	\centering
	\includegraphics[width=\linewidth]{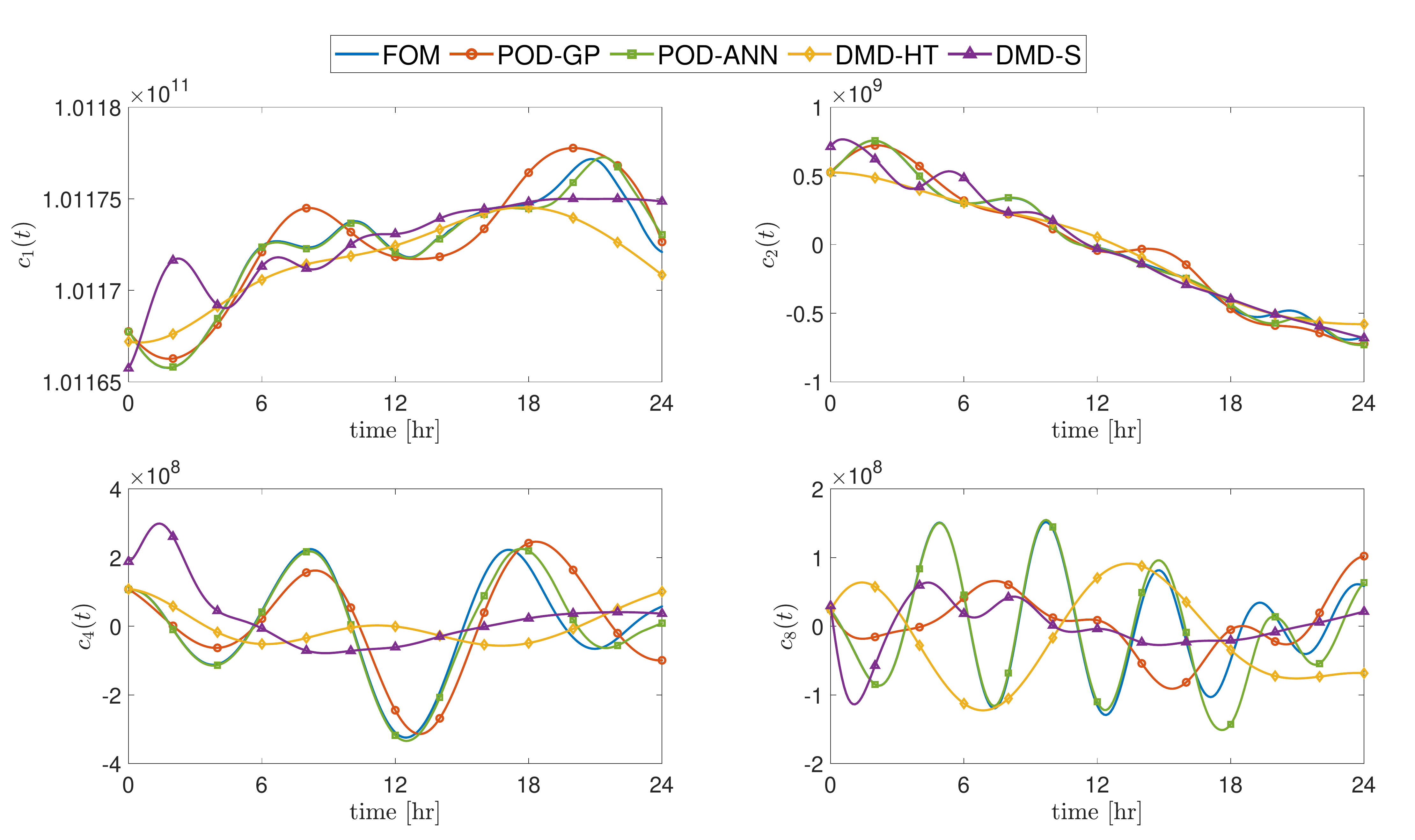}
	\caption{Time series prediction using the first 8 modes from Experiment 2 - obtained without mean subtraction}
	\label{fig:c8_exp2NO}
\end{figure}

\begin{figure}[ht]
	\centering
	\includegraphics[width=\linewidth]{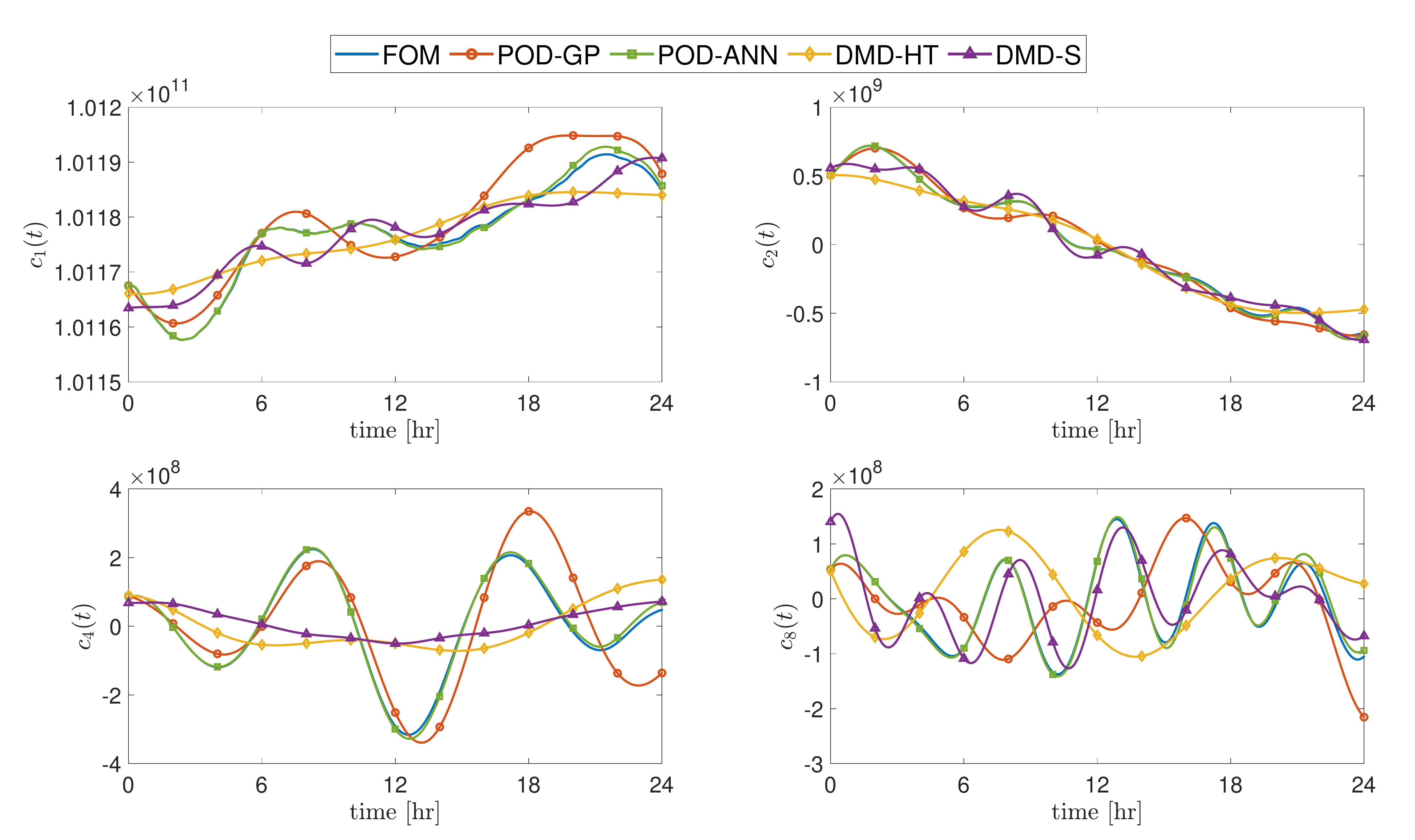}
	\caption{Time series prediction using the first 8 modes from Experiment 3 - obtained without mean subtraction}
	\label{fig:c8_exp3NO}
\end{figure}

\begin{figure}[ht]
	\centering
	\includegraphics[width=\linewidth]{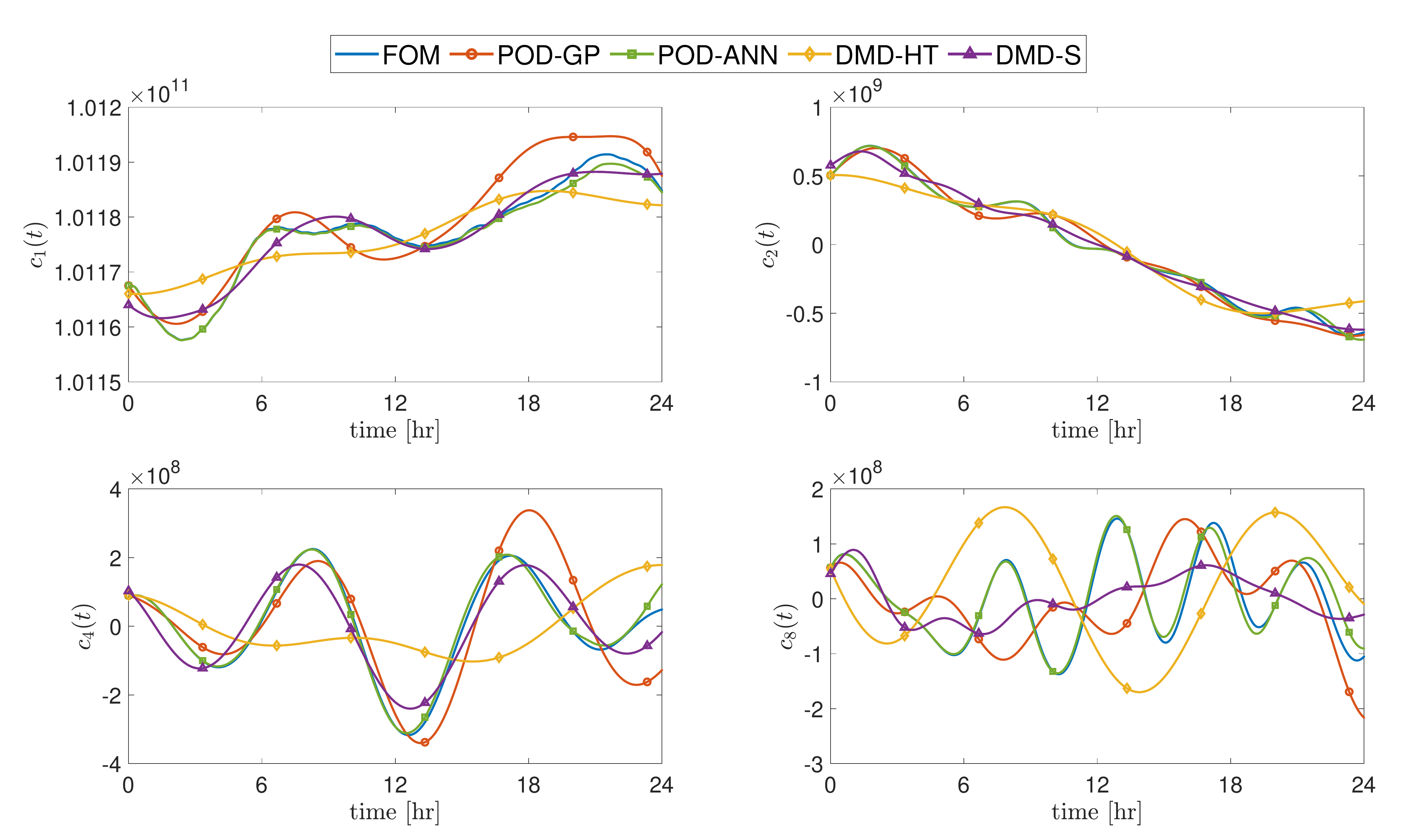}
	\caption{Time series prediction using the first 8 modes from Experiment 4 - obtained without mean subtraction}
	\label{fig:c8_exp4NO}
\end{figure}

\begin{figure}[ht]
	\centering
	\includegraphics[width=\linewidth]{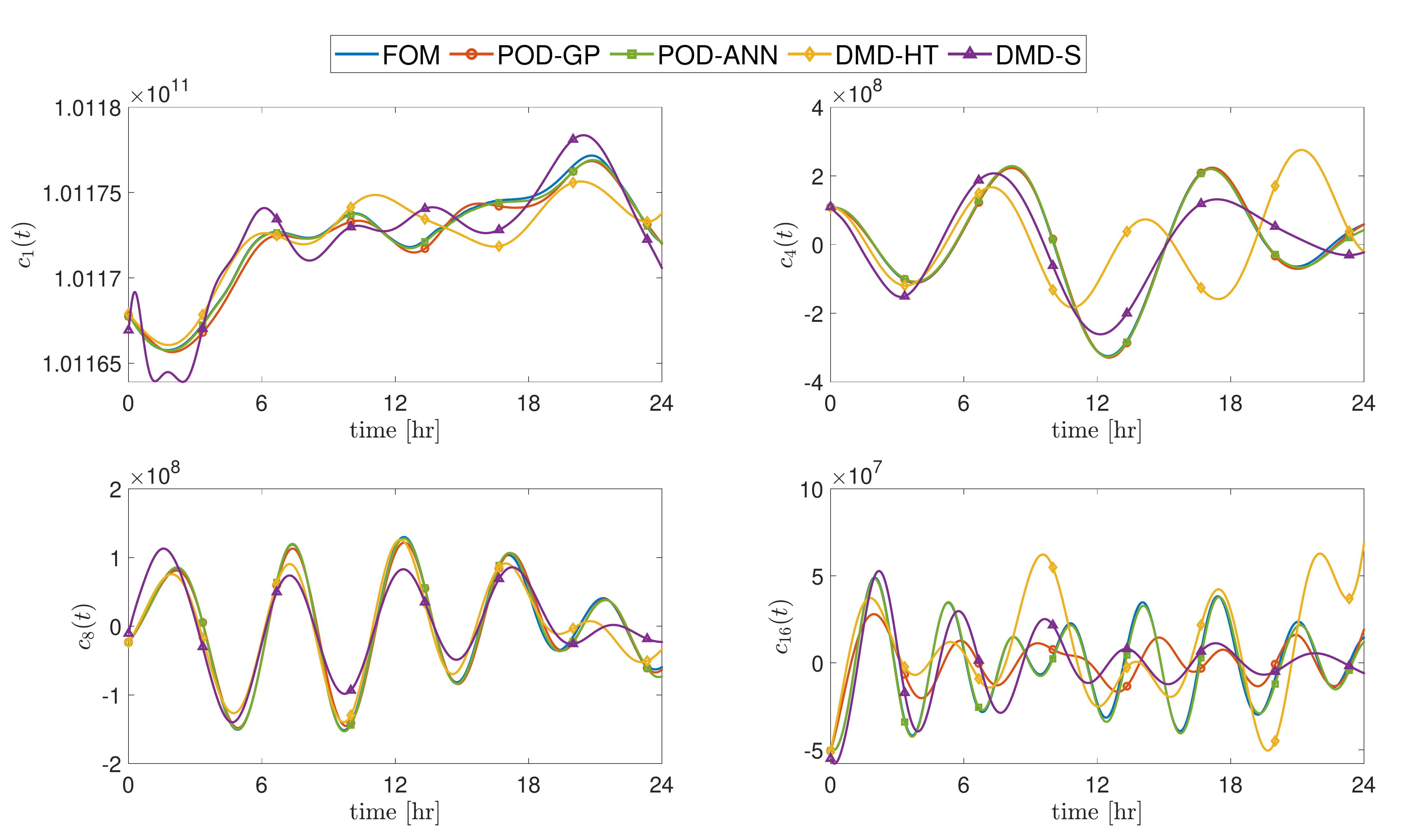}
	\caption{Time series prediction using the first 16 modes from Experiment 1 - obtained without mean subtraction}
	\label{fig:c16_exp1NO}
\end{figure}

\begin{figure}[ht]
	\centering
	\includegraphics[width=\linewidth]{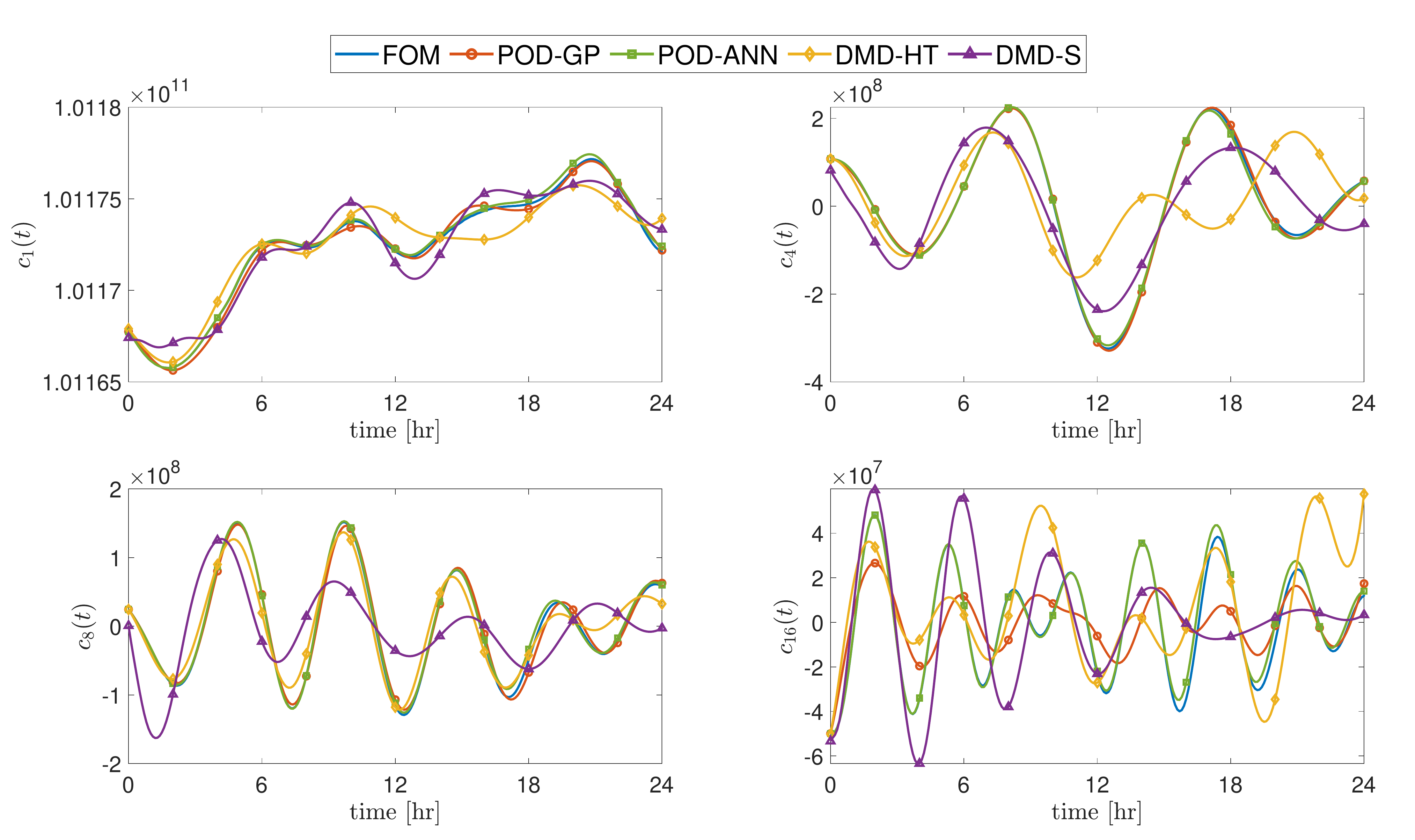}
	\caption{Time series prediction using the first 16 modes from Experiment 2 - obtained without mean subtraction}
	\label{fig:c16_exp2NO}
\end{figure}

\begin{figure}[ht]
	\centering
	\includegraphics[width=\linewidth]{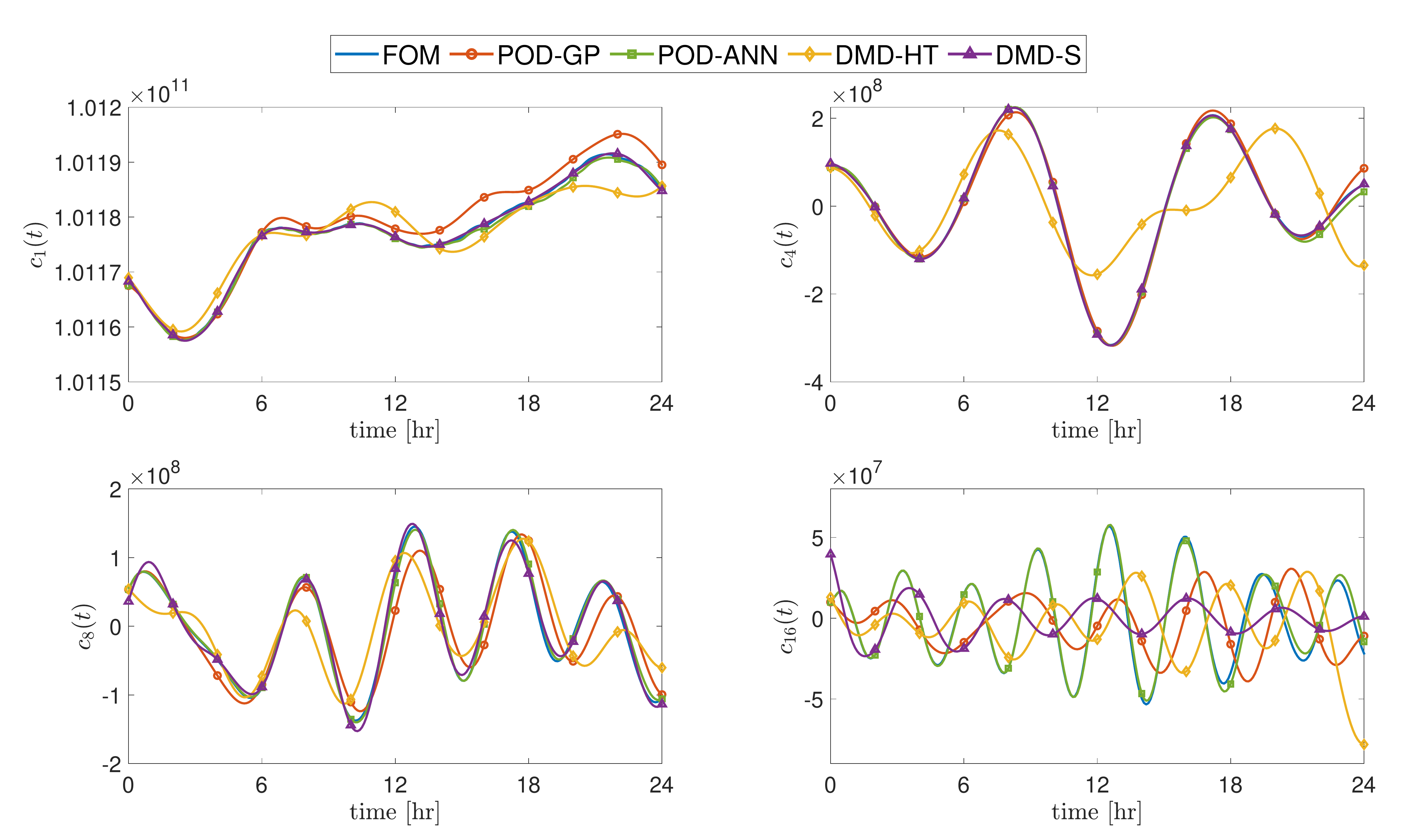}
	\caption{Time series prediction using the first 16 modes from Experiment 3 - obtained without mean subtraction}
	\label{fig:c16_exp3NO}
\end{figure}

\begin{figure}[ht]
	\centering
	\includegraphics[width=\linewidth]{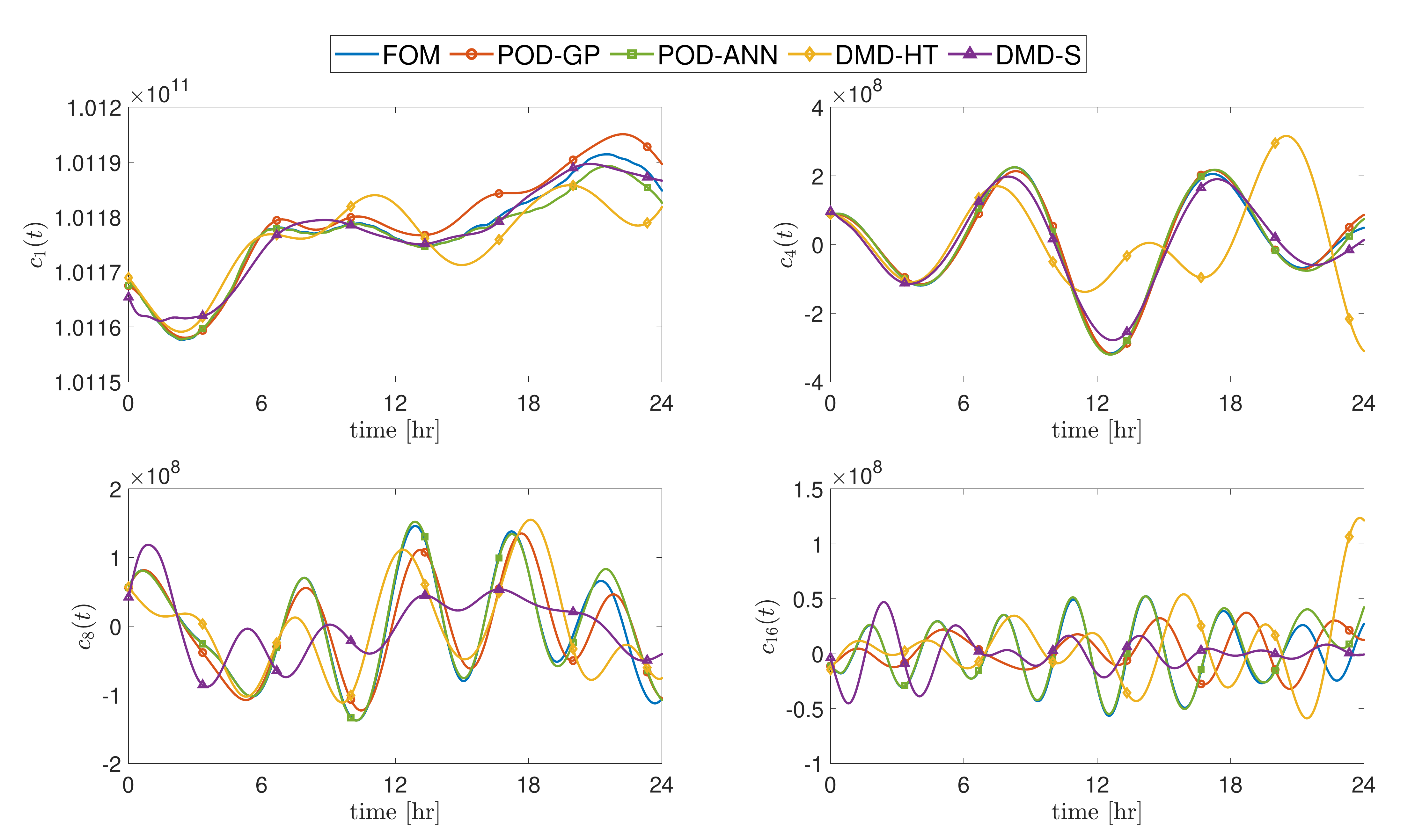}
	\caption{Time series prediction using the first 16 modes from Experiment 4 - obtained without mean subtraction}
	\label{fig:c16_exp4NO}
\end{figure}

On the other hand, the DMD results are dramatically affected by this subtraction. This can be easily seen from the reconstructed geopotential height fields, shown in Figures~\ref{fig:h_exp1no}-\ref{fig:h_exp4no}, compared to Figures~\ref{fig:h_exp1}-\ref{fig:h_exp4}. Also, the $RMSE$ for cases with and without mean subtraction is shown in Table~\ref{table:RMSEdmd}. Similar observations can be obtained from plotting the timeseries prediction, Figures~\ref{fig:c8_exp1NO}-\ref{fig:c16_exp4NO}.

\begin{table}[ht!]
	\caption{RMSE for DMD at $t=24$ hours}
	\centering
	\begin{tabular}[t]{l  c  c  c  c} 
		\hline
		\multirow{2}{*}{$R$} & \multicolumn{2}{c}{\underline{\  \quad \quad \quad DMD [Hard Threshold] \  \quad \quad \quad}}  & \multicolumn{2}{c}{\underline{\quad \quad \quad \quad \quad DMD [Sorted]\quad \quad \quad \quad \quad}}  \\ [0.8ex] 
		& With Subtraction & Without Subtraction & With Subtraction & Without Subtraction \\  [0.8ex] 
		\hline
		\multicolumn{5}{l}{\emph{Experiment 1}}\\
		4  & 304.80  & 61.83  & 152.40  & 83.56  \\ 	
		8  & 543.08  & 70.84  & 138.01  & 62.10  \\ 
		16 & 381.83  & 62.75  & 128.53  & 53.98 \medskip \\ 
		\multicolumn{5}{l}{\emph{Experiment 2}}\\
		4  &  247.22 &  62.69  &  150.09  &  44.20  \\ 	
		8  &  477.84 &  63.82  &  133.08  &  44.05   \\ 
		16 &  61.50  &  52.06  &  119.75  &  36.34 \medskip \\ 
		\multicolumn{5}{l}{\emph{Experiment 3}}\\
		4  &  233.21  &  58.70  &  143.68  & 47.62   \\ 	
		8  &  245.70  &  66.36  &  128.04  & 38.86   \\ 
		16 &  67.27   &  68.31  &  114.69  & 16.32 \medskip \\ 
		\multicolumn{5}{l}{\emph{Experiment 4}}\\
		4  &  296.31  & 57.79   &  145.92  &  48.22   \\ 	
		8  &  262.01  & 75.38   &  132.81  &  42.40  \\ 
		16 &  100.68  & 107.44  &  124.07  & 35.46   \\ 
		\hline
	\end{tabular}
	
	\label{table:RMSEdmd}
\end{table}

\begin{figure}[ht]
	\centering
	\includegraphics[width=\linewidth]{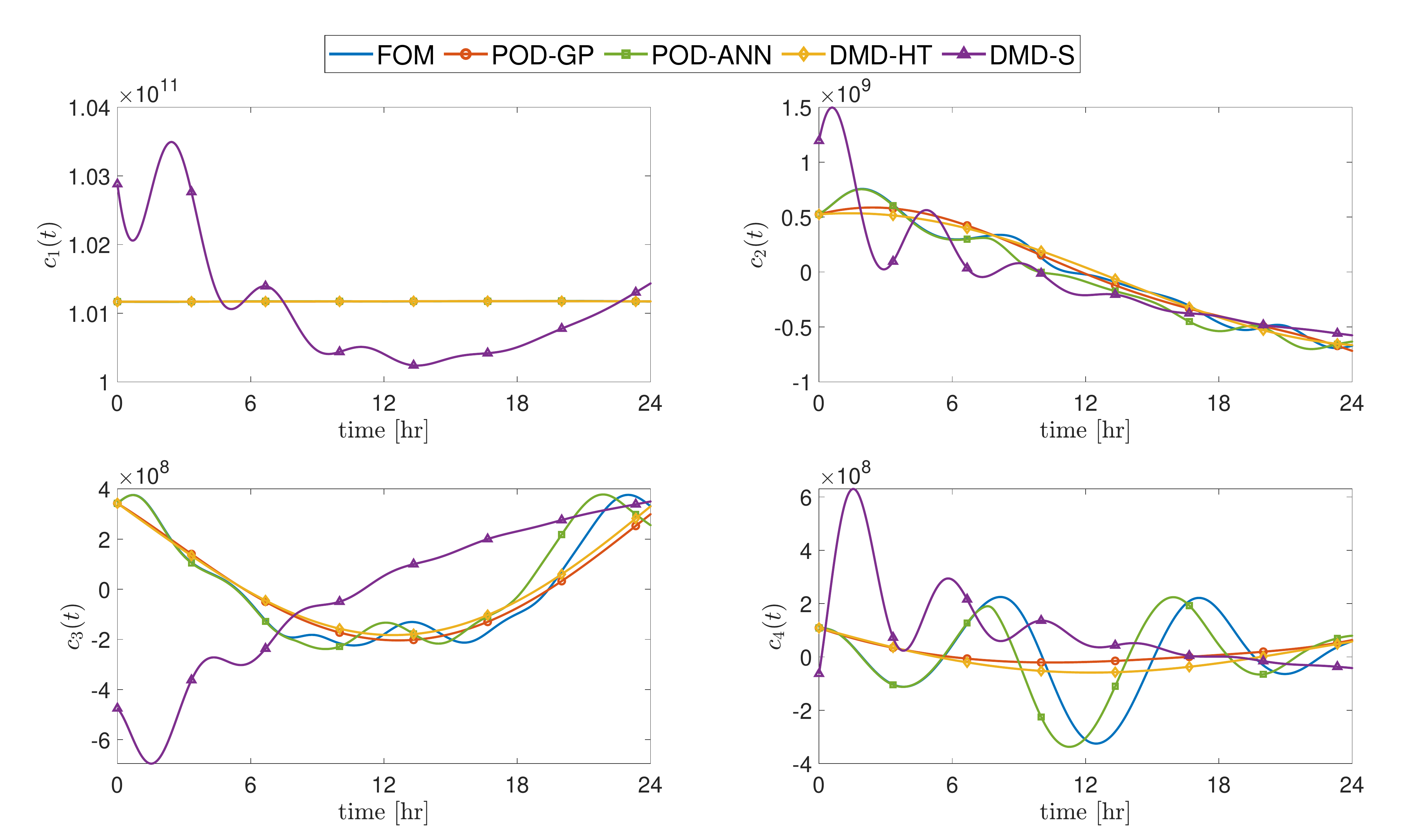}
	\caption{Time series prediction using the first 4 modes from Experiment 1 - obtained without mean subtraction}
	\label{fig:c4_exp1NO}
\end{figure}

\begin{figure}[ht]
	\centering
	\includegraphics[width=\linewidth]{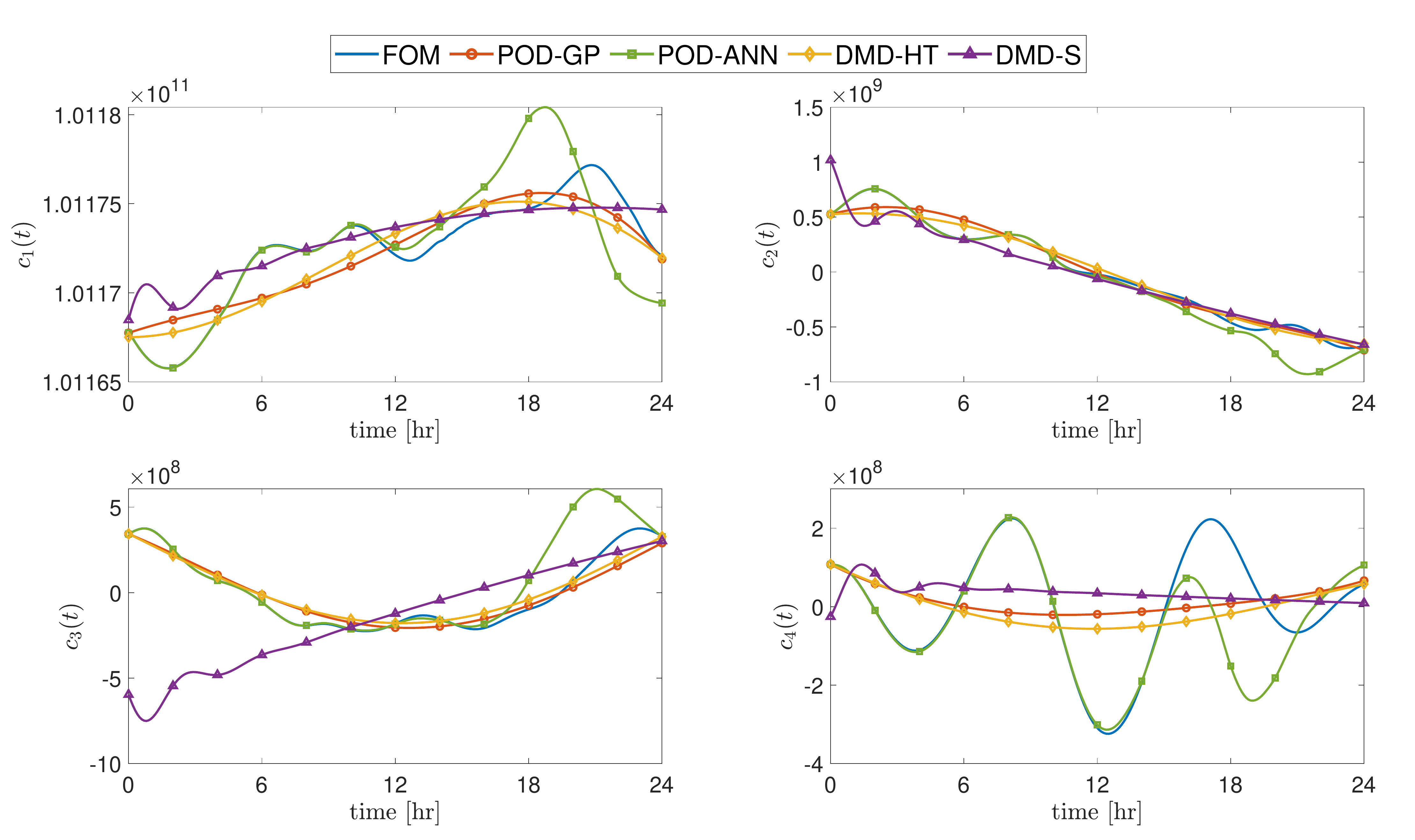}
	\caption{Time series prediction using the first 4 modes from Experiment 2 - obtained without mean subtraction}
	\label{fig:c4_exp2NO}
\end{figure}

\begin{figure}[ht]
	\centering
	\includegraphics[width=\linewidth]{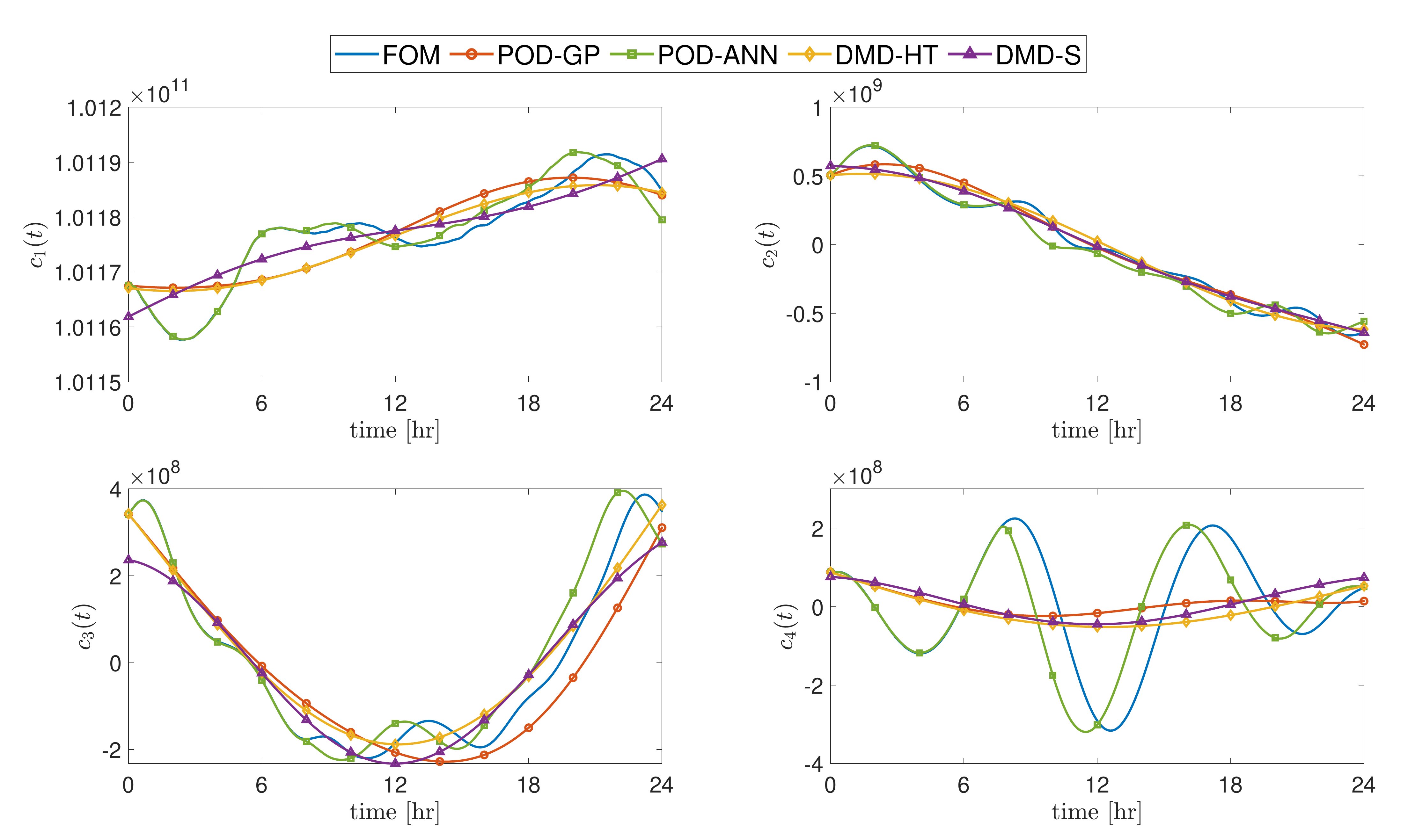}
	\caption{Time series prediction using the first 4 modes from Experiment 3 - obtained without mean subtraction}
	\label{fig:c4_exp3NO}
\end{figure}

\begin{figure}[ht]
	\centering
	\includegraphics[width=\linewidth]{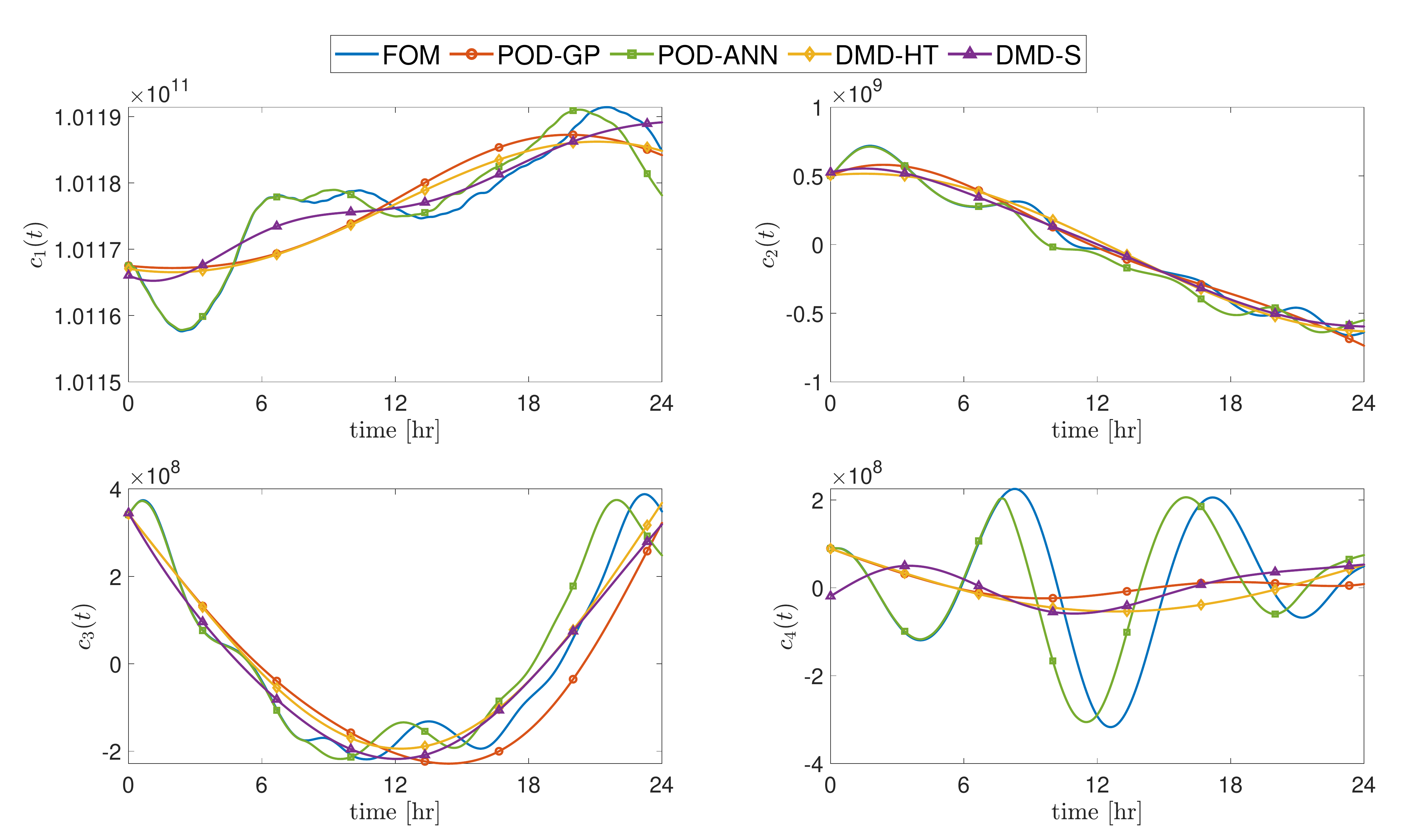}
	\caption{Time series prediction using the first 4 modes from Experiment 4 - obtained without mean subtraction}
	\label{fig:c4_exp4NO}
\end{figure}


\section{Concluding Remarks} \label{sec:conclusion}
In this work, we investigate the influences of training data on different ROM frameworks. We tested two popular energy-based and frequency-based decomposition techniques, namely POD and DMD, respectively. We used four datasets spanning a wide range of resolutions (about $2700$ and $42,600$ grid points) and sampling rates ($240$ and $1440$ snapshots) generated from solving the inviscid 2D SWEs. Different ROM frameworks were studied; POD-GP as a classical intrusive technique, and DMD and POD-ANN as evolving non-intrusive ROMs. We have compared the final results reconstructed in full order space, as geopotential height field using 4, 8, and 16 modes. Furthermore, we investigated the POD modal time series predictions from these frameworks. For comparison purposes, we artificially generated DMD time series by projecting the DMD field predictions onto POD space.

From the aforementioned experiments, we can conclude that DMD is more sensitive to the training datasets, and more dense information (i.e., lower sampling rates and resolutions) is favored. A very high sampling rate results in overfitting problems in approximating Koopman operator. Resolution has negligible effect on DMD convergence, provided that it carries sufficient amount of information. Also, coarse grid can serve as a filter to mitigate overfitting resulting from high sampling rates, especially in DMD-HT algorithm. On the other hand, POD modes have proved to be more robust than DMD ones, showing a very stable and converging behavior for all experiments. 

Moreover, we investigated the effect of centering data through mean-subtraction. We demonstrated that constructing DMD without subtracting the mean geopotential height field provides much better predictions for both DMD-HT and DMD-S algorithms. Therefore, we believe that although centering data in periodic flows might offer a valuable pre-processing step, it might be unfavorable in convective flow situations, like SWE systems. In terms of POD, centering data and applying POD on the anomaly field was found to yield better decomposition of flow fields. However, this could result in slight amplification of noise while solving the ROM from POD-GP.

POD-ANN was shown to provide a valuable non-intrusive ROM, benefiting from the optimality, order, and robustness of POD modes, as well as the mapping and predictive capabilities of artificial neural networks. Unlike DMD, which requires either further treatment and analysis of the system and its invariants to better select the most effective modes or a priori knowledge of essential frequencies to adjust sampling rate, the selection criterion in POD-based frameworks is trivial and automated based on the modal eigenvalues. Table~\ref{tab:compare} gives a summary of the main findings of the present study and provides some insights on sensitivity of decomposition techniques to the training data as well as possible enhancements. Although, we only used ANN in the present study, many other time series prediction tools (e.g., LSTM, Gaussian Processes, etc) can be incorporated in the same framework.

\begin{table}[ht!]
\centering
\caption{Modeling perspectives on investigated approaches}
\begin{tabular}{|p{2.0cm}|p{4.0cm}|p{4.0cm}|p{4.0cm}|}  
\hline
  & Strengths & Weaknesses & Mitigation strategies \\ [0.5ex]  
\hline
\begin{itemize}[leftmargin=*]
\item[] POD-GP
\end{itemize} & 
\begin{itemize}[leftmargin=*]
  \item Bases are orthonormal in space, ordered and optimal in $L_2$ sense
  \item Harness the power of physics and often rely on the underlying equations
  \item Arguably first hybrid physics-data driven approach for complex dynamical systems
  \item Less prone to overfitting due to sampling rate
  \end{itemize}
   & 
  \begin{itemize}[leftmargin=*]
  \item Lack of conservation
  \item Online prediction cost with $O(R^3)$ for fluid dynamics applications (quadratic nonlinearity)
  \item Intrusive and needs access to the physical processes and subcomponents
  \item Its intrinsic global nature adds modal deformation, especially for convective flows
\end{itemize}
   & 
  \begin{itemize}[leftmargin=*]
  \item Many closure modeling ideas have been developed
  \item Online basis adaption is applicable with increasing computational cost
  \item Frequency domain forms of POD are available 
\end{itemize}
\\
\hline
\begin{itemize}[leftmargin=*]
\item[] DMD
\end{itemize}  & 
\begin{itemize}[leftmargin=*]
  \item Fully non-intrusive and does not need knowledge of governing equations
  \item DMD modes are temporally orthogonal (i.e., a frequency based approach)
  \item Projection-free approach
  \item Online prediction is very fast
  \end{itemize}
   & 
  \begin{itemize}[leftmargin=*]
  \item Basis functions are not orthogonal in space
  \item Usually requires more basis functions compared to POD
  \item Highly dependent on the sampling procedure and decay/growth rate
  \item Intrinsically interpolatory, and hard to generalize to parametric systems
  \item Basis functions are not sorted automatically 
\end{itemize}
   & 
  \begin{itemize}[leftmargin=*]
  \item Dynamic streaming approaches are applicable
  \item Additional optimization procedures can be applied to select better representative basis functions
  \item Sparsity promoting, multi-scale, multi-resolution, and recursive approaches are available to improve accuracy and computational efficiency
\end{itemize}
\\
\hline
\begin{itemize}[leftmargin=*]
\item[] POD-ANN
\end{itemize} & 
\begin{itemize}[leftmargin=*]
  \item Fully non-intrusive and does not need knowledge of governing equations
  \item Harnesses the power of emerging machine learning tools
  \item Quite modular, and different time series prediction tools can be applied
  \item Basis functions are orthogonal and optimal
  \item Basis are also sorted trivially
  \item Online prediction is fast
  \end{itemize}
   & 
  \begin{itemize}[leftmargin=*]
  \item Prone to overfitting
  \item Intrinsically interpolatory
  \item Deeper architectures are required for larger number of modes
  \item Works well if there is scale separation (i.e., with higher decay rate in eigenspectrum or lower Kolmogorov-width)
  \item Hyper-parameter tuning is usually required
  \item Requires a black-box integrator
\end{itemize}
   & 
  \begin{itemize}[leftmargin=*]
  \item Partitioning or local approaches can be applied
  \item Gaussian processes can be used to estimate uncertainties
  \item Closure models can be developed for the problems with higher Kolmogorov width (i.e., smaller decay rate in eigenspectrum)
\end{itemize}
\\
\hline

\end{tabular}

\label{tab:compare}
\end{table}

\section*{Acknowledgements}
This material is based upon work supported by the U.S. Department of Energy, Office of Science, Office of Advanced Scientific Computing Research under Award Number DE-SC0019290. 
OS gratefully acknowledges their support. Disclaimer: This report was prepared as an account of work sponsored by an agency of the United States Government. Neither the United States Government nor any agency thereof, nor any of their employees, makes any warranty, express or implied, or assumes any legal liability or responsibility for the accuracy, completeness, or usefulness of any information, apparatus, product, or process disclosed, or represents that its use would not infringe privately owned rights. Reference herein to any specific commercial product, process, or service by trade name, trademark, manufacturer, or otherwise does not necessarily constitute or imply its endorsement, recommendation, or favoring by the United States Government or any agency thereof. The views and opinions of authors expressed herein do not necessarily state or reflect those of the United States Government or any agency thereof.

\bibliographystyle{unsrt} 
\bibliography{ref}   

\end{document}